\pdfoutput=1
\documentclass[twocolumn,journal]{IEEEtran}
\usepackage[T1]{fontenc}
\usepackage[latin9]{inputenc}
\usepackage{float}
\usepackage{textcomp}
\usepackage{mathrsfs}
\usepackage{amsmath}
\usepackage{graphicx}
\usepackage[unicode=true,
 bookmarks=true,bookmarksnumbered=true,bookmarksopen=true,bookmarksopenlevel=1,
 breaklinks=false,pdfborder={0 0 0},pdfborderstyle={},backref=false,colorlinks=false]
 {hyperref}
\hypersetup{pdftitle={Your Title},
 pdfauthor={Your Name},
 pdfpagelayout=OneColumn, pdfnewwindow=true, pdfstartview=XYZ, plainpages=false}

\makeatletter

\providecommand{\tabularnewline}{\\}
\floatstyle{ruled}
\newfloat{algorithm}{tbp}{loa}
\providecommand{\algorithmname}{Algorithm}
\floatname{algorithm}{\protect\algorithmname}

\usepackage[caption=false,font=footnotesize]{subfig}

\makeatother

\begin{document}
\title{Beamforming Optimization for MIMO Wireless Power Transfer with Nonlinear
Energy Harvesting: RF Combining versus DC Combining}
\author{Shanpu~Shen,~\IEEEmembership{Member,~IEEE,} and~Bruno~Clerckx,~\IEEEmembership{Senior~Member,~IEEE}\thanks{Manuscript received; This work was supported in part by the EPSRC
of U.K. under Grant EP/P003885/1 and EP/R511547/1. \textit{(Corresponding
author: Shanpu Shen.)}}\thanks{The authors are with the Department of Electrical and Electronic Engineering,
Imperial College London, London SW7 2AZ, U.K. (e-mail: s.shen@imperial.ac.uk;
b.clerckx@imperial.ac.uk).}}
\maketitle
\begin{abstract}
In this paper, we study the multiple-input and multiple-output (MIMO)
wireless power transfer (WPT) system so as to enhance the output DC
power of the rectennas. To that end, we revisit the rectenna nonlinearity
considering multiple receive antennas. Two combining schemes for multiple
rectennas at the receiver, DC and RF combinings, are modeled and analyzed.
For DC combining, we optimize the transmit beamforming, adaptive to
the channel state information (CSI), so as to maximize the total output
DC power. For RF combining, we compute a closed-form solution of the
optimal transmit and receive beamforming. In addition, we propose
a practical RF combining circuit using RF phase shifter and RF power
combiner and also optimize the analog receive beamforming adaptive
to CSI. We also analytically derive the scaling laws of the output
DC power as a function of the number of transmit and receive antennas.
Those scaling laws confirm the benefits of using multiple antennas
at the transmitter or receiver. They also highlight that RF combining
significantly outperforms DC combining since it leverages the rectenna
nonlinearity more efficiently. Two types of performance evaluations,
based on the nonlinear rectenna model and based on realistic and accurate
rectenna circuit simulations, are provided. The evaluations demonstrate
that the output DC power can be linearly increased by using multiple
rectennas at the receiver and that the relative gain of RF combining
versus DC combining in terms of the output DC power level is very
significant, of the order of 240\% in a one-transmit antenna ten-receive
antenna setup.
\end{abstract}


\section{Introduction}

\IEEEPARstart{T}{HE} Internet of Things (IoT) connects sensors, actuators,
machines, and other objects to Internet so that processes and services
such as manufacturing, monitoring, transportation, and healthcare
can be enhanced \cite{zorzi2010today}. Wireless Sensor Networks (WSN)
and radio frequency identification (RFID) network can be loosely viewed
as possible parts of the IoT. However, the devices of IoT might be
deployed in unreachable or hazard environment such that battery replacement
or recharging becomes inconvenient. Besides, battery replacement or
recharging becomes prohibitive and unsustainable for a large number
of IoT devices. Therefore, powering the devices of IoT in a reliable,
controllable, user-friendly, and cost-effective manner remains a challenging
issue.

Far-field wireless power transfer (WPT) via radio-frequency has a
long history and nowadays it attracts more and more attention as a
promising technology for overcoming this issue. WPT utilizes a dedicated
source to radiate electromagnetic energy through a wireless channel
and a rectifying antenna (rectenna) at the receiver to receive and
convert this energy into DC power. The major challenge of far-field
WPT is to increase the DC power level at the output of the rectenna
without increasing the transmit power, and to power devices located
tens to hundreds of meters away from the transmitter. To overcome
this challenge, the vast majority of the technical efforts in the
literature have been devoted to the design of efficient rectenna \cite{2004TMTT_EH_Popovic}\nocite{2013TMTT_EH_AmbientEH}-\cite{2013_ProcIEEE_RF_WSN}.

Another promising approach to increase the output DC power level is
to design efficient WPT signals \cite{2017_TOC_WPT_YZeng_Bruno_RZhang}.
Interestingly, it was observed through RF measurements in the RF literature
that the RF-to-DC conversion efficiency is a function of the input
waveforms. In \cite{2009_IntConfRFID}, \cite{2011_IMS_Multisine},
a multisine signal excitation is shown through analysis, simulations
and measurements to enhance the DC power and RF-to-DC conversion efficiency
over a single sinewave signal. In \cite{2014_MWCL_WPT_Optimal_Waveform},
various input waveforms (OFDM, white noise, chaotic) are considered
and experiments show that waveforms with high peak to average power
ratio (PAPR) increase RF-to-DC conversion efficiency. However, the
main limitation of those methods is not only the lack of a systematic
approach to design waveforms, but also the fact that they operate
without Channel State Information (CSI) at the Transmitter (CSIT)
and Receiver (CSIR). Inspired by communications, CSI is also very
helpful to WPT to adjust dynamically the transmit signal as a function
of the channel state and the multipath fading. The first systematic
analysis, design and optimization of waveforms for WPT was conducted
in \cite{2016_TSP_WPT_Bruno_Waveform}. Those waveforms are adaptive
to the CSI and jointly exploit a beamforming gain, the frequency-selectivity
of the channel and the rectenna nonlinearity so as to maximize the
amount of harvested DC power. Since then, further enhancements have
been made to waveform optimization adaptive to CSI with the objective
to reduce the complexity of the design and extend to large scale multi-antenna
multi-sine WPT \cite{2017_TSP_WPT_Bruno_Yang_Large}-\nocite{2017_AWPL_WPT_Bruno_LowComplexity}\cite{2017_IEEE_SPAWC_Waveform},
to account for limited feedback \cite{2018_TWC_WPT_Bruno_HYang_LimitedFeedback},
to energize multiple devices (multi-user setting) \cite{2017_TSP_WPT_Bruno_Yang_Large},
\cite{2019_JSAC_Waveform_MultipleDevice}, to transfer information
and power simultaneously \cite{2018_TSP_WIPT_Bruno_WIPT}-\nocite{2019_TOC_SWIPT_nonZeroInput}\cite{2019_TWC_WIPT_AsymmetricModulation},
and enable efficient wireless powered communications \cite{2017_CL_WPBackscatterComm},
\cite{2019_TWC_MuWPScatteredComm}.

In addition to optimizing the rectenna circuit and waveform, the use
of multiple rectennas, also known as multiport rectennas, at the receiver
can increase the output DC power level. In \cite{ShanpuShen2016_TAP_Impedancematching}-\nocite{ShanpuShen2017_AWPL_DPTB}\nocite{ShanpuShen2018_EuCap_QPDP}\nocite{ShanpuShen2017_TAP_EHPIXEL}\cite{ShanpuShen2019_TMTT_Freqdepend},
multiport rectennas have been designed for ambient RF energy harvesting,
which is similar to WPT but does not have a controllable and dedicated
transmitter. It was shown that using multiport rectennas can linearly
increase the output DC power level with the number of rectennas at
the receiver while keeping a compact multiport rectenna size as single
rectenna. Two combinings, DC and RF combinings, for the multiple rectennas
at the receiver has been investigated in \cite{2011AWPL_EH_InvestRectArray}.
However, the investigation is at the level of RF circuit design and
does not consider the impact on communication and signal designs,
including CSI acquisition, as well as adaptive waveform and beamforming
optimization.

In this paper, we utilize multiple rectennas at the receiver for WPT
to increase the output DC power level. Together with the multiple
antennas available at the transmitter, a multiple-input and multiple-output
(MIMO) WPT system is formed. The contributions of the paper are summarized
as follow.

\textit{First}, we analyze the model of MIMO WPT system with the rectenna
nonlinearity. Two combining schemes for multiple rectenna at the receiver,
DC and RF combinings, are modeled and analyzed. This contrasts with
prior works \cite{2016_TSP_WPT_Bruno_Waveform}-\nocite{2017_TSP_WPT_Bruno_Yang_Large}\nocite{2017_AWPL_WPT_Bruno_LowComplexity}\nocite{2017_IEEE_SPAWC_Waveform}\nocite{2018_TWC_WPT_Bruno_HYang_LimitedFeedback}\nocite{2019_JSAC_Waveform_MultipleDevice}\nocite{2018_TSP_WIPT_Bruno_WIPT}\nocite{2019_TOC_SWIPT_nonZeroInput}\nocite{2019_TWC_WIPT_AsymmetricModulation}\nocite{2017_CL_WPBackscatterComm}\cite{2019_TWC_MuWPScatteredComm}
that assumed a single rectenna per device.

\textit{Second}, for DC combining, assuming perfect CSIT can be attained
and making use of the rectenna model, we optimize the beamforming
for multiple antennas at the transmitter in MIMO WPT system. We formulate
an optimization problem to adaptively change the beamforming as a
function of the CSIT so as to maximize the total DC power of all rectenna
outputs. By solving a non-convex posynomial maximization problem with
semi-definite relaxation (SDR), we optimize the beamforming and we
also numerically show that the proposed algorithm finds a stationary
point of the problem for the tested channel realizations.

\textit{Third}, for RF combining, assuming perfect CSIT and CSIR and
making use of the rectenna model, we optimize the beamformers at both
the transmitter and the receiver in the MIMO WPT system. The global
optimal beamforming weights at both the transmitter and the receiver
are obtained in closed form.\textit{ }Additionally, a practical RF
combining circuit using RF phase shifter and RF power combiner is
proposed for the MIMO WPT system. Assuming perfect CSIT and CSIR and
making use of the rectenna model, the analog receive beamforming is
optimized by solving a non-convex optimization problem with SDR. We
numerically show that SDR is very tight and the proposed algorithm
can find nearly the global optimal solution for the tested channel
realizations.

\textit{Fourth}, scaling laws of the output DC power for multiple-input
single-output (MISO) WPT system and single-input multiple-output (SIMO)
with DC and RF combinings are analytically derived as a function of
the number of transmit antennas $M$ and the number of receive antennas
$Q$. Those scaling laws confirm the benefits of using multiple antennas
at the transmitter or receiver and show that different combining schemes
leads to different output DC power. They also highlight that RF combining
significantly outperforms DC combining since the receive beamforming
in RF combining leverages the rectenna nonlinearity more efficiently.

\textit{Fifth}, the beamforming for DC and RF combinings, adaptive
to the CSI and accounting for the rectenna nonlinearity, are shown
through realistic circuit evaluations to boost the output DC power
level. Moreover, the circuit simulations show that RF combining outperforms
DC combining in terms of the output DC power level since the receive
beamforming in RF combining leverages the rectenna nonlinearity more
efficiently.

\textit{Sixth}, the impact of DC combining versus RF combining on
the WPT system design is also discussed and a comprehensive comparison
of DC and RF combinings is provided.

It is also worth contrasting our contributions with the recent MIMO
WPT systems proposed in \cite{2016_TWC_RZhang_MIMOWPT_Limitedfeed},
\cite{2019_SPL_MIMO_WPT}. Our work is different from \cite{2016_TWC_RZhang_MIMOWPT_Limitedfeed}
in several aspects: 1) the nonlinearity of the rectenna is not considered
in \cite{2016_TWC_RZhang_MIMOWPT_Limitedfeed} while this work shows
that exploiting such nonlinearity is important to boost the output
DC power, and 2) only DC combining is considered in \cite{2016_TWC_RZhang_MIMOWPT_Limitedfeed}
while this work considers both DC and RF combinings and shows that
RF combining can boost the output DC power by leveraging the nonlinearity
of the rectenna. On the other hand, our work is different from \cite{2019_SPL_MIMO_WPT}
in several aspects: 1) this work focuses on the low power (e.g. below
mW input power) WPT scenario so that a nonlinear rectenna model based
on Taylor expansion of the diode I-V characteristics is used while
\cite{2019_SPL_MIMO_WPT} focuses on medium and large power WPT scenario
so that a sigmoidal function-based rectenna model which reflects the
property of saturated output dc power is used, 2) the generic architecture
in \cite{2019_SPL_MIMO_WPT} uses many power splitters and power combiners,
which in reality causes high insertion loss and is not suitable for
low power WPT scenario, so this work focuses on the two practical
combining schemes with low complexity for the low power WPT MIMO system,
and 3) this work proposes a more practical RF combining circuit consisting
of phase shifters and RF power combiner and the optimization for the
phase shifts in the practical RF combining is also provided.

In addition, we would like to clarify the differences between the
beamforming design in this work and the waveform design in \cite{2016_TSP_WPT_Bruno_Waveform}.
Both of them are effective approaches to increase the output DC power
in WPT system. However, waveform design focuses on how to allocate
power (with adaptive magnitude and phase) at different frequency tones
while beamforming design focuses on how to allocate power (with adaptive
magnitude and phase) at different transmit antennas and how to combine
power from different receive antennas in RF combining. In other words,
waveform design is considered from the perspective of frequency domain
while beamforming design is from spatial domain. Besides, waveform
design is performed at the transmitter. However, in MIMO WPT system,
beamforming design is not only performed at the transmitter but can
be also performed at the receiver (in the case of RF combining). Interestingly,
it remains a future work to jointly optimize the waveform design and
the beamforming design to further improve the output DC power.

As a takeaway message, this paper again shows the crucial role played
by the rectenna nonlinearity. It is well understood from \cite{2016_TSP_WPT_Bruno_Waveform},
\cite{2018_TSP_WIPT_Bruno_WIPT}, \cite{2019_JSAC_WIPT_Bruno_RZhang_RSchober_DIKim_HVPoor}
that nonlinearity favors a different waveform, modulation, input distribution
and transceiver architecture as well as a different use of the RF
spectrum in WPT, and wireless information and power transfer (WIPT).
What this paper further highlights is that nonlinearity also changes
how to make use of multiple receive antennas in WPT, and therefore
how to design the corresponding beamformers. As shown in this paper,
if we ignore the nonlinearity of the rectenna and assume the (inaccurate)
linear model of the rectenna \cite{2016_TWC_RZhang_MIMOWPT_Limitedfeed},
\cite{2013_TWC_SWIPT_RZhang}, DC combining and RF combining would
lead to the same output DC power.

\textit{Organization}: Section II introduces the MIMO WPT system model
and Section III briefly revisits the rectenna models. Section IV and
Section V tackle the beamforming optimization for DC and RF combinings,
respectively. Section VI demonstrates the beneficial role of the rectenna
nonlinearity. Section VII analytically derives the scaling laws for
the MIMO WPT system. Section VIII evaluates the performance and Section
IX discusses the impact of DC and RF combinings on the WPT system
design. Section X concludes the work.

\textit{Notations}: Bold lower and upper case letters stand for vectors
and matrices, respectively. A symbol not in bold font represents a
scalar. $\mathscr{\mathcal{E}}\left\{ \cdot\right\} $ refers to the
expectation/averaging operator. $\Re\left\{ x\right\} $, $\Im\left\{ x\right\} $,
and $\left|x\right|$ refer to the real part, imaginary part, and
modulus of a complex number $x$. $\left\Vert \mathbf{x}\right\Vert $
and $\left\Vert \mathbf{x}\right\Vert _{1}$ refer to the $l_{2}$-norm
and $l_{1}$-norm of a vector $\mathbf{x}$, respectively. $\arg\left(\mathbf{x}\right)$
refers to a vector with each element being the phase of the corresponding
element in a vector $\mathbf{x}$. $\mathbf{X}^{T}$, $\mathbf{X}^{H}$,
$\left[\mathbf{X}\right]_{ij}$, $\mathrm{Tr}\left(\mathbf{X}\right)$,
and $\mathrm{rank}\left(\mathbf{X}\right)$ refer to the transpose,
conjugate transpose, $\left(i,j\right)$th element, trace, and rank
of a matrix $\mathbf{X}$, respectively. $\mathbf{X}\succeq0$ means
that $\mathbf{X}$ is positive semi-definite. $\chi_{k}^{2}$ denotes
the chi-square distribution with $k$ degrees of freedom. $\mathcal{CN}\left(\mathbf{0},\mathbf{\Sigma}\right)$
denotes the distribution of a circularly symmetric complex Gaussian
(CSCG) random vector with mean vector $\mathbf{0}$ and covariance
matrix $\mathbf{\Sigma}$ and $\sim$ stands for \textquotedblleft distributed
as\textquotedblright . $\mathbf{I}$ and $\mathbf{0}$ denote an identity
matrix and an all-zero vector, respectively. log is in base $e$.

\section{MIMO WPT System Model}

We consider a point-to-point MIMO WPT system. There are $M$ antennas
at the transmitter and $Q$ antennas at the receiver. The transmitted
signal at time $t$ on the $m$th transmit antenna can be expressed
by 
\begin{equation}
x_{m}\left(t\right)=\Re\left\{ w_{T,m}e^{j\omega_{c}t}\right\} =s_{T,m}\cos\left(\omega_{c}t+\phi_{T,m}\right),
\end{equation}
where $\omega_{c}$ denotes the center frequency, $w_{T,m}=s_{T,m}e^{j\phi_{T,m}}$
denotes the complex weight with $s_{T,m}$ and $\phi_{T,m}$ referring
to the amplitude and phase of the signal on the $m$th transmit antenna,
and we take the real part of $w_{T,m}e^{j\omega_{c}t}$ to convert
a phasor into a sinusoidal function of time $t$. Stacking up all
transmit signals, we can write the transmit signal vector as 
\begin{equation}
\mathbf{x}\left(t\right)=\Re\left\{ \mathbf{w}_{T}e^{j\omega_{c}t}\right\} ,
\end{equation}
where the complex weights are collected into a vector $\mathbf{w}_{T}=\left[w_{T,1},w_{T,2},...,w_{T,M}\right]^{T}$.
The transmitter is subject to a transmit power constraint given by
\begin{equation}
\frac{1}{2}\left\Vert \mathbf{w}_{T}\right\Vert ^{2}\leq P,
\end{equation}
where $P$ denotes the transmit power and the scaler $\frac{1}{2}$
is due to that the average power of the sinusoidal function $x_{m}\left(t\right)$
is $\frac{1}{2}\left|w_{T,m}\right|^{2}$.

The signals transmitted by the multiple transmit antennas propagate
through a wireless channel. The received signal at the $q$-th receive
antenna can be expressed as

\begin{align}
y_{q}\left(t\right) & =\Re\left\{ \mathbf{h}_{q}\mathbf{w}_{T}e^{j\omega_{c}t}\right\} ,
\end{align}
where $\mathbf{h}_{q}=\left[h_{q1},h_{q2},...,h_{qM}\right]$ denotes
the channel vector (a row vector) for the $q$-th receive antenna
with $h_{qm}$ referring to the complex channel gain between the $m$th
transmit antenna and the $q$th receive antenna. We collect all $\mathbf{h}_{q}$
into a matrix $\mathbf{H}=\left[\mathbf{h}_{1}^{T},\mathbf{h}_{2}^{T},...,\mathbf{h}_{Q}^{T}\right]^{T}$
where $\mathbf{H}$ represents the $Q\times M$ channel matrix of
the MIMO WPT system. We assume that the channel matrix $\mathbf{H}$
is perfectly known to the transmitter.

\section{Rectenna Model}

We briefly revisit two simple and tractable models of the rectenna
circuit derived in the past literatures \cite{2016_TSP_WPT_Bruno_Waveform},
\cite{2017_TSP_WPT_Bruno_Yang_Large}. The goal of describing two
different models is to emphasize the rectenna nonlinearity in WPT
systems. Those two models account for the rectenna nonlinearity through
the higher order terms in the Taylor expansion of the diode I-V characteristics
while having a simple and tractable expression. Interestingly, those
two models are shown to be equivalent in terms of optimization even
though they rely on different physical assumptions \cite{2017_TSP_WPT_Bruno_Yang_Large}.

\begin{figure}[t]
\begin{centering}
\includegraphics[scale=0.18]{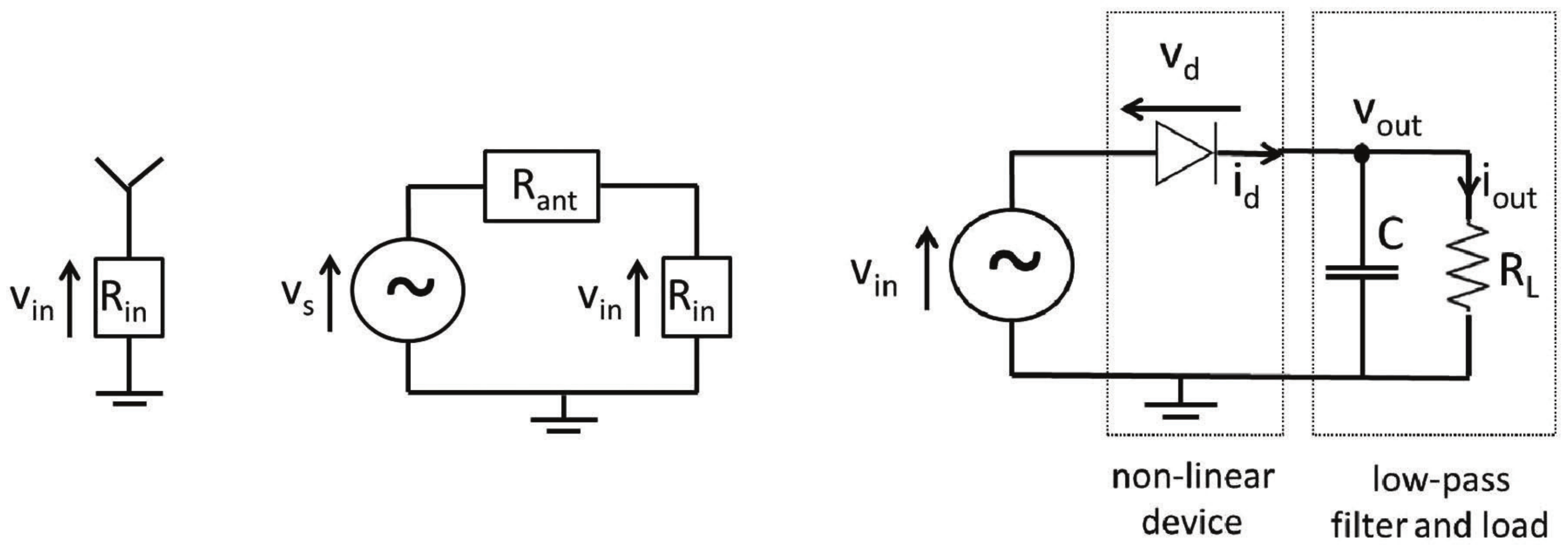}
\par\end{centering}
\caption{\label{fig:Antenna-equivalent-circuit}Antenna equivalent circuit
(left) and a single diode rectifier (right).}
\end{figure}

Consider a rectifier with input impedance $R_{\mathrm{in}}$ connected
to a receive antenna as shown in Fig. \ref{fig:Antenna-equivalent-circuit}.
The signal $y\left(t\right)$ impinging on the antenna has an average
power $P_{\mathrm{av}}=\mathscr{\mathcal{E}}\left\{ y\left(t\right)^{2}\right\} $.
The receive antenna is assumed lossless and modeled as an equivalent
voltage source $v_{s}\left(t\right)$ in series with an impedance
$R_{\mathrm{ant}}=50\,\Omega$ as shown in Fig. \ref{fig:Antenna-equivalent-circuit}.
With perfect matching ($R_{\mathrm{in}}=R_{\mathrm{ant}}$), the input
voltage of the rectifier $v_{\mathrm{in}}\left(t\right)$ can be related
to the received signal $y\left(t\right)$ by $v_{\mathrm{in}}\left(t\right)=y\left(t\right)\sqrt{R_{\mathrm{ant}}}$.

A rectifier is always made of a nonlinear rectifying component such
as diode followed by a low pass filter with load \cite{2004TMTT_EH_Popovic},
\cite{2013TMTT_EH_AmbientEH}, \cite{ShanpuShen2017_AWPL_DPTB}-\nocite{ShanpuShen2018_EuCap_QPDP}\nocite{ShanpuShen2017_TAP_EHPIXEL}\cite{ShanpuShen2019_TMTT_Freqdepend}
as shown in Fig. \ref{fig:Antenna-equivalent-circuit}. The current
$i_{d}\left(t\right)$ flowing through an ideal diode (neglecting
its series resistance) relates to the voltage drop across the diode
$v_{d}\left(t\right)$ = $v_{\mathrm{in}}\left(t\right)$ \textminus{}
$v_{\mathrm{out}}\left(t\right)$ as $i_{d}\left(t\right)=i_{s}\left(e^{\frac{v_{d}\left(t\right)}{nv_{t}}}-1\right)$
where $i_{s}$ is the reverse bias saturation current, $v_{t}$ is
the thermal voltage, $n$ is the ideality factor (assumed equal to
1.05). Based on the Taylor expansion of the diode I-V characteristics
and some physical assumptions, the analysis of the rectifier circuit
is simplified and therefore two simple and tractable models are provided
as shown in the following subsections.

\subsection{Current Model}

In \cite{2016_TSP_WPT_Bruno_Waveform}, by taking the Taylor expansion
of the diode I-V characteristics around the negative of the output
DC voltage $-v_{\mathrm{out}}$, the output DC current of the rectifier
$i_{\mathrm{out}}$ is approximated as

\begin{equation}
i_{\mathrm{out}}\approx k_{0}\left(i_{\mathrm{out}}\right)+\sum_{i\,\mathrm{even},\,i\geq2}^{n_{0}}k_{i}\left(i_{\mathrm{out}}\right)R_{\mathrm{ant}}^{i/2}\mathscr{\mathcal{E}}\left\{ y\left(t\right)^{i}\right\} ,
\end{equation}
where the Taylor expansion is truncated to the $n_{0}$th order term.
It is shown that the maximization of $i_{\mathrm{out}}$ is equivalent
to maximizing the quantity

\begin{equation}
z_{\mathrm{DC}}=\sum_{i\,\mathrm{even},\,i\geq2}^{n_{0}}\kappa_{i}\mathscr{\mathcal{E}}\left\{ y\left(t\right)^{i}\right\} ,\label{eq:zdc}
\end{equation}
where $\kappa_{i}=\frac{i_{s}R_{\mathrm{ant}}^{i/2}}{i!\left(nv_{t}\right)^{i}}$
and $n_{0}=4$ is a good choice \cite{2016_TSP_WPT_Bruno_Waveform}.

\subsection{Voltage Model}

In \cite{2017_TSP_WPT_Bruno_Yang_Large}, by assuming zero output
DC current and taking Taylor expansion at zero quiescent point to
the $n_{0}$th-order term, the output DC voltage of the rectifier
$v_{\mathrm{out}}$ is approximated as

\begin{align}
v_{\mathrm{out}} & =\sum_{i\,\mathrm{even},\,i\geq2}^{n_{0}}\beta_{i}\mathscr{\mathcal{E}}\left\{ y\left(t\right)^{i}\right\} ,\label{eq:VoutRectennaModel}
\end{align}
where $\beta_{i}=\frac{R_{\mathrm{ant}}^{i/2}}{i!\left(nv_{t}\right)^{i-1}}$
and $n_{0}=4$ is a good choice \cite{2017_TSP_WPT_Bruno_Yang_Large}.
Multiplying $v_{\mathrm{out}}$ by $\frac{i_{s}}{nv_{t}}$ achieves
the same model $z_{\mathrm{DC}}$ (i.e. $\kappa_{i}=\frac{i_{s}}{nv_{t}}\cdot\beta_{i}$)
in \eqref{eq:zdc} so those two models are equivalent. In the following
Sections, we mainly use the voltage model since it has a straightforward
physical meaning. It should be noted that this voltage model is derived
based on some simplifications and assumptions (detailed in \cite{2017_TSP_WPT_Bruno_Yang_Large})
so that it can characterize in a simple and tractable manner the dependence
of the rectenna nonlinearity on the input signal properties. However,
this does not mean that the model in \eqref{eq:VoutRectennaModel}
is accurate enough to predict the rectifier output DC power using
$P_{\mathrm{out}}=v_{\mathrm{out}}^{2}/R_{L}$ where $R_{L}$ refers
to the load resistance. Nevertheless, the model and its benefits in
deriving optimized signals have been validated by circuit simulations
in \cite{2016_TSP_WPT_Bruno_Waveform}, \cite{2017_AWPL_WPT_Bruno_LowComplexity},
\cite{2018_TWC_WPT_Bruno_Transmit_Diversity} and experimentally in
\cite{2018_TWC_WPT_Bruno_Transmit_Diversity}, \cite{2019_Junghoon_Prototyping}.

\section{Beamforming Optimization with DC Combining}

\begin{figure}[t]
\begin{centering}
\includegraphics[scale=0.64]{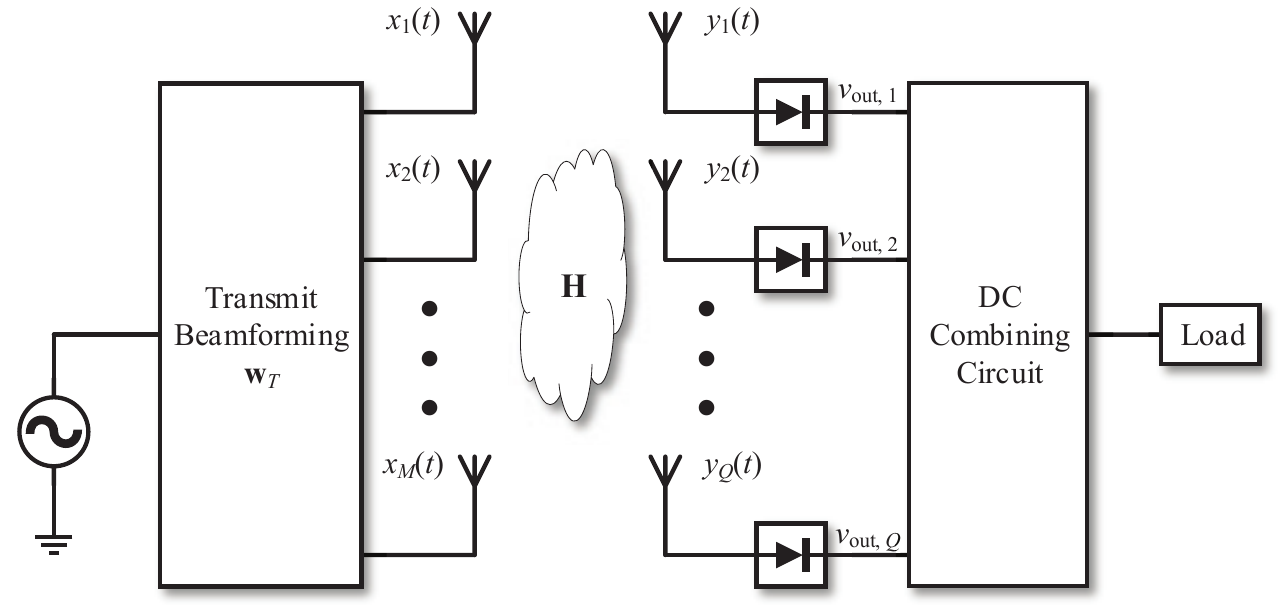}
\par\end{centering}
\caption{\label{fig:DC combinging}Schematic of the MIMO WPT system with DC
combining in the receiver.}
\end{figure}

Consider the DC combining scheme for the multiple receive antenna
system as shown in Fig. \ref{fig:DC combinging}. Each receive antenna
is connected to a rectifier so that the RF signal received by each
antenna is individually rectified. Using the aforementioned voltage
model of the nonlinear rectenna, the output DC voltage of the $q$th
rectifier (connected to the $q$th receive antenna) is given by

\begin{align}
v_{\mathrm{out},q} & =\sum_{i\,\mathrm{even},\,i\geq2}^{n_{0}}\beta_{i}\mathscr{\mathcal{E}}\left\{ y_{q}\left(t\right)^{i}\right\} ,
\end{align}
where $\mathscr{\mathcal{E}}\left\{ y_{q}\left(t\right)^{i}\right\} $
is given by

\begin{align}
\mathscr{\mathcal{E}}\left\{ y_{q}\left(t\right)^{i}\right\}  & =\zeta_{i}\left|\mathbf{h}_{q}\mathbf{w}_{T}\right|^{i},\label{eq:yq(t)constant}
\end{align}
with $\zeta_{i}=\frac{1}{2\pi}\int_{0}^{2\pi}\sin^{i}t\:\textrm{d}t$
and specifically $\zeta_{2}=\frac{1}{2}$, $\zeta_{4}=\frac{3}{8}$,
and $\zeta_{6}=\frac{5}{16}$.

The output DC power of all rectifiers are combined together by a DC
combining circuit such as multiple-input and multiple-output (MIMO)
switching DC-DC converter \cite{KKKKI}-\nocite{MIMO_Convertor_ISSCC}\cite{MIMO_Convertor_JSSC}
as shown in Fig. \ref{fig:DC combinging}. The total output DC power
is then given by $P_{\mathrm{out}}=\sum_{q=1}^{Q}v_{\mathrm{out},q}^{2}/R_{L}$
where we assume each rectifier has the same load $R_{L}$. Therefore,
we aim to maximize the total output DC power subject to the transmit
power constraint, which can be formulated as

\begin{align}
\underset{\mathbf{w}_{T}}{\mathrm{max}}\;\;\; & \sum_{q=1}^{Q}\frac{v_{\mathrm{out},q}^{2}}{R_{L}}\label{eq:OP1-MaxPoutStP-1}\\
\mathrm{s.t.}\;\;\; & \frac{1}{2}\left\Vert \mathbf{w}_{T}\right\Vert ^{2}\leq P.\label{eq:OP1-MaxPoutStP-2}
\end{align}
Observing the expression of the objective function $P_{\mathrm{out}}$,
we find that it is hard to determine whether $P_{\mathrm{out}}$ monotonically
increases/decreases with $\Re\left\{ w_{T,m}\right\} $ or $\Im\left\{ w_{T,m}\right\} $.
Hence, we introduce the auxiliary variables $r_{q}=\left|\mathbf{h}_{q}\mathbf{w}_{T}\right|^{2}$
where $r_{q}>0$ and we equivalently rewrite the total output DC power
as 
\begin{equation}
P_{\mathrm{out}}=\frac{1}{R_{L}}\sum_{q=1}^{Q}\left(\sum_{i\,\mathrm{even},\,i\geq2}^{n_{0}}\beta_{i}\zeta_{i}r_{q}^{\frac{i}{2}}\right)^{2},\label{eq:Poutisposynomial}
\end{equation}
which is a posynomial so that we express it in a compact form that
$P_{\mathrm{out}}=\sum_{k=1}^{K}g_{k}\left(\mathbf{r}\right)$ where
$K$ is the number of monomials in the posynomial and $g_{k}\left(\mathbf{r}\right)=\rho_{k}r_{1}^{\xi_{1k}}r_{2}^{\xi_{2k}}\cdots r_{Q}^{\xi_{Qk}}$
is the $k$th monomial with $\mathbf{r}=\left[r_{1},r_{2},...r_{Q}\right]^{T}$
where $\xi_{1k}$, ..., $\xi_{Qk}$, and $\rho_{k}$ are constants
and $\rho_{k}>0$. When $n_{0}=4$ (which is considered in Section
VIII), we have $P_{\mathrm{out}}=\frac{1}{R_{L}}\sum_{q=1}^{Q}\left(\beta_{2}^{2}\zeta_{2}^{2}r_{q}^{2}+2\beta_{2}\beta_{4}\zeta_{2}\zeta_{4}r_{q}^{3}+\beta_{4}^{2}\zeta_{4}^{2}r_{q}^{4}\right)$
and there are $K=3Q$ monomials. Therefore, we equivalently rewrite
the problem \eqref{eq:OP1-MaxPoutStP-1}-\eqref{eq:OP1-MaxPoutStP-2}
as

\begin{align}
\underset{\mathbf{w}_{T},r_{q}}{\mathrm{max}}\;\;\; & \sum_{k=1}^{K}g_{k}\left(\mathbf{r}\right)\label{eq:OP2-Maxr2r3r4StP-1}\\
\mathrm{s.t.}\;\;\;\; & \frac{1}{2}\left\Vert \mathbf{w}_{T}\right\Vert ^{2}\leq P,\label{eq:OP2-Maxr2r3r4StP-2}\\
 & r_{q}=\left|\mathbf{h}_{q}\mathbf{w}_{T}\right|^{2},\:1\leq q\leq Q.\label{eq:OP2-Maxr2r3r4StP-3}
\end{align}
We find that the objective function \eqref{eq:OP2-Maxr2r3r4StP-1}
monotonically increases with $r_{q}$. Hence, the constraint $r_{q}=\left|\mathbf{h}_{q}\mathbf{w}_{T}\right|^{2}$
can be replaced with $r_{q}\leq\left|\mathbf{h}_{q}\mathbf{w}_{T}\right|^{2}$
without affecting the optimal solution of the problem \eqref{eq:OP2-Maxr2r3r4StP-1}-\eqref{eq:OP2-Maxr2r3r4StP-3}.
However, $r_{q}\leq\left|\mathbf{h}_{q}\mathbf{w}_{T}\right|^{2}$
is still a non-convex constraint so we use SDR to transform it to
a convex constraint \cite{2010_SPMag_SDR}. By introducing an auxiliary
positive semi-definite matrix variable $\mathbf{W}_{T}=\mathbf{w}_{T}\mathbf{w}_{T}^{H}$,
we equivalently rewrite the problem \eqref{eq:OP2-Maxr2r3r4StP-1}-\eqref{eq:OP2-Maxr2r3r4StP-3}
as

\begin{align}
\underset{\mathbf{W}_{T},r_{q}}{\mathrm{max}}\;\;\; & \sum_{k=1}^{K}g_{k}\left(\mathbf{r}\right)\label{eq:OP3-Maxr2r3r4StP-SDR-1}\\
\mathrm{s.t.}\;\;\;\; & \mathrm{Tr}\left(\mathbf{W}_{T}\right)\leq2P,\label{eq:OP3-Maxr2r3r4StP-SDR-2}\\
 & r_{q}-\mathrm{Tr}\left(\mathbf{h}_{q}^{H}\mathbf{h}_{q}\mathbf{W}_{T}\right)\leq0,\:1\leq q\leq Q,\label{eq:OP3-Maxr2r3r4StP-SDR-3}\\
 & \mathbf{W}_{T}\succeq0,\label{eq:OP3-Maxr2r3r4StP-SDR-4}\\
 & \mathrm{rank}\left(\mathbf{W}_{T}\right)=1.\label{eq:OP3-Maxr2r3r4StP-SDR-5}
\end{align}

We use SDR to relax the rank-1 constraint \eqref{eq:OP3-Maxr2r3r4StP-SDR-5},
but the relaxed problem \eqref{eq:OP3-Maxr2r3r4StP-SDR-1}-\eqref{eq:OP3-Maxr2r3r4StP-SDR-4}
is still a non-convex optimization problem. Therefore, we introduce
an auxiliary variable $t_{0}>0$ and equivalently rewrite the relaxed
problem \eqref{eq:OP3-Maxr2r3r4StP-SDR-1}-\eqref{eq:OP3-Maxr2r3r4StP-SDR-4}
as

\begin{align}
\underset{\mathbf{W}_{T},r_{q},t_{0}}{\mathsf{\mathrm{min}}}\;\;\; & \frac{1}{t_{0}}\label{eq:OP4-Min1t0-SDR-1}\\
\mathrm{s.t.}\;\;\;\;\;\; & \mathrm{Tr}\left(\mathbf{W}_{T}\right)\leq2P,\label{eq:OP4-Min1t0-SDR-2}\\
 & r_{q}-\mathrm{Tr}\left(\mathbf{h}_{q}^{H}\mathbf{h}_{q}\mathbf{W}_{T}\right)\leq0,\:1\leq q\leq Q,\label{eq:OP4-Min1t0-SDR-3}\\
 & \mathbf{W}_{T}\succeq0,\label{eq:OP4-Min1t0-SDR-4}\\
 & \frac{t_{0}}{\sum_{k=1}^{K}g_{k}\left(\mathbf{r}\right)}\leq1.\label{eq:OP4-Min1t0-SDR-5}
\end{align}
However, $1/\sum_{k=1}^{K}g_{k}\left(\mathbf{r}\right)$ is not a
posynomial which prevents the transformation to a convex constraint.
Therefore, the idea is to upper bound $1/\sum_{k=1}^{K}g_{k}\left(\mathbf{r}\right)$
by a monomial. The choice of the upper bound relies on the fact that
an arithmetic mean (AM) is greater or equal to the geometric mean
(GM). Hence, we have that

\begin{equation}
\frac{1}{\sum_{k=1}^{K}g_{k}\left(\mathbf{r}\right)}\leq\frac{1}{\prod{}_{k=1}^{K}\left(\frac{g_{k}\left(\mathbf{r}\right)}{\gamma_{k}}\right)^{\gamma_{k}}},\label{eq:AM-GM-bound}
\end{equation}
where $\gamma_{k}\geq0$ and $\sum_{k=1}^{K}\gamma_{k}=1$. We replace
the constraint \eqref{eq:OP4-Min1t0-SDR-5} with $t_{0}\prod{}_{k=1}^{K}\left(\frac{g_{k}\left(\mathbf{r}\right)}{\gamma_{k}}\right)^{-\gamma_{k}}\leq1$
in a conservative way. For a given choice of $\left\{ \gamma_{k}\right\} $,
the problem \eqref{eq:OP4-Min1t0-SDR-1}-\eqref{eq:OP4-Min1t0-SDR-5}
is now replaced by

\begin{align}
\underset{\mathbf{W}_{T},r_{q},t_{0}}{\mathsf{\mathrm{min}}}\;\;\; & \frac{1}{t_{0}}\label{eq:OP5-Min1t0-SDR-GP-1}\\
\mathsf{\mathrm{s.t.}}\;\;\;\;\;\; & \mathrm{Tr}\left(\mathbf{W}_{T}\right)\leq2P,\label{eq:OP5-Min1t0-SDR-GP-2}\\
 & r_{q}-\mathrm{Tr}\left(\mathbf{h}_{q}^{H}\mathbf{h}_{q}\mathbf{W}_{T}\right)\leq0,\:1\leq q\leq Q,\label{eq:OP5-Min1t0-SDR-GP-3}\\
 & \mathbf{W}_{T}\succeq0,\label{eq:OP5-Min1t0-SDR-GP-4}\\
 & t_{0}\prod{}_{k=1}^{K}\left(\frac{g_{k}\left(\mathbf{r}\right)}{\gamma_{k}}\right)^{-\gamma_{k}}\leq1,\label{eq:OP5-Min1t0-SDR-GP-5}
\end{align}
which looks like a GP problem but actually it is not a standard GP
problem because the left sides of the constraints \eqref{eq:OP5-Min1t0-SDR-GP-2},
\eqref{eq:OP5-Min1t0-SDR-GP-3}, and \eqref{eq:OP5-Min1t0-SDR-GP-4}
are not posynomials. To see the details, we rewrite the monomial term
as $t_{0}\prod{}_{k=1}^{K}\left(\frac{g_{k}\left(\mathbf{r}\right)}{\gamma_{k}}\right)^{-\gamma_{k}}=c_{1}t_{0}r_{1}^{\alpha_{1}}r_{2}^{\alpha_{2}}\cdots r_{Q}^{\alpha_{Q}}$
where $c_{1}=\prod{}_{k=1}^{K}\left(\frac{\rho_{k}}{\gamma_{k}}\right)^{-\gamma_{k}}$
and $\alpha_{q}=-\sum{}_{k=1}^{K}\xi_{qk}\gamma_{k}$ for $q=1,...,Q$.
We introduce auxiliary variables $\tilde{t}_{0}=\log t_{0}$ and $\tilde{r}_{q}=\log r_{q}$
(so that $e^{\tilde{t}_{0}}=t_{0}$ and $e^{\tilde{r}_{q}}=r_{q}$
). We use the logarithmic transformation for the objective function
\eqref{eq:OP5-Min1t0-SDR-GP-1} and the constraints \eqref{eq:OP5-Min1t0-SDR-GP-3}
and \eqref{eq:OP5-Min1t0-SDR-GP-5} so we can equivalently rewrite
the problem \eqref{eq:OP5-Min1t0-SDR-GP-1}-\eqref{eq:OP5-Min1t0-SDR-GP-5}
as
\begin{align}
\underset{\mathbf{W}_{T},\tilde{r}_{q},\tilde{t}_{0}}{\mathsf{\mathrm{min}}}\;\;\; & -\tilde{t}_{0}\label{eq:OP6-Min-t0-log-1}\\
\mathsf{\mathrm{s.t.}}\;\;\;\;\;\; & \mathrm{Tr}\left(\mathbf{W}_{T}\right)\leq2P,\label{eq:OP6-Min-t0-log-2}\\
 & e^{\tilde{r}_{q}}-\mathrm{Tr}\left(\mathbf{h}_{q}^{H}\mathbf{h}_{q}\mathbf{W}_{T}\right)\leq0,\:1\leq q\leq Q,\label{eq:OP6-Min-t0-log-3}\\
 & \mathbf{W}_{T}\succeq0,\label{eq:OP6-Min-t0-log-4}\\
 & \log c_{1}+\tilde{t}_{0}+\sum_{q=1}^{Q}\alpha_{q}\tilde{r}_{q}\leq0,\label{eq:OP6-Min-t0-log-5}
\end{align}
which is a convex problem that can be solved within polynomial time
by CVX \cite{grant2008cvx} which adopts interior point method.

Note that the tightness of the upper bound \eqref{eq:AM-GM-bound}
heavily depends on the choice of $\left\{ \gamma_{k}\right\} $. Following
\cite{chiang2005geometric_programming}, \cite{Applied_Geometric_Programming},
an iterative procedure can be used to tighten the bound, where at
each iteration the problem \eqref{eq:OP5-Min1t0-SDR-GP-1}-\eqref{eq:OP5-Min1t0-SDR-GP-5}
is solved for an updated set of $\left\{ \gamma_{k}\right\} $. Assuming
a feasible $\mathbf{r}^{\left(i-1\right)}$ at iteration $i-1$, compute
$\gamma_{k}=g_{k}\left(\mathbf{r}^{\left(i-1\right)}\right)/\sum_{k=1}^{K}g_{k}\left(\mathbf{r}^{\left(i-1\right)}\right)\:\forall k$
at iteration $i$ and solve problem \eqref{eq:OP5-Min1t0-SDR-GP-1}-\eqref{eq:OP5-Min1t0-SDR-GP-5}
to obtain $\mathbf{r}^{\left(i\right)}$. Repeat the iterations till
convergence. Note that the successive approximation method that we
used is also known as a successive convex approximation or inner approximation
method \cite{OperRes_1978_InnerApprox}. It cannot guarantee to converge
to the global optimal solution of the problem \eqref{eq:OP4-Min1t0-SDR-1}-\eqref{eq:OP4-Min1t0-SDR-5},
but it converges to a stationary point \cite{OperRes_1978_InnerApprox},
\cite{2007_TWC_GP_Daniel} which is denoted as $\mathbf{W}_{T}^{\star}$.
Due to the equivalence, $\mathbf{W}_{T}^{\star}$ is also a stationary
point of the problem \eqref{eq:OP3-Maxr2r3r4StP-SDR-1}-\eqref{eq:OP3-Maxr2r3r4StP-SDR-4}.

If $\mathrm{rank}\left(\mathbf{W}_{T}^{\star}\right)=1$, the SDR
is tight so that $\mathbf{W}_{T}^{\star}$ is a stationary point of
the problem \eqref{eq:OP3-Maxr2r3r4StP-SDR-1}-\eqref{eq:OP3-Maxr2r3r4StP-SDR-5}.
Using eigenvalue decomposition (EVD), we obtain $\mathbf{W}_{T}^{\star}=\mathbf{w}_{T}^{\star}\mathbf{w}_{T}^{\star H}$
and therefore $\mathbf{w}_{T}^{\star}$ is a stationary point of the
problem \eqref{eq:OP1-MaxPoutStP-1}-\eqref{eq:OP1-MaxPoutStP-2}.
Otherwise, for the case of $\mathrm{rank}\left(\mathbf{W}_{T}^{\star}\right)>1$,
we aim to extract a suboptimal rank-1 solution from $\mathbf{W}_{T}^{\star}$.
A commonly adopted approach is the so-called Gaussian randomization
method \cite{2010_SPMag_SDR}. Particularly, we first use the EVD
to decompose $\mathbf{W}_{T}^{\star}=\mathbf{U}_{T}\mathbf{\Sigma}_{T}\mathbf{U}_{T}^{H}$
where $\mathbf{U}_{T}$ and $\mathbf{\Sigma}_{T}$ are $M\times M$
unitary and diagonal matrices, respectively. Then, we generate $M$-dimensional
random vectors $\mathbf{n}^{\left(l\right)}\sim\mathcal{CN}\left(\mathbf{0},\mathbf{I}\right)$
($l=1,\ldots,L$), multiply \textbf{$\mathbf{n}^{\left(l\right)}$}
by $\mathbf{U}_{T}\mathbf{\Sigma}_{T}^{1/2}$, and scale the norm
of $\mathbf{U}_{T}\mathbf{\Sigma}_{T}^{1/2}\mathbf{n}^{\left(l\right)}$
to $\sqrt{2P}$ ($P$ refers to the transmit power), so that we obtain
$\mathbf{w}_{T}^{\left(l\right)}=\sqrt{2P}\frac{\mathbf{U}_{T}\mathbf{\Sigma}_{T}^{1/2}\mathbf{n}^{\left(l\right)}}{\left\Vert \mathbf{U}_{T}\mathbf{\Sigma}_{T}^{1/2}\mathbf{n}^{\left(l\right)}\right\Vert }$
($l=1,\ldots,L$). Finally, we evaluate the output DC power $P_{\mathrm{out}}$
for each $\mathbf{w}_{T}^{\left(l\right)}$ and choose the best one
as the final transmit beamforming weight vector $\mathbf{w}_{T}^{\star}$,
which has good performance even though it is not guaranteed to be
a stationary point of the problem \eqref{eq:OP1-MaxPoutStP-1}-\eqref{eq:OP1-MaxPoutStP-2}.
Algorithm \ref{alg:DC-Combining-Optimization.} summarizes the procedure
for optimizing the beamforming with DC combining. In Section VIII,
we numerically show that $\mathbf{W}_{T}^{\star}$ is rank-1 so that
Algorithm \ref{alg:DC-Combining-Optimization.} finds a stationary
point of the problem \eqref{eq:OP1-MaxPoutStP-1}-\eqref{eq:OP1-MaxPoutStP-2}
for all tested channel realizations.

Lastly, it is worth investigating the choice of the truncation order
$n_{0}$. Our approach for beamforming optimization with DC combining
as shown in Algorithm \ref{alg:DC-Combining-Optimization.} is applicable
to any truncation order $n_{0}$. However, choosing a large truncation
order $n_{0}$ is not practical since the optimization complexity
exponentially increases with $n_{0}$. In the simplest case $n_{0}=2$,
we have that $v_{\mathrm{out},q}=\beta_{2}\mathscr{\mathcal{E}}\left\{ y_{q}\left(t\right)^{2}\right\} $.
Therefore, in the MISO WPT system (where $Q=1$), truncating to the
second order is equivalent to the linear rectenna model \cite{2016_TWC_RZhang_MIMOWPT_Limitedfeed},
\cite{2013_TWC_SWIPT_RZhang}. This has also been shown in \cite{2016_TSP_WPT_Bruno_Waveform}.
However, when it comes to the MIMO WPT system with DC combining, truncating
to the second order is not equivalent to the linear rectenna model
\cite{2016_TWC_RZhang_MIMOWPT_Limitedfeed}, \cite{2013_TWC_SWIPT_RZhang}
due to the DC combining operation $\sum_{q=1}^{Q}v_{\mathrm{out},q}^{2}/R_{L}$.
On the other hand, it is shown in \cite{2016_TSP_WPT_Bruno_Waveform}
and \cite{2017_TSP_WPT_Bruno_Yang_Large} that $n_{0}=4$ is a good
choice which effectively characterizes the rectenna nonlinearity while
the complexity is not high. In Section VII, we show the numerical
experimental results of the beamforming optimization with DC combining
based on the rectenna model truncated to the 4th order.

\begin{algorithm}[t]
\begin{enumerate}
\item \textbf{Initialize}: $i=0$, $\mathbf{r}^{\left(0\right)}$, $t_{0}^{\left(0\right)}$
\item \textbf{do}
\item ~~~~$i=i+1$
\item ~~~~$\gamma_{k}=g_{k}\left(\mathbf{r}^{\left(i-1\right)}\right)/\sum_{k=1}^{K}g_{k}\left(\mathbf{r}^{\left(i-1\right)}\right)$,
$k=1,\ldots,K$
\item ~~~~Obtain $\mathbf{r}^{\left(i\right)}$, $t_{0}^{\left(i\right)}$,
$\mathbf{W}_{T}^{\left(i\right)}$ by solving problem \eqref{eq:OP5-Min1t0-SDR-GP-1}-\eqref{eq:OP5-Min1t0-SDR-GP-5}
\item \textbf{until} $\left|t_{0}^{\left(i\right)}-t_{0}^{\left(i-1\right)}\right|<\epsilon$
or $i=i_{\mathrm{max}}$
\item Set $\mathbf{W}_{T}^{\star}=\mathbf{W}_{T}^{\left(i\right)}$
\item \textbf{if} $\mathrm{rank}\left(\mathbf{W}_{T}^{\star}\right)=1$
\item ~~~~Obtain $\mathbf{w}_{T}^{\star}$ by $\mathbf{W}_{T}^{\star}=\mathbf{w}_{T}^{\star}\mathbf{w}_{T}^{\star H}$
\item \textbf{else}
\item ~~~~Obtain $\mathbf{U}_{T}$ and $\mathbf{\Sigma}_{T}$ by $\mathbf{U}_{T}\mathbf{\Sigma}_{T}\mathbf{U}_{T}^{H}=\mathbf{W}_{T}^{\star}$
\item ~~~~\textbf{for} $l=1$ to $L$ \textbf{do}
\item ~~~~~~~~Generate $\mathbf{n}^{\left(l\right)}\sim\mathcal{CN}\left(\mathbf{0},\mathbf{I}\right)$
\item ~~~~~~~~Obtain $\mathbf{w}_{T}^{\left(l\right)}=\sqrt{2P}\frac{\mathbf{U}_{T}\mathbf{\Sigma}_{T}^{1/2}\mathbf{n}^{\left(l\right)}}{\left\Vert \mathbf{U}_{T}\mathbf{\Sigma}_{T}^{1/2}\mathbf{n}^{\left(l\right)}\right\Vert }$
\item ~~~~\textbf{end}
\item ~~~~Set $l^{\star}=\underset{l=1,\ldots,L}{\arg\max}\:P_{\mathrm{out}}\left(\mathbf{w}_{T}^{\left(l\right)}\right)$,
$\mathbf{w}_{T}^{\star}=\mathbf{w}_{T}^{\left(l^{\star}\right)}$
\item \textbf{end}
\end{enumerate}
\caption{\label{alg:DC-Combining-Optimization.}Beamforming Optimization with
DC Combining.}
\end{algorithm}

\section{Beamforming Optimization with RF Combining}

\begin{figure}[t]
\begin{centering}
\includegraphics[scale=0.64]{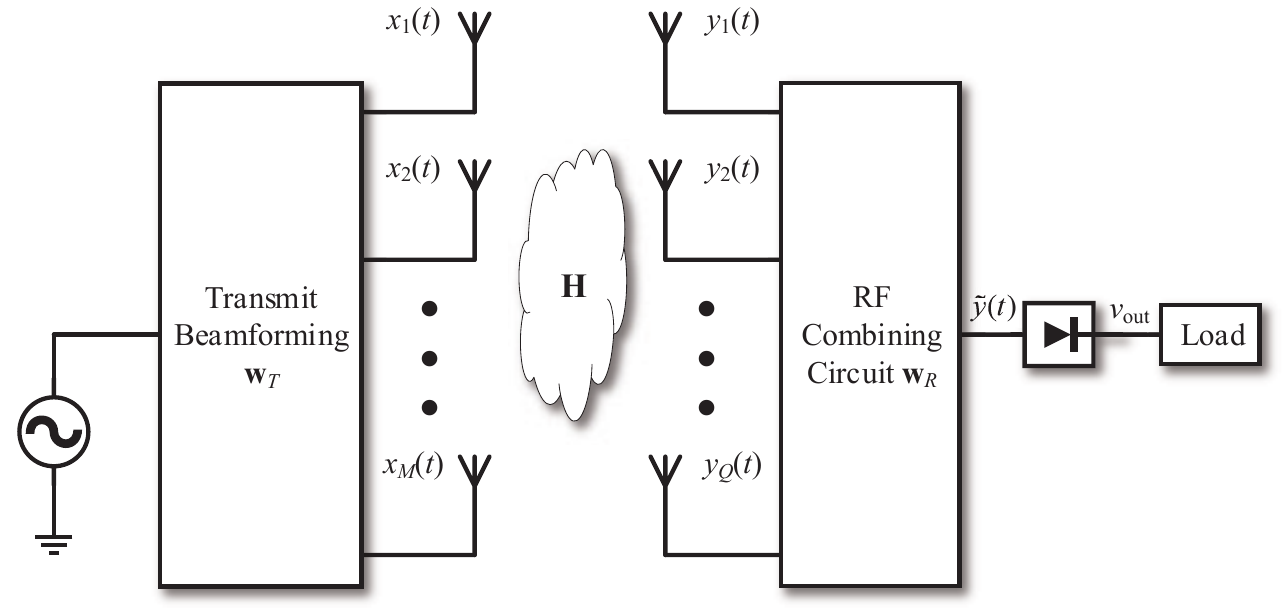}
\par\end{centering}
\caption{\label{fig:RF combinging}Schematic of the MIMO WPT system with RF
combining in the receiver.}
\end{figure}

Consider the RF combining scheme for the multiple antennas as shown
in Fig. \ref{fig:RF combinging}. All receive antennas are connected
to an RF combining circuit. The RF combining circuit includes fixed
RF power combiner such as T-junction and reconfigurable power combiner
with variable power ratio and phase shifts \cite{ReconfiguarablePowerCombiner}
to adapt to the channel. The received signal at all receive antennas
are combined together so that the RF combined signal $\widetilde{y}\left(t\right)$
can be expressed as

\begin{align}
\widetilde{y}\left(t\right) & =\Re\left\{ \mathbf{w}_{R}^{H}\mathbf{H}\mathbf{w}_{T}e^{j\omega_{c}t}\right\} ,
\end{align}
where $\mathbf{w}_{R}$ denotes the receive beamforming weight vector.
Since the RF combining circuit is passive, the output power of the
RF combining circuit should be equal or less than the input power,
which results in a constraint of $\left\Vert \mathbf{w}_{R}\right\Vert ^{2}\leq1$.
The RF combined signal $\widetilde{y}\left(t\right)$ is then rectified
by the only one rectifier in the RF combining scheme as shown in Fig.
\ref{fig:RF combinging}. Using the aforementioned voltage model of
the nonlinear rectenna, the output DC voltage of the one rectifier
is given by

\begin{align}
v_{\mathrm{out}} & =\sum_{i\,\mathrm{even},\,i\geq2}^{n_{0}}\beta_{i}\mathscr{\mathcal{E}}\left\{ \widetilde{y}\left(t\right)^{i}\right\} ,\label{a}
\end{align}
where $\mathscr{\mathcal{E}}\left\{ \widetilde{y}\left(t\right)^{i}\right\} $
is given by
\begin{align}
\mathscr{\mathcal{E}}\left\{ \widetilde{y}\left(t\right)^{i}\right\}  & =\zeta_{i}\left|\mathbf{w}_{R}^{H}\mathbf{H}\mathbf{w}_{T}\right|^{i},
\end{align}
with $\zeta_{i}=\frac{1}{2\pi}\int_{0}^{2\pi}\sin^{i}t\:\textrm{d}t$
(same as \eqref{eq:yq(t)constant}).

We aim to maximize the output DC power $P_{\mathrm{out}}=v_{\mathrm{out}}^{2}/R_{L}$
subject to the transmit power constraint. Noticing that $P_{\mathrm{out}}$
monotonically increases with $\left|\mathbf{w}_{R}^{H}\mathbf{H}\mathbf{w}_{T}\right|^{2}$
irrespectively of the truncation order $n_{0}$, we have the following
equivalent problem

\begin{align}
\underset{\mathbf{w}_{T},\mathbf{w}_{R}}{\mathrm{max}}\;\;\; & \left|\mathbf{w}_{R}^{H}\mathbf{H}\mathbf{w}_{T}\right|^{2}\label{eq:OP7-MaxwrHwtStP-1}\\
\mathsf{\mathrm{s.t.}}\;\;\;\;\; & \frac{1}{2}\left\Vert \mathbf{w}_{T}\right\Vert ^{2}\leq P,\label{eq:OP7-MaxwrHwtStP-2}\\
 & \:\:\:\left\Vert \mathbf{w}_{R}\right\Vert ^{2}\leq1.\label{eq:OP7-MaxwrHwtStP-3}
\end{align}
Therefore, the choice of the truncation order $n_{0}$ has no effect
on the beamforming optimization with RF combining in the MIMO WPT
system, which is different from the case of DC combining.

\subsection{General Receive Beamforming}

The optimal transmit and receive beamforming for the problem \eqref{eq:OP7-MaxwrHwtStP-1}-\eqref{eq:OP7-MaxwrHwtStP-3}
has closed-form solutions by using singular value decomposition (SVD).
Using SVD, we can express the channel matrix as $\mathbf{H}=\mathbf{U}\mathbf{\Sigma}\mathbf{V}^{H}$
where $\mathbf{U}$ is a $Q\times Q$ unitary matrix, $\mathbf{V}$
is a $M\times M$ unitary matrix, and $\mathbf{\Sigma}$ is a $Q\times M$
diagonal matrix. Therefore, the optimal transmit and receive beamformers
are given by

\begin{align}
\mathbf{w}_{T}^{\star} & =\mathbf{v}_{1}\sqrt{2P},\label{WT SVD}\\
\mathbf{w}_{R}^{\star} & =\mathbf{u}_{1},
\end{align}
where $\mathbf{v}_{1}$ and $\mathbf{u}_{1}$ denote the vectors in
$\mathbf{V}$ and $\mathbf{U}$ corresponding to the maximum singular
value $\sigma_{1}$. Therefore, the maximum value of $\left|\mathbf{w}_{R}^{H}\mathbf{H}\mathbf{w}_{T}\right|^{2}$
is $\sigma_{1}^{2}$.

\subsection{Analog Receive Beamforming}

\begin{figure}[t]
\begin{centering}
\includegraphics[scale=0.64]{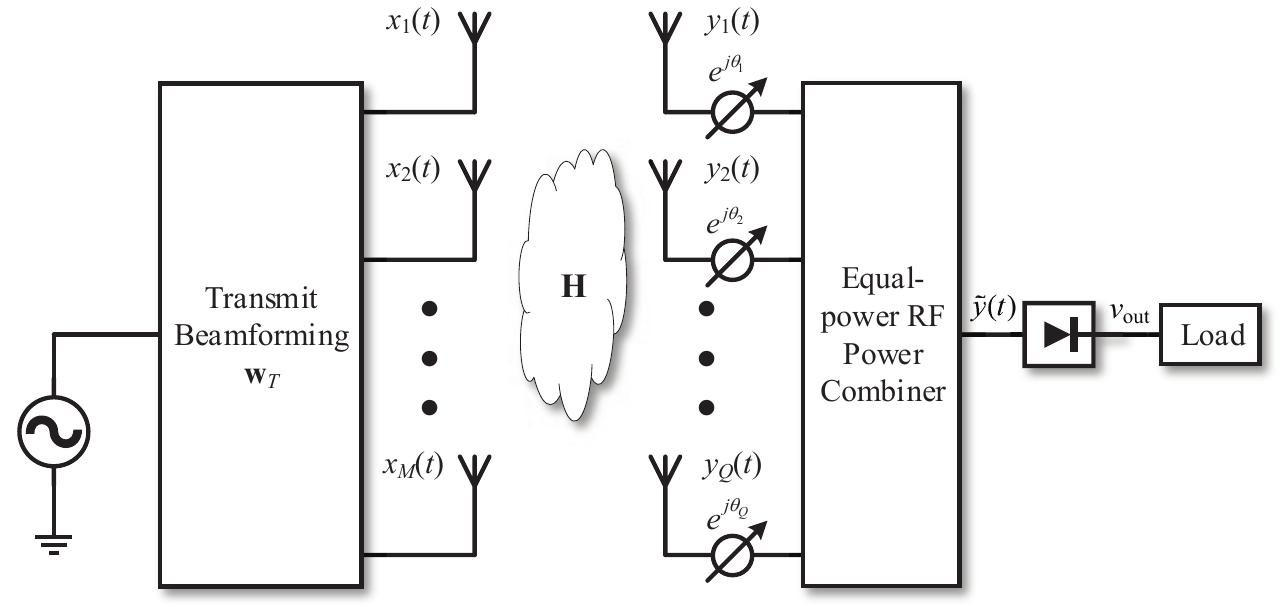}
\par\end{centering}
\caption{\label{fig:RF combinging-ABF}Schematic of the MIMO WPT system with
analog receive beamforming in the receiver.}
\end{figure}

We consider a practical RF combining circuit for the RF combining
scheme as shown in Fig. \ref{fig:RF combinging-ABF}. It consists
of an equal-power RF power combiner and $Q$ phase shifters. Each
receive antenna is connected to a phase shifter and the outputs of
the $Q$ phase shifters are connected to the RF power combiner. Therefore,
we refer to it as analog receive beamforming and the analog receive
beamforming weight vector satisfies the constraint that

\begin{equation}
\mathbf{w}_{R}=\frac{1}{\sqrt{Q}}\left[e^{-j\theta_{1}},e^{-j\theta_{2}},...,e^{-j\theta_{Q}}\right]^{T},\label{eq:Phaseshift-constraint-1}
\end{equation}

\begin{equation}
-\pi\leq\theta_{q}<\pi,\:1\leq q\leq Q,\label{eq:Phaseshift-constraint-2}
\end{equation}
where $\theta_{q}$ denotes the phase shift and the coefficient $\frac{1}{\sqrt{Q}}$
comes from the equal-power RF power combiner (it can be also explained
by the constraint $\left\Vert \mathbf{w}_{R}\right\Vert ^{2}\leq1$).
We replace the constraint \eqref{eq:OP7-MaxwrHwtStP-3} with the constraints
\eqref{eq:Phaseshift-constraint-1} and \eqref{eq:Phaseshift-constraint-2}
so that we have the following problem

\begin{align}
\underset{\mathbf{w}_{T},\mathbf{w}_{R},\theta_{1},...,\theta_{Q}}{\mathrm{max}} & \left|\mathbf{w}_{R}^{H}\mathbf{H}\mathbf{w}_{T}\right|^{2}\label{eq:OP8-MaxwrHwtStP-analogBF-1}\\
\mathrm{s.t.}\;\;\quad\; & \frac{1}{2}\left\Vert \mathbf{w}_{T}\right\Vert ^{2}\leq P,\label{eq:OP8-MaxwrHwtStP-analogBF-2}\\
 & \mathbf{w}_{R}=\frac{1}{\sqrt{Q}}\left[e^{-j\theta_{1}},e^{-j\theta_{2}},...,e^{-j\theta_{Q}}\right]^{T},\label{eq:OP8-MaxwrHwtStP-analogBF-3}\\
 & -\pi\leq\theta_{q}<\pi,\:1\leq q\leq Q,\label{eq:OP8-MaxwrHwtStP-analogBF-4}
\end{align}
which is a non-convex optimization problem due to the non-convex constraint
of $\mathbf{w}_{R}$ \eqref{eq:OP8-MaxwrHwtStP-analogBF-3} and \eqref{eq:OP8-MaxwrHwtStP-analogBF-4}.
The problem \eqref{eq:OP8-MaxwrHwtStP-analogBF-1}-\eqref{eq:OP8-MaxwrHwtStP-analogBF-4}
can be solved by the following two stages.

\subsubsection{Maximum Ratio Transmission}

In the first stage, for any given $\mathbf{w}_{R}$, it is known that
the maximum ratio transmission (MRT), $\mathbf{w}_{T}=\sqrt{2P}\frac{\left(\mathbf{w}_{R}^{H}\mathbf{H}\right)^{H}}{\left\Vert \mathbf{w}_{R}^{H}\mathbf{H}\right\Vert }$,
is the optimal transmit beamforming solution so that we have $\left|\mathbf{w}_{R}^{H}\mathbf{H}\mathbf{w}_{T}\right|^{2}=2P\left\Vert \mathbf{w}_{R}^{H}\mathbf{H}\right\Vert ^{2}$
and the problem \eqref{eq:OP8-MaxwrHwtStP-analogBF-1}-\eqref{eq:OP8-MaxwrHwtStP-analogBF-4}
can be equivalently rewritten as

\begin{align}
\underset{\mathbf{w}_{R},\theta_{1},...,\theta_{Q}}{\mathrm{max}} & \left\Vert \mathbf{w}_{R}^{H}\mathbf{H}\right\Vert ^{2}\label{OP9-MaxwrH2-analogBF-1}\\
\mathsf{\mathrm{s.t.}}\;\quad & \mathbf{w}_{R}=\frac{1}{\sqrt{Q}}\left[e^{-j\theta_{1}},e^{-j\theta_{2}},...,e^{-j\theta_{Q}}\right]^{T},\label{OP9-MaxwrH2-analogBF-2}\\
 & -\pi\leq\theta_{q}<\pi,\:1\leq q\leq Q,\label{OP9-MaxwrH2-analogBF-3}
\end{align}
which is still a non-convex optimization problem due to the non-convex
constraints \eqref{OP9-MaxwrH2-analogBF-2}, \eqref{OP9-MaxwrH2-analogBF-3}.

\subsubsection{Semi-definite Relaxation}

In the second stage, the non-convex constraints \eqref{OP9-MaxwrH2-analogBF-2}
and \eqref{OP9-MaxwrH2-analogBF-3} are transformed into convex constraints
by using SDR. By introducing an auxiliary positive semi-definite matrix
variable $\mathbf{W}_{R}=\mathbf{w}_{R}\mathbf{w}_{R}^{H}$, the problem
\eqref{OP9-MaxwrH2-analogBF-1}-\eqref{OP9-MaxwrH2-analogBF-3} is
equivalently rewritten as

\begin{align}
\underset{\mathbf{W}_{R}}{\mathrm{max}}\;\;\; & \mathrm{Tr}\left(\mathbf{H}\mathbf{H}^{H}\mathbf{W}_{R}\right)\label{eq:OP10-MaxTrHHWR-SDR-1}\\
\mathsf{\mathrm{s.t.}}\;\;\; & \left[\mathbf{W}_{R}\right]_{qq}=\frac{1}{Q},\:1\leq q\leq Q,\label{eq:OP10-MaxTrHHWR-SDR-2}\\
 & \mathbf{W}_{R}\succeq0,\label{eq:OP10-MaxTrHHWR-SDR-3}\\
 & \mathrm{rank}\left(\mathbf{W}_{R}\right)=1.\label{eq:OP10-MaxTrHHWR-SDR-4}
\end{align}
We use SDR to relax the rank-1 constraint \eqref{eq:OP10-MaxTrHHWR-SDR-4}
so that the relaxed problem \eqref{eq:OP10-MaxTrHHWR-SDR-1}-\eqref{eq:OP10-MaxTrHHWR-SDR-3}
becomes a semi-definite program (SDP), which can be solved by CVX
with complexity of $\mathcal{O}\left(Q^{4.5}\log\frac{1}{\epsilon}\right)$
($\epsilon$ is the solution accuracy) \cite{2010_SPMag_SDR} to achieve
the global optimal solution. We denote $\mathbf{W}_{R}^{\star}$ as
the global optimal solution of the problem \eqref{eq:OP10-MaxTrHHWR-SDR-1}-\eqref{eq:OP10-MaxTrHHWR-SDR-3}.
If $\mathrm{rank}\left(\mathbf{W}_{R}^{\star}\right)=1$, the SDR
is tight so that $\mathbf{W}_{R}^{\star}$ is also the global optimal
solution of the problem \eqref{eq:OP10-MaxTrHHWR-SDR-1}-\eqref{eq:OP10-MaxTrHHWR-SDR-4}.
Using EVD, we obtain $\mathbf{W}_{R}^{\star}=\mathbf{w}_{R}^{\star}\mathbf{w}_{R}^{\star H}$
and therefore $\mathbf{w}_{R}^{\star}$ is the global optimal solution
of the problem \eqref{OP9-MaxwrH2-analogBF-1}-\eqref{OP9-MaxwrH2-analogBF-3}.
Otherwise, for the case of $\mathrm{rank}\left(\mathbf{W}_{R}^{\star}\right)>1$,
we aim to extract a suboptimal rank-1 solution from $\mathbf{W}_{R}^{\star}$
by the Gaussian randomization method as shown in Section IV. Particularly,
we first use the EVD to decompose $\mathbf{W}_{R}^{\star}=\mathbf{U}_{R}\mathbf{\Sigma}_{R}\mathbf{U}_{R}^{H}$
where $\mathbf{U}_{R}$ and $\mathbf{\Sigma}_{R}$ are $Q\times Q$
unitary and diagonal matrices, respectively. Then, we generate $Q$-dimensional
random vectors $\mathbf{n}^{\left(l\right)}\sim\mathcal{CN}\left(\mathbf{0},\mathbf{I}\right)$
($l=1,\ldots,L$), multiply \textbf{$\mathbf{n}^{\left(l\right)}$}
by $\mathbf{U}_{R}\mathbf{\Sigma}_{R}^{1/2}$, and extract its phase
as the phase shift so that we obtain $\mathbf{w}_{R}^{\left(l\right)}=\frac{1}{\sqrt{Q}}e^{j\arg\left(\mathbf{U}_{R}\mathbf{\Sigma}_{R}^{1/2}\mathbf{n}^{\left(l\right)}\right)}$
($l=1,\ldots,L$). Finally, we evaluate $\left\Vert \mathbf{w}_{R}^{\left(l\right)H}\mathbf{H}\right\Vert ^{2}$
for each $\mathbf{w}_{R}^{\left(l\right)}$ and choose the best one
as the final receive beamforming weight vector $\mathbf{w}_{R}^{\star}$.
The Gaussian randomization method guarantees at least a $\frac{\pi}{4}$-approximation
of the optimal objective value for the problem \eqref{OP9-MaxwrH2-analogBF-1}-\eqref{OP9-MaxwrH2-analogBF-3}
\cite{2010_SPMag_SDR}. Once we obtain $\mathbf{w}_{R}^{\star}$,
the optimal transmit beamforming weight vector $\mathbf{w}_{T}^{\star}$
can be found by the aforementioned MRT. Algorithm \ref{alg:RF-Combining-Optimization}
summarizes the procedure for optimizing the analog receive beamforming.
In Section VIII, we numerically show that $\mathbf{W}_{R}^{\star}$
is rank-1 matrix for most of the tested channel realizations and therefore
Algorithm \ref{alg:RF-Combining-Optimization} finds nearly the global
optimal solution of the problem \eqref{OP9-MaxwrH2-analogBF-1}-\eqref{OP9-MaxwrH2-analogBF-3}
for the tested channel realizations\footnote{Other algorithm \cite{2012_TSP_Nonsmooth} has also been considered
and implemented for solving the problem \eqref{eq:OP10-MaxTrHHWR-SDR-1}-\eqref{eq:OP10-MaxTrHHWR-SDR-4},
but the proposed algorithm was found to be more efficient for the
tested channel realizations.}.

\begin{algorithm}[t]
\begin{enumerate}
\item Obtain $\mathbf{W}_{R}^{\star}$ by solving problem \eqref{eq:OP10-MaxTrHHWR-SDR-1}-\eqref{eq:OP10-MaxTrHHWR-SDR-4}
\item \textbf{if} $\mathrm{rank}\left(\mathbf{W}_{R}^{\star}\right)=1$
\item ~~~~Obtain $\mathbf{w}_{R}^{\star}$ by $\mathbf{W}_{R}^{\star}=\mathbf{w}_{R}^{\star}\mathbf{w}_{R}^{\star H}$
\item \textbf{else}
\item ~~~~Obtain $\mathbf{U}_{R}$ and $\mathbf{\Sigma}_{R}$ by $\mathbf{U}_{R}\mathbf{\Sigma}_{R}\mathbf{U}_{R}^{H}=\mathbf{W}_{R}^{\star}$
\item ~~~~\textbf{for} $l=1$ to $L$ \textbf{do}
\item ~~~~~~~~Generate $\mathbf{n}^{\left(l\right)}\sim\mathcal{CN}\left(\mathbf{0},\mathbf{I}\right)$
\item ~~~~~~~~Obtain $\mathbf{w}_{R}^{\left(l\right)}=\frac{1}{\sqrt{Q}}e^{j\arg\left(\mathbf{U}_{R}\mathbf{\Sigma}_{R}^{1/2}\mathbf{n}^{\left(l\right)}\right)}$
\item ~~~~\textbf{end}
\item ~~~~Set $l^{\star}=\underset{l=1,\ldots,L}{\arg\max}\:\left\Vert \mathbf{w}_{R}^{\left(l\right)H}\mathbf{H}\right\Vert ^{2}$,
$\mathbf{w}_{R}^{\star}=\mathbf{w}_{R}^{\left(l^{\star}\right)}$
\item \textbf{end}
\item Set $\mathbf{w}_{T}^{\star}=\sqrt{2P}\frac{\left(\mathbf{w}_{R}^{\star H}\mathbf{H}\right)^{H}}{\left\Vert \mathbf{w}_{R}^{\star H}\mathbf{H}\right\Vert }$
\end{enumerate}
\caption{\label{alg:RF-Combining-Optimization}Beamforming Optimization with
RF Combining.}
\end{algorithm}

\section{Beneficial Role of Rectenna Nonlinearity}

In this section, we provide some insights into the role of nonlinearity.
This is crucial to understand why RF combining outperforms DC combining
in the output DC power level. The voltage model $v_{\mathrm{out}}=\sum_{i\,\mathrm{even},\,i\geq2}^{n_{0}}\beta_{i}\mathscr{\mathcal{E}}\left\{ y\left(t\right)^{i}\right\} $
characterizes the rectenna nonlinearity which is crucial in WPT system.
Using Jensen\textquoteright s inequality, we have $\mathscr{\mathcal{E}}\left\{ y\left(t\right)^{i}\right\} \geq\left(\mathscr{\mathcal{E}}\left\{ y\left(t\right)^{2}\right\} \right)^{\frac{i}{2}}=\left(P_{\mathrm{RF}}\right)^{\frac{i}{2}}$
($P_{\mathrm{RF}}$ is the input RF power) for $i$ is even and $i\geq2$
so that $v_{\mathrm{out}}\geq\sum_{i\,\mathrm{even},\,i\geq2}^{n_{0}}\beta_{i}\left(P_{\mathrm{RF}}\right)^{\frac{i}{2}}$,
which implies that the RF-to-DC efficiency increases with the input
RF power. This is consistent with the measured RF-to-DC efficiency
of rectifier circuits in \cite{ShanpuShen2017_AWPL_DPTB}-\nocite{ShanpuShen2018_EuCap_QPDP}\nocite{ShanpuShen2017_TAP_EHPIXEL}\cite{ShanpuShen2019_TMTT_Freqdepend},
\cite{2018_TWC_WPT_Bruno_Transmit_Diversity}.

To understand the beneficial role of rectenna nonlinearity, we first
consider the MIMO WPT system with linear rectenna model which has
a constant RF-to-DC conversion efficiency $\eta$. In DC combining,
the RF power received by the $q$th antenna is $P_{\mathrm{RF},q}=\frac{1}{2}\left|\mathbf{h}_{q}\mathbf{w}_{T}\right|^{2}$.
Each receive antenna is connected to a rectifier so that the output
DC power at the $q$th rectifier is given by $P_{\mathrm{out},q}=\frac{1}{2}\left|\mathbf{h}_{q}\mathbf{w}_{T}\right|^{2}\eta$.
Therefore, the total output DC power is given by $P_{\mathrm{out}}^{\mathrm{DC\:Combining}}=\frac{1}{2}\left\Vert \mathbf{H}\mathbf{w}_{T}\right\Vert ^{2}\eta$
where $\mathbf{H}=\left[\mathbf{h}_{1}^{T},\mathbf{h}_{2}^{T},...,\mathbf{h}_{Q}^{T}\right]^{T}$.
On the other hand, in RF combining, the RF power of the combined RF
signal is $P_{\mathrm{RF}}=\frac{1}{2}\left|\mathbf{w}_{R}^{H}\mathbf{H}\mathbf{w}_{T}\right|^{2}$.
The combined RF signal is input into the single rectifier so that
the output DC power is given by $P_{\mathrm{out}}^{\mathrm{RF\:Combining}}=\frac{1}{2}\left|\mathbf{w}_{R}^{H}\mathbf{H}\mathbf{w}_{T}\right|^{2}\eta$.
It is obvious that the optimal receive beamforming is $\mathbf{w}_{R}^{\star}=\frac{\mathbf{H}\mathbf{w}_{T}}{\left\Vert \mathbf{H}\mathbf{w}_{T}\right\Vert }$
so that we have $P_{\mathrm{out}}^{\mathrm{RF\:Combining}}=\frac{1}{2}\left\Vert \mathbf{H}\mathbf{w}_{T}\right\Vert ^{2}\eta$
which is equal to the total output DC power in DC Combining. Therefore,
if we use the linear rectenna model, RF combining has the same performance
as DC combining even though the optimal receive beamforming $\mathbf{w}_{R}^{\star}$
is used in RF combining.

However, rectenna is a nonlinear device and the rectenna nonlinearity
cannot be ignored. The RF-to-DC conversion efficiency $\eta$ is a
function of the input signal waveform and the input power level. For
single-tone input signal, $\eta$ increases with the input RF power
level, which is denoted as $\eta\left(P_{\mathrm{RF}}\right)$. Considering
the rectenna nonlinearity, the total output DC power of DC and RF
combining will be different. In DC combining, the output DC power
is given by 
\begin{equation}
P_{\mathrm{out}}^{\mathrm{DC\:Combining}}=\frac{1}{2}\sum_{q=1}^{Q}\left|\mathbf{h}_{q}\mathbf{w}_{T}\right|^{2}\eta\left(\frac{1}{2}\left|\mathbf{h}_{q}\mathbf{w}_{T}\right|^{2}\right),
\end{equation}
where $\eta\left(\frac{1}{2}\left|\mathbf{h}_{q}\mathbf{w}_{T}\right|^{2}\right)$
denotes the RF-to-DC efficiency at the input RF power level of $\frac{1}{2}\left|\mathbf{h}_{q}\mathbf{w}_{T}\right|^{2}$.
Therefore, nonlinearity exists in DC combining so that it can be leveraged
to increase the output DC power. On the other hand, in RF combining,
the output DC power is given by $P_{\mathrm{out}}^{\mathrm{RF\:Combining}}=\frac{1}{2}\left|\mathbf{w}_{R}^{H}\mathbf{H}\mathbf{w}_{T}\right|^{2}\eta\left(\frac{1}{2}\left|\mathbf{w}_{R}^{H}\mathbf{H}\mathbf{w}_{T}\right|^{2}\right)$
where $\eta\left(\frac{1}{2}\left|\mathbf{w}_{R}^{H}\mathbf{H}\mathbf{w}_{T}\right|^{2}\right)$denotes
the RF-to-DC efficiency at the input RF power level of $\frac{1}{2}\left|\mathbf{w}_{R}^{H}\mathbf{H}\mathbf{w}_{T}\right|^{2}$.
The optimal receive beamforming is still $\mathbf{w}_{R}^{\star}=\frac{\mathbf{H}\mathbf{w}_{T}}{\left\Vert \mathbf{H}\mathbf{w}_{T}\right\Vert }$
which simultaneously maximize the terms $\left|\mathbf{w}_{R}^{H}\mathbf{H}\mathbf{w}_{T}\right|^{2}$
and $\eta\left(\frac{1}{2}\left|\mathbf{w}_{R}^{H}\mathbf{H}\mathbf{w}_{T}\right|^{2}\right)$.
Then, the output DC power with $\mathbf{w}_{R}^{\star}$ is given
by 
\begin{equation}
P_{\mathrm{out}}^{\mathrm{RF\:Combining}}=\frac{1}{2}\left\Vert \mathbf{H}\mathbf{w}_{T}\right\Vert ^{2}\eta\left(\frac{1}{2}\left\Vert \mathbf{H}\mathbf{w}_{T}\right\Vert ^{2}\right),
\end{equation}
which shows that nonlinearity also exists in RF combining so that
it can be leveraged. Therefore, we can conclude that rectenna nonlinearity
plays a beneficial role in both DC and RF combinings to increase the
output DC power. However, the receive beamforming in RF combining
can leverage the nonlinearity more efficiently than DC combining which
has no receive beamforming. In DC combining, the efficiency for each
rectenna is $\eta\left(\frac{1}{2}\left|\mathbf{h}_{q}\mathbf{w}_{T}\right|^{2}\right)$
while in RF combining the efficiency is $\eta\left(\frac{1}{2}\left\Vert \mathbf{H}\mathbf{w}_{T}\right\Vert ^{2}\right)$.
Since the efficiency increases with the input RF power, we have $\eta\left(\frac{1}{2}\left\Vert \mathbf{H}\mathbf{w}_{T}\right\Vert ^{2}\right)\geq\eta\left(\frac{1}{2}\left|\mathbf{h}_{q}\mathbf{w}_{T}\right|^{2}\right)$
so that $P_{\mathrm{out}}^{\mathrm{RF\:Combining}}\geq P_{\mathrm{out}}^{\mathrm{DC\:Combining}}$.
Hence, we can conclude that RF combining outperforms DC combining
because the receive beamforming in RF combining can leverage the nonlinearity
more efficiently than DC combining. This conclusion is also verified
in the next sections of scaling laws analysis and performance evaluations.

\section{Scaling Laws}

In order to get insights into the fundamental limits of MIMO WPT system
and get insights into the role of the combining strategy, we want
to quantify how the output DC power $P_{\mathrm{out}}$ scales as
a function of the number of transmit antennas $M$ and the number
of receive antennas $Q$. In the following, we consider MISO WPT system
and SIMO WPT system, respectively, and for SIMO WPT system we consider
the DC combining and RF combining, respectively. We mainly consider
the scaling laws with the truncation order $n_{0}=4$. We assume that
the channel gains $h$ are modeled as i.i.d. CSCG random variables
with zero mean and unit variance. In this Section, we also assume
that $R_{L}=1\;\Omega$ for simplicity.

\subsection{MISO WPT System}

We first consider the MISO system. Because there is only one antenna
at the receiver, there is no combining issue. The output DC voltage
of the single rectifier is given by $v_{\mathrm{out}}=\frac{\beta_{2}}{2}\left|\mathbf{h}\mathbf{w}_{T}\right|^{2}+\frac{3\beta_{4}}{8}\left|\mathbf{h}\mathbf{w}_{T}\right|^{4}$
where $\mathbf{h}=\left[h_{1},h_{2},\ldots,h_{M}\right]$ refers to
the channel vector (a row vector) for the MISO system. It is obvious
that the MRT $\mathbf{w}_{T}=\sqrt{2P}\frac{\mathbf{h}^{H}}{\left\Vert \mathbf{h}\right\Vert }$
gives the maximum output DC voltage $v_{\mathrm{out,\:MRT}}=\beta_{2}P\left\Vert \mathbf{h}\right\Vert ^{2}+\frac{3\beta_{4}P^{2}}{2}\left\Vert \mathbf{h}\right\Vert ^{4}$
so that we have the maximum output DC power given by $P_{\mathrm{out,\:MRT}}=\beta_{2}^{2}P^{2}\left\Vert \mathbf{h}\right\Vert ^{4}+3\beta_{2}\beta_{4}P^{3}\left\Vert \mathbf{h}\right\Vert ^{6}+\frac{9\beta_{4}^{2}P^{4}}{4}\left\Vert \mathbf{h}\right\Vert ^{8}.$
Therefore, the average output DC power is given by $\bar{P}_{\mathrm{out,\:MRT}}=\:\:\beta_{2}^{2}P^{2}\mathscr{\mathcal{E}}\left\{ \left\Vert \mathbf{h}\right\Vert ^{4}\right\} +3\beta_{2}\beta_{4}P^{3}\mathscr{\mathcal{E}}\left\{ \left\Vert \mathbf{h}\right\Vert ^{6}\right\} +\frac{9\beta_{4}^{2}P^{4}}{4}\mathscr{\mathcal{E}}\left\{ \left\Vert \mathbf{h}\right\Vert ^{8}\right\} .$
Making use of the moments of a $\chi_{2M}^{2}$ random variable, we
have that $\mathscr{\mathcal{E}}\left\{ \left\Vert \mathbf{h}\right\Vert ^{2n}\right\} =\frac{\left(M+n-1\right)!}{\left(M-1\right)!}$
so that the average output DC power is given by 
\begin{align}
\bar{P}_{\mathrm{out,\:MRT}}=\: & \:\beta_{2}^{2}P^{2}M\left(M+1\right)\nonumber \\
 & +3\beta_{2}\beta_{4}P^{3}M\left(M+1\right)\left(M+2\right)\nonumber \\
 & +\frac{9\beta_{4}^{2}P^{4}}{4}M\left(M+1\right)\left(M+2\right)\left(M+3\right).\label{PoutMRT}
\end{align}

Equation \eqref{PoutMRT} shows that $\bar{P}_{\mathrm{out,\:MRT}}$
increases with $M$ in the order of $M^{4}$ when the number of transmit
antennas $M$ is large, which demonstrates that by adapting to CSI
using multiple antennas at the transmitter can effectively increase
the output DC power level at the rectenna.

\subsection{SIMO WPT System}

We then consider the SIMO system. There is no transmit beamforming
due to the only one antenna at the transmitter. In the presence of
multiple receive antennas, we consider the DC combining and RF combining
in the following.

\subsubsection{DC Combining}

The received signal at the $q$-th receive antenna is given by $y_{q}\left(t\right)=\Re\left\{ \sqrt{2P}h_{q}e^{j\omega_{c}t}\right\} $
where $h_{q}$ refers to the channel gain for the $q$-th receive
antenna, so that the output DC voltage of the $q$th rectifier is
given by $v_{\mathrm{out},q}^{\mathrm{DC\:Combining}}=\beta_{2}P\left|h_{q}\right|^{2}+\frac{3\beta_{4}P^{2}}{2}\left|h_{q}\right|^{4}.$The
total output DC power is given by $P_{\mathrm{out}}^{\mathrm{DC\:Combining}}=\sum_{q=1}^{Q}\left(\beta_{2}^{2}P^{2}\left|h_{q}\right|^{4}+3\beta_{2}\beta_{4}P^{3}\left|h_{q}\right|^{6}+\frac{9\beta_{4}^{2}P^{4}}{4}\left|h_{q}\right|^{8}\right).$
Therefore we can find the average total output DC power, which is
given by $\bar{P}_{\mathrm{out}}^{\mathrm{DC\:Combining}}=\sum_{q=1}^{Q}(\beta_{2}^{2}P^{2}\mathscr{\mathcal{E}}\left\{ \left|h_{q}\right|^{4}\right\} +3\beta_{2}\beta_{4}P^{3}\mathscr{\mathcal{E}}\left\{ \left|h_{q}\right|^{6}\right\} +\frac{9\beta_{4}^{2}P^{4}}{4}\mathscr{\mathcal{E}}\left\{ \left|h_{q}\right|^{8}\right\} ).$
Making use of the moments of the exponential distribution, we have
$\mathscr{\mathcal{E}}\left\{ \left|h_{q}\right|^{4}\right\} =2$,
$\mathscr{\mathcal{E}}\left\{ \left|h_{q}\right|^{6}\right\} =6$,
$\mathscr{\mathcal{E}}\left\{ \left|h_{q}\right|^{8}\right\} =24$,
so that the average output DC power is given by
\begin{align}
\bar{P}_{\mathrm{out}}^{\mathrm{DC\:Combining}} & =\left(2\beta_{2}^{2}P^{2}+18\beta_{2}\beta_{4}P^{3}+54\beta_{4}^{2}P^{4}\right)Q,\label{PoutDC combining}
\end{align}
which shows that $\bar{P}_{\mathrm{out}}^{\mathrm{DC\:Combining}}$
linearly increases with $Q$.

\subsubsection{RF Combining}

The RF combined signal is $\widetilde{y}\left(t\right)=\Re\left\{ \sqrt{2P}\mathbf{w}_{R}^{H}\mathbf{h}e^{j\omega_{c}t}\right\} $
where $\mathbf{h}=\left[h_{1},h_{2},\ldots,h_{Q}\right]^{T}$ refers
to the SIMO channel vector. The output DC voltage of the single rectifier
is given by $v_{\mathrm{out}}^{\mathrm{RF\:Combining}}=\beta_{2}P\left|\mathbf{w}_{R}^{H}\mathbf{h}\right|^{2}+\frac{3\beta_{4}P^{2}}{2}\left|\mathbf{w}_{R}^{H}\mathbf{h}\right|^{4}.$
It is obvious that the maximum ratio combining (MRC) $\mathbf{w}_{R}=\frac{\mathbf{h}}{\left\Vert \mathbf{h}\right\Vert }$
gives the maximum output DC voltage $v_{\mathrm{out,\:MRC}}^{\mathrm{RF\:Combining}}=\beta_{2}P\left\Vert \mathbf{h}\right\Vert ^{2}+\frac{3\beta_{4}P^{2}}{2}\left\Vert \mathbf{h}\right\Vert ^{4}.$
Similar to the MISO WPT case, the average output DC power in this
case is given by 
\begin{align}
\bar{P}_{\mathrm{out,\:MRC}}^{\mathrm{RF\:Combining}}=\: & \:\beta_{2}^{2}P^{2}Q\left(Q+1\right)\nonumber \\
 & +3\beta_{2}\beta_{4}P^{3}Q\left(Q+1\right)\left(Q+2\right)\nonumber \\
 & +\frac{9\beta_{4}^{2}P^{4}}{4}Q\left(Q+1\right)\left(Q+2\right)\left(Q+3\right),\label{PoutRF combining}
\end{align}
which shows that the average output DC power $\bar{P}_{\mathrm{out,\:MRC}}^{\mathrm{RF\:Combining}}$
increases with $Q$ in the order of $Q^{4}$ when the number of receive
antennas $Q$ is large.

\begin{figure}[tbh]
\begin{centering}
\includegraphics[scale=0.6]{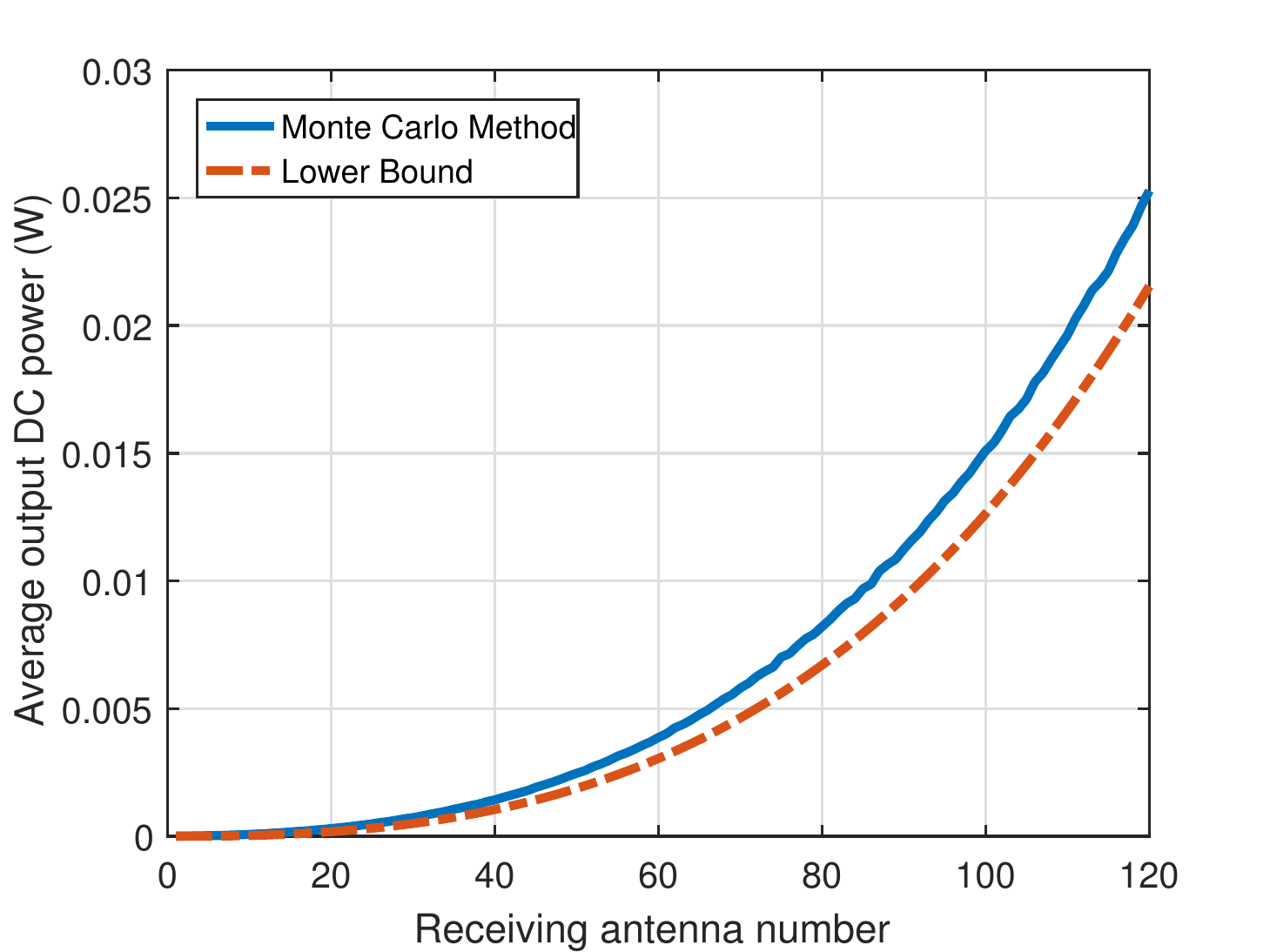}
\par\end{centering}
\caption{\label{low bound}Comparison of $\bar{P}_{\mathrm{out,\:Analog}}^{\mathrm{RF\:Combining}}$
and the lower bound in \eqref{PoutRF combining analog}.}
\end{figure}

Now we consider the analog receive beamforming where the receive beamforming
weight is constrained by \eqref{eq:Phaseshift-constraint-1}, \eqref{eq:Phaseshift-constraint-2}.
It is obvious that the optimal analog receive beamforming weight is
$\mathbf{w}_{R}=\frac{1}{\sqrt{Q}}e^{j\arg\left(\mathbf{h}\right)}$
so that the output DC voltage is given by $v_{\mathrm{out,\:Analog}}^{\mathrm{RF\:Combining}}=\beta_{2}P\frac{\left\Vert \mathbf{h}\right\Vert _{1}^{2}}{Q}+\frac{3\beta_{4}P^{2}}{2}\frac{\left\Vert \mathbf{h}\right\Vert _{1}^{4}}{Q^{2}}.$
Therefore, the average output DC power is given by

\begin{align}
\bar{P}_{\mathrm{out,\:Analog}}^{\mathrm{RF\:Combining}}=\: & \:\beta_{2}^{2}P^{2}\frac{\mathscr{\mathcal{E}}\left\{ \left\Vert \mathbf{h}\right\Vert _{1}^{4}\right\} }{Q^{2}}+3\beta_{2}\beta_{4}P^{3}\frac{\mathscr{\mathcal{E}}\left\{ \left\Vert \mathbf{h}\right\Vert _{1}^{6}\right\} }{Q^{3}}\nonumber \\
 & +\frac{9\beta_{4}^{2}P^{4}}{4}\frac{\mathscr{\mathcal{E}}\left\{ \left\Vert \mathbf{h}\right\Vert _{1}^{8}\right\} }{Q^{4}}.
\end{align}
Finding the closed-form expressions of the moments of random variable
$\left\Vert \mathbf{h}\right\Vert _{1}$ is complicated. For simplicity,
we give a tight lower bound (for large $Q$) as $\mathscr{\mathcal{E}}\left\{ \left\Vert \mathbf{h}\right\Vert _{1}^{n}\right\} \geq\frac{Q!}{\left(Q-n\right)!}\mathscr{\mathcal{E}}\left\{ \left|h_{q}\right|\right\} ^{n}=\frac{Q!}{\left(Q-n\right)!}\Gamma\left(1.5\right)^{n}$
where $\Gamma(x)$ refers to the gamma function. Therefore, we have
the lower bound for the average output DC power as 
\begin{align}
\bar{P}_{\mathrm{out,\:Analog}}^{\mathrm{RF\:Combining}}\geq\: & \:\frac{\beta_{2}^{2}P^{2}\Gamma\left(1.5\right)^{4}Q!}{Q^{2}\left(Q-4\right)!}+\frac{3\beta_{2}\beta_{4}P^{3}\Gamma\left(1.5\right)^{6}Q!}{Q^{3}\left(Q-6\right)!}\nonumber \\
 & +\frac{9\beta_{4}^{2}P^{4}\Gamma\left(1.5\right)^{8}Q!}{4Q^{4}\left(Q-8\right)!},\label{PoutRF combining analog}
\end{align}
which shows that the average output DC power $\bar{P}_{\mathrm{out,\:Analog}}^{\mathrm{RF\:Combining}}$
increases with $Q$ in the order of $Q^{4}$ when the number of receive
antennas $Q$ is large in spite of the lower bound. We also use Monte
Carlo method to find $\bar{P}_{\mathrm{out,\:Analog}}^{\mathrm{RF\:Combining}}$,
and compare it with the lower bound in \eqref{PoutRF combining analog}
as shown in Fig. \ref{low bound}, which shows that the lower bound
in \eqref{PoutRF combining analog} is tight.

To conclude, equation \eqref{PoutDC combining} suggests that the
average output DC power linearly increases with $Q$ in the DC combining
while equation \eqref{PoutRF combining} and \eqref{PoutRF combining analog}
suggest that the average output DC power increases with $Q$ in the
order of $Q^{4}$ for large $Q$ in the RF combining. Therefore, using
multiple antennas at the receiver can effectively increase the average
output DC power for both combinings, but the RF combining outperforms
the DC combining since the receive beamforming in RF combining leverages
the rectenna nonlinearity more efficiently than DC combining which
has no receive beamforming (see Section VI).

Compared with the MRT MISO WPT system \eqref{PoutMRT}, the SIMO WPT
system using the MRC RF combining can achieve the same scaling law
\eqref{PoutRF combining} while using the analog receive beamforming
achieves a slightly lower (but same order) scaling law \eqref{PoutRF combining analog}.
Besides, the SIMO WPT system with DC combining \eqref{PoutDC combining}
achieves a much lower scaling law than the MRT MISO WPT system. Therefore,
overall the MISO WPT system is more beneficial than the SIMO system.
However, using the MIMO WPT system can simultaneously exploit the
benefits of MISO and SIMO system to boost the output DC power.

Table \ref{tab:Summary-of-Scaling} summarizes the scaling laws for
MISO and SIMO systems with DC and RF combinings. The scaling laws
for MISO and SIMO systems with DC and RF combinings for $n_{0}=2$
are also provided in Table \ref{tab:Summary-of-Scaling}. Interestingly,
the scaling laws for $n_{0}=2$ can be easily derived from the scaling
laws for $n_{0}=4$ by setting $\beta_{4}=0$. We notice that the
scaling laws change with the truncation order. For example, in MISO
WPT system the average output DC power increases with $M$ in the
order of $M^{2}$, instead of $M^{4}$, for large $M$. In addition,
RF combining still outperforms the DC combining because $n_{0}=2$
is not equivalent to the linear model in MIMO WPT system with DC combining
as discussed in Section IV.

\begin{table*}[tbh]
\caption{\label{tab:Summary-of-Scaling}Summary of Scaling Laws}

\centering{}%
\begin{tabular}{|c|c|c|c|}
\hline 
\textbf{\tiny{}Beamforming} & {\tiny{}$M$, $Q$} & {\tiny{}$n_{0}=2$} & {\tiny{}$n_{0}=4$}\tabularnewline
\hline 
\hline 
\textbf{\tiny{}MISO} &  &  & \tabularnewline
\hline 
{\tiny{}$\bar{P}_{\mathrm{out,\:MRT}}$} & {\tiny{}$M\geq1$, $Q=1$} & {\tiny{}$\beta_{2}^{2}P^{2}M\left(M+1\right)$} & {\tiny{}$\beta_{2}^{2}P^{2}M\left(M+1\right)+3\beta_{2}\beta_{4}P^{3}M\left(M+1\right)\left(M+2\right)+\frac{9\beta_{4}^{2}P^{4}}{4}M\left(M+1\right)\left(M+2\right)\left(M+3\right)$}\tabularnewline
\hline 
\textbf{\tiny{}SIMO} &  &  & \tabularnewline
\hline 
{\tiny{}$\bar{P}_{\mathrm{out}}^{\mathrm{DC\:Combining}}$} & {\tiny{}$M=1$, $Q\geq1$} & {\tiny{}$2\beta_{2}^{2}P^{2}Q$} & {\tiny{}$\left(2\beta_{2}^{2}P^{2}+18\beta_{2}\beta_{4}P^{3}+54\beta_{4}^{2}P^{4}\right)Q$}\tabularnewline
\hline 
{\tiny{}$\bar{P}_{\mathrm{out,\:MRC}}^{\mathrm{RF\:Combining}}$} & {\tiny{}$M=1$, $Q\geq1$} & {\tiny{}$\beta_{2}^{2}P^{2}Q\left(Q+1\right)$} & {\tiny{}$\beta_{2}^{2}P^{2}Q\left(Q+1\right)+3\beta_{2}\beta_{4}P^{3}Q\left(Q+1\right)\left(Q+2\right)+\frac{9\beta_{4}^{2}P^{4}}{4}Q\left(Q+1\right)\left(Q+2\right)\left(Q+3\right)$}\tabularnewline
\hline 
{\tiny{}$\bar{P}_{\mathrm{out,\:Analog}}^{\mathrm{RF\:Combining}}$} & {\tiny{}$M=1$, $Q\geq1$} & {\tiny{}$\geq\frac{\beta_{2}^{2}P^{2}\Gamma\left(1.5\right)^{4}Q!}{Q^{2}\left(Q-4\right)!}$} & {\tiny{}$\geq\frac{\beta_{2}^{2}P^{2}\Gamma\left(1.5\right)^{4}Q!}{Q^{2}\left(Q-4\right)!}+\frac{3\beta_{2}\beta_{4}P^{3}\Gamma\left(1.5\right)^{6}Q!}{Q^{3}\left(Q-6\right)!}+\frac{9\beta_{4}^{2}P^{4}\Gamma\left(1.5\right)^{8}Q!}{4Q^{4}\left(Q-8\right)!}$}\tabularnewline
\hline 
\end{tabular}
\end{table*}

\section{Performance Evaluations}

We consider two types of performance evaluations, the first one is
based on the simplified and tractable nonlinear rectenna model truncated
to the 4th order as introduced in Section III, while the second one
relies on an actual and accurate modeling of the rectenna in the circuit
simulation solver Advanced Design System (ADS).

\subsection{Nonlinear Model-Based Performance Evaluations}

The first type of evaluations displays $P_{\mathrm{out}}$ averaged
over many channel realizations for DC and RF combinings. We assume
$v_{t}=25.86$ mV, $n=1.05$, and $R_{L}=5000\;\Omega$.

We now evaluate the performance of the MIMO WPT system with DC and
RF combinings in a scenario representative of a WiFi-like environment
at a center frequency of 2.45 GHz with a 36 dBm transmit power and
66 dB path loss with a Rayleigh fading. The elements of the channel
matrix $\mathbf{H}$ are modeled as i.i.d. circularly symmetric complex
Gaussian random variables and the average received power is -30 dBm.

For DC combining, we evaluate the adaptive optimized (OPT) transmit
beamforming as shown in Algorithm 1 versus a benchmark: a transmit
beamforming scheme based on SVD. Specially, such transmit beamforming
weight vector is given by $\mathbf{w}_{T}=\mathbf{v}_{1}\sqrt{2P}$
which is the same as \eqref{WT SVD} in that $\mathbf{v}_{1}$ denotes
the vector in $\mathbf{V}$ corresponding to the maximum singular
value $\sigma_{1}$ and $\mathbf{V}$ is obtained from SVD $\mathbf{H}=\mathbf{U}\mathbf{\Sigma}\mathbf{V}^{H}$.
This transmit beamforming is optimal for maximizing the output DC
power when the linear model of rectenna is considered \cite{2013_TWC_SWIPT_RZhang}.
For RF combining, we evaluate the general receive beamforming based
on SVD and the analog receive beamforming (ABF) as shown in Algorithm
2. Other baselines for DC and RF combinings have been simulated but
we omit them in the paper since they have worse performance than the
chosen baselines. We also assume that $L=100$ for the Gaussian randomization
method in Algorithm \ref{alg:DC-Combining-Optimization.} and \ref{alg:RF-Combining-Optimization}.

\begin{figure}[t]
\begin{centering}
\includegraphics[scale=0.31]{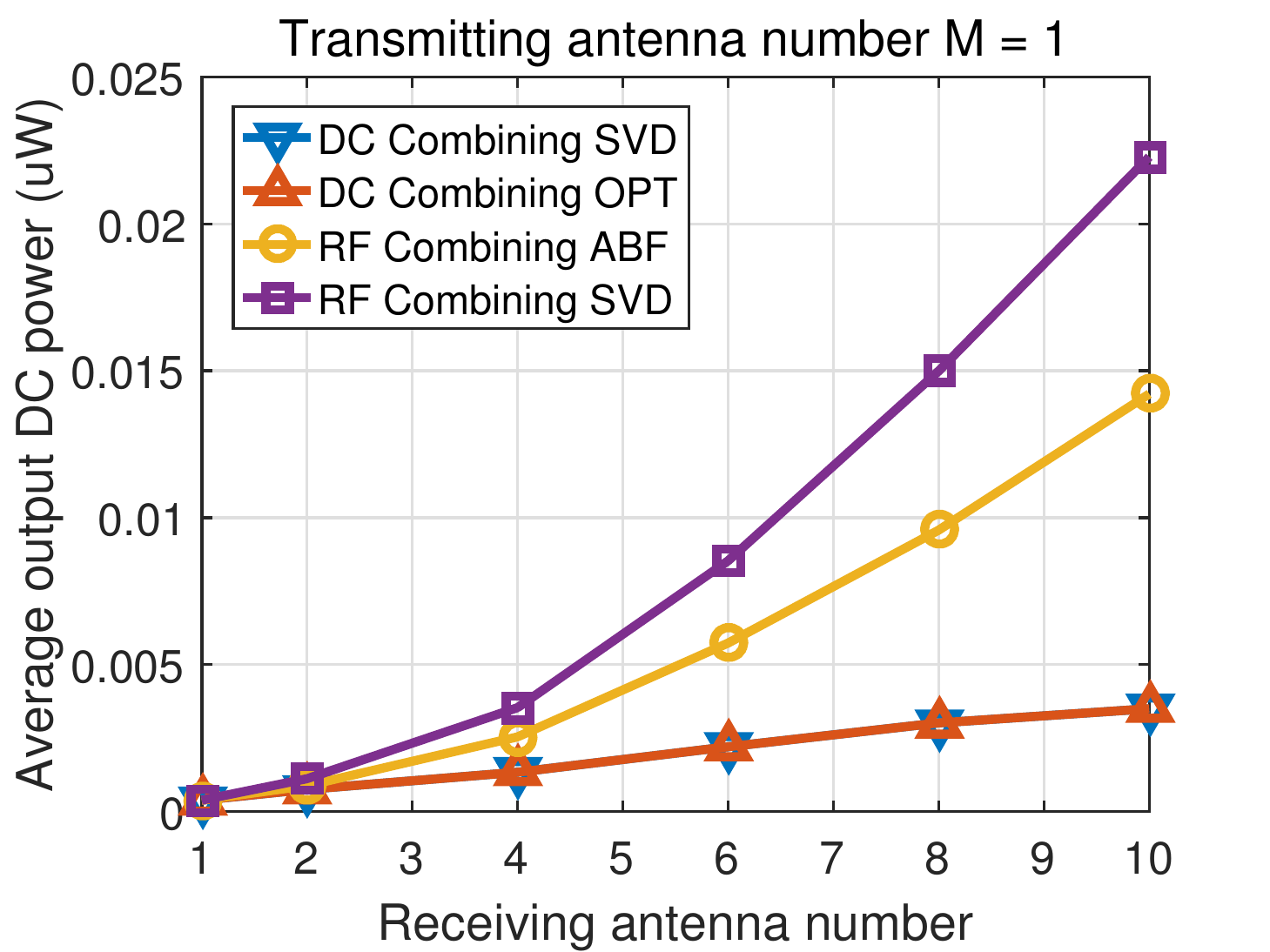}\includegraphics[scale=0.31]{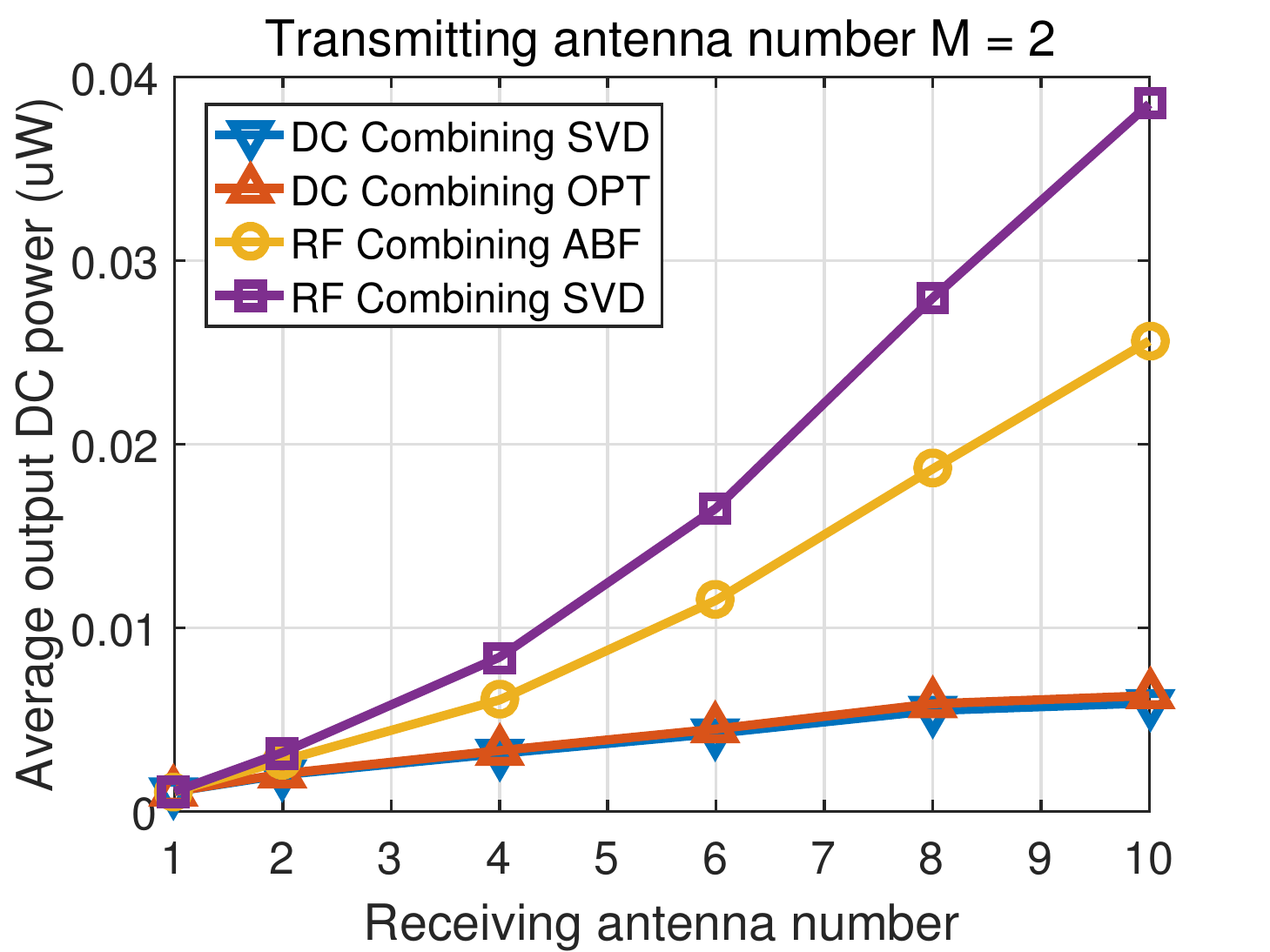}
\par\end{centering}
\begin{centering}
\includegraphics[scale=0.31]{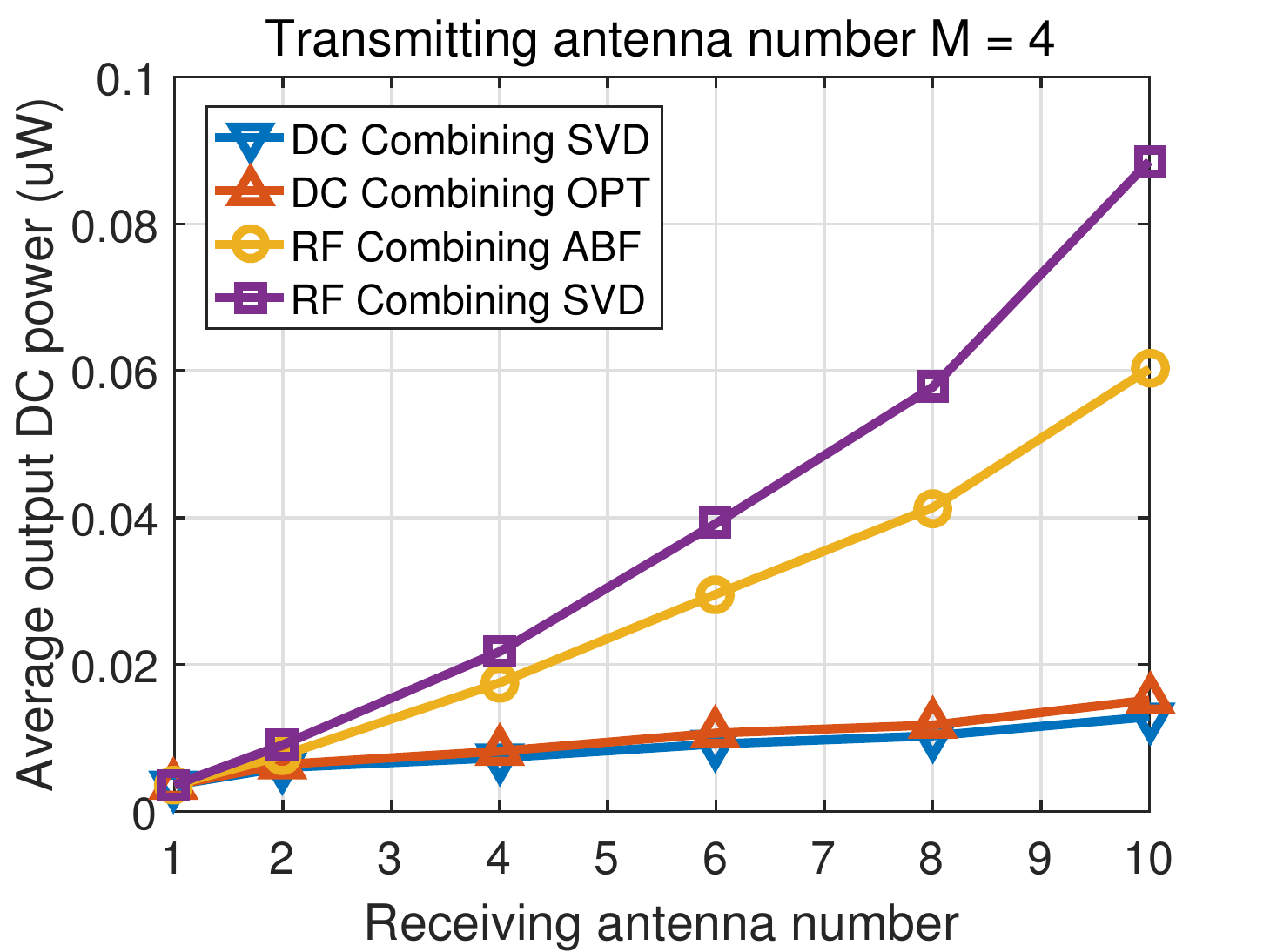}\includegraphics[scale=0.31]{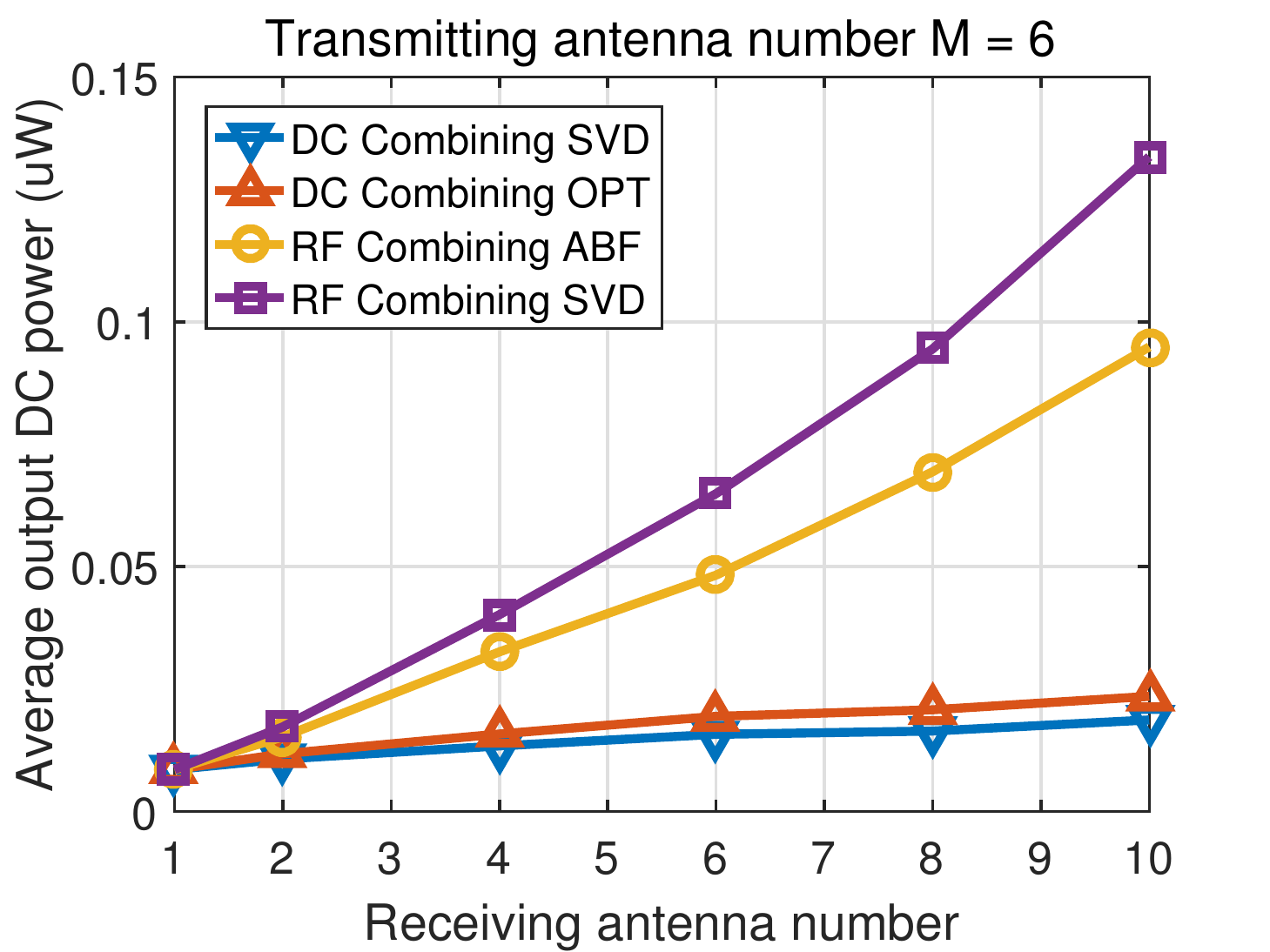}
\par\end{centering}
\begin{centering}
\includegraphics[scale=0.31]{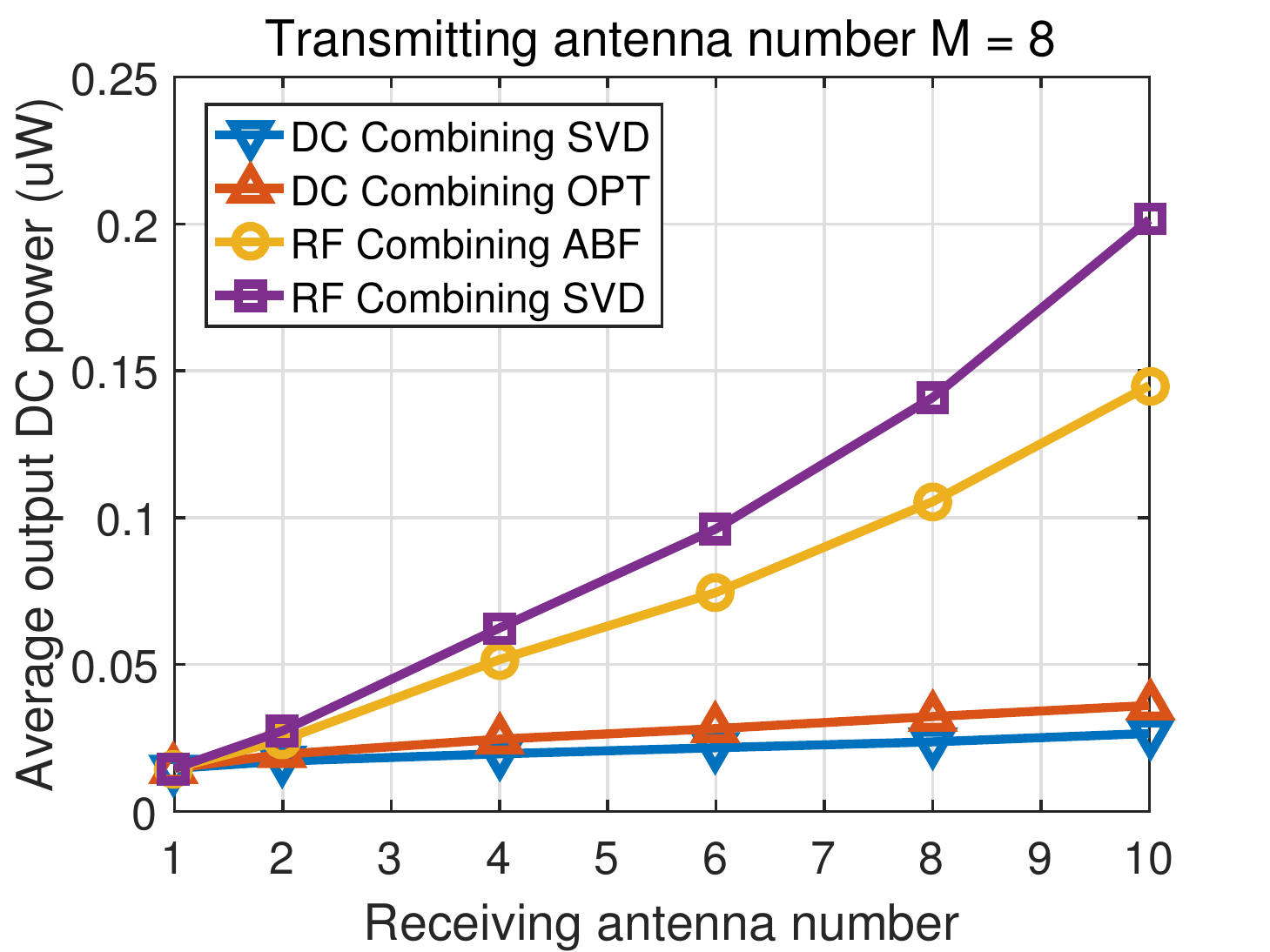}\includegraphics[scale=0.31]{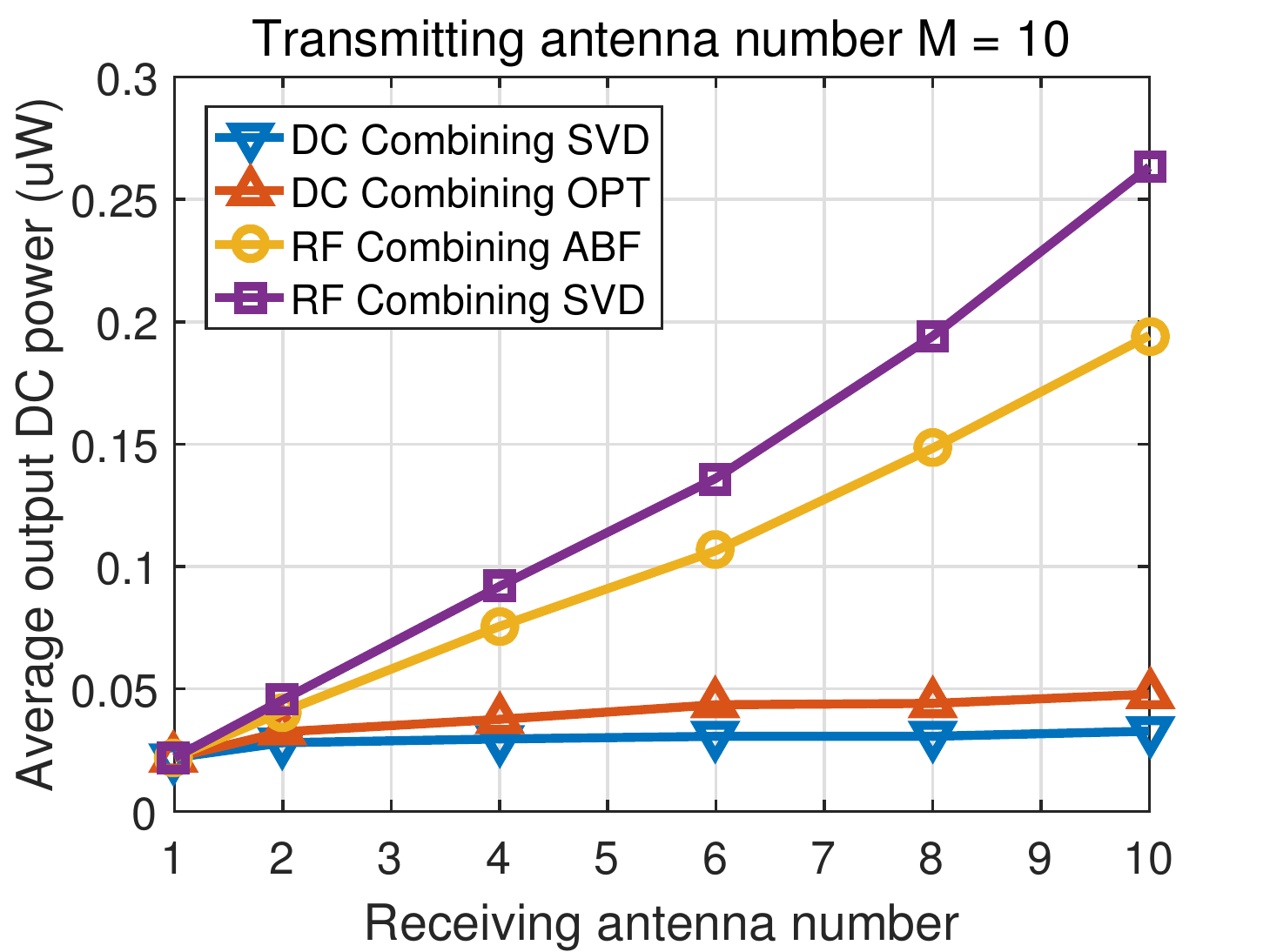}
\par\end{centering}
\caption{\label{fig:Average-output-DC-Rectenna Model}Average output DC power
versus the number of receive antennas for different numbers of transmit
antennas based on the rectenna model.}
\end{figure}

Fig. \ref{fig:Average-output-DC-Rectenna Model} displays the output
DC power $P_{\mathrm{out}}$ averaged over many channel realizations
versus the number of receive antennas for different numbers of transmit
antennas. We make the following observations. \textsl{First}, the
output DC power increases with the number of transmit/receive antennas
for the four beamformings with DC and RF combinings, showing that
the output DC power can be effectively increased by using multiple
antennas at the transmitter/receiver. \textsl{Second}, for DC combining,
the OPT beamforming achieves higher output DC power than the beamforming
based on SVD. This is because the OPT beamforming leverages the rectenna
nonlinearity that the RF-to-DC conversion efficiency increases with
the input RF power while the beamforming based on SVD ignores the
rectenna nonlinearity. \textsl{Third}, for RF combining, the general
receive beamforming based on SVD outperforms the analog receive beamforming.
This is because the constraints of the analog receive beamforming
\eqref{eq:Phaseshift-constraint-1}, \eqref{eq:Phaseshift-constraint-2}
restrict the received RF power. \textsl{Fourth}, RF combining outperforms
DC combining, especially when the number of receive antennas goes
large. This is because the receive beamforming in RF combining leverages
the rectenna nonlinearity more efficiently by inputting the combined
RF signal into a single rectifier (see Section VI). Since the RF-to-DC
conversion efficiency increases with the input RF power, RF combining
can achieve a higher RF-to-DC conversion efficiency than DC combining
which inputs each RF signal to each rectifier, so that RF combinings
has a higher output DC power.

\begin{figure}[t]
\begin{centering}
\includegraphics[scale=0.31]{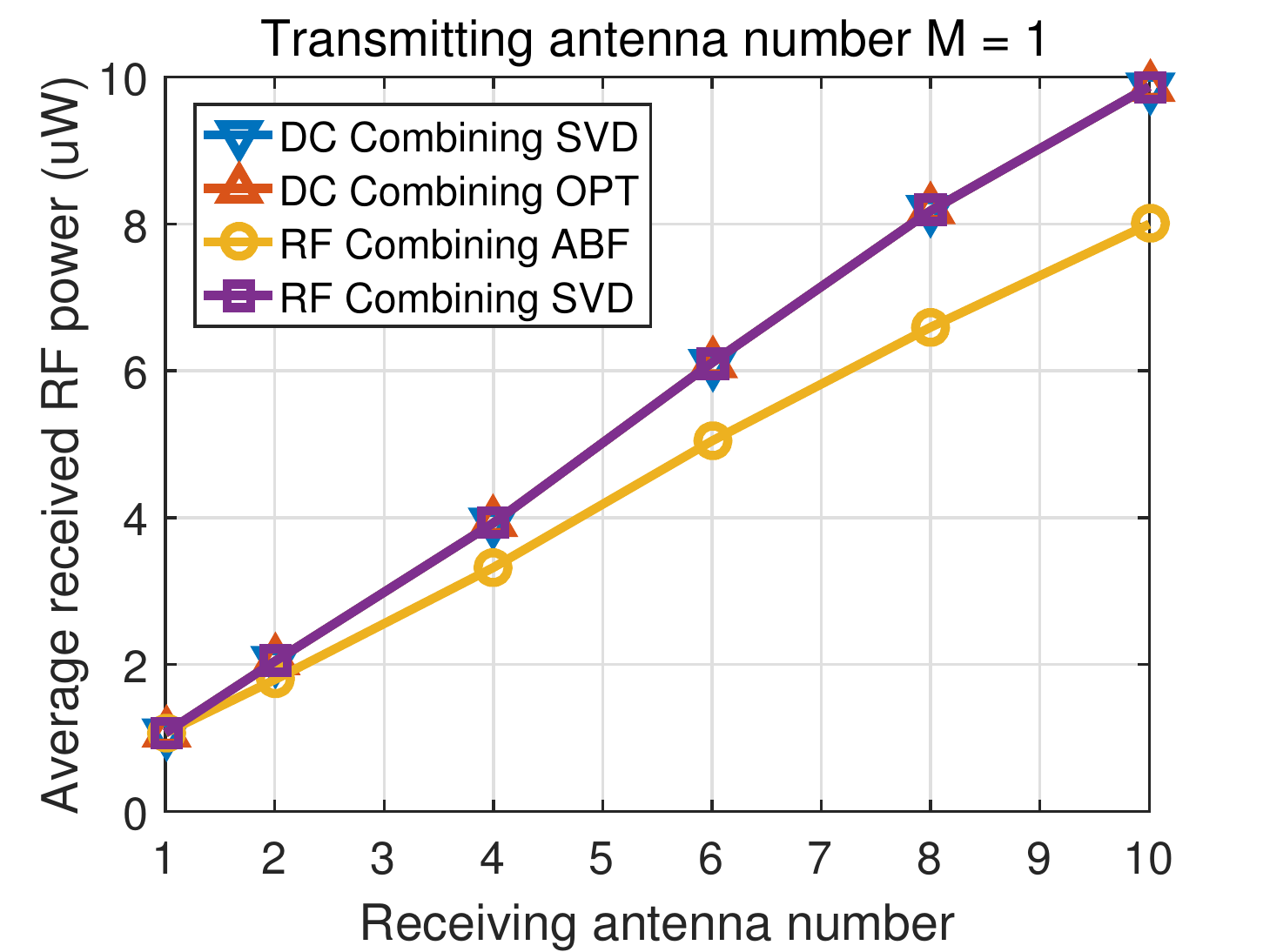}\includegraphics[scale=0.31]{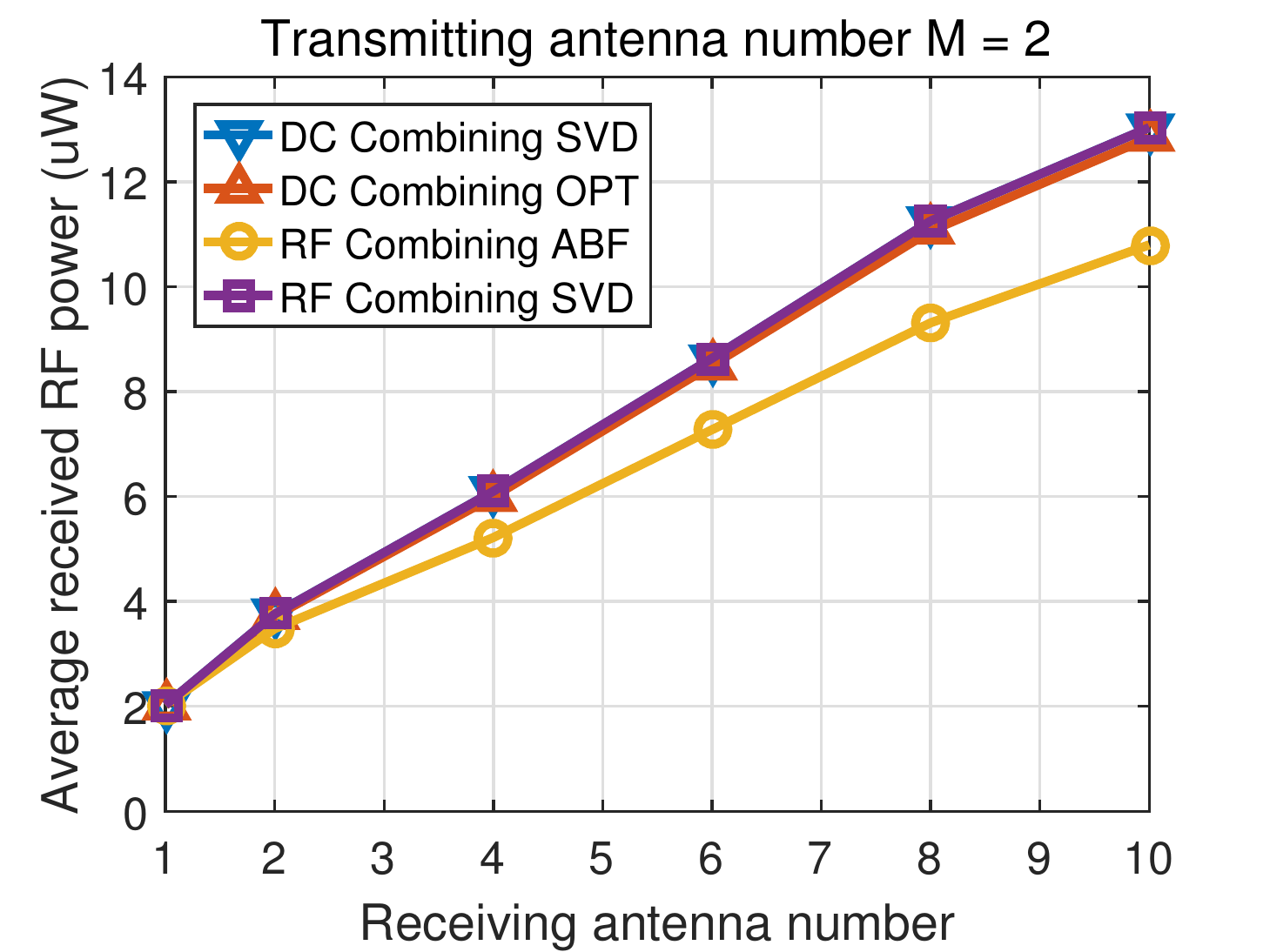}
\par\end{centering}
\begin{centering}
\includegraphics[scale=0.31]{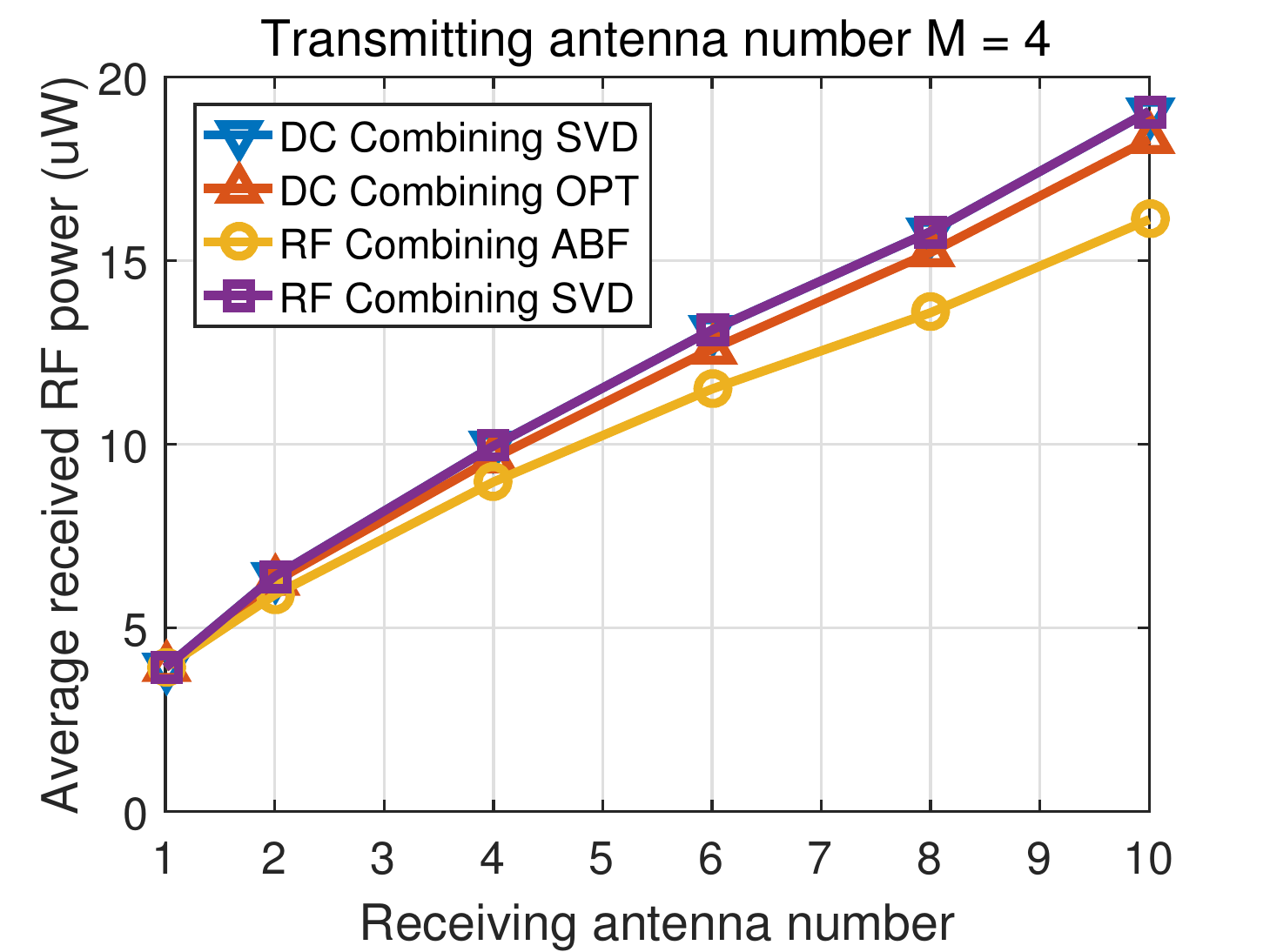}\includegraphics[scale=0.31]{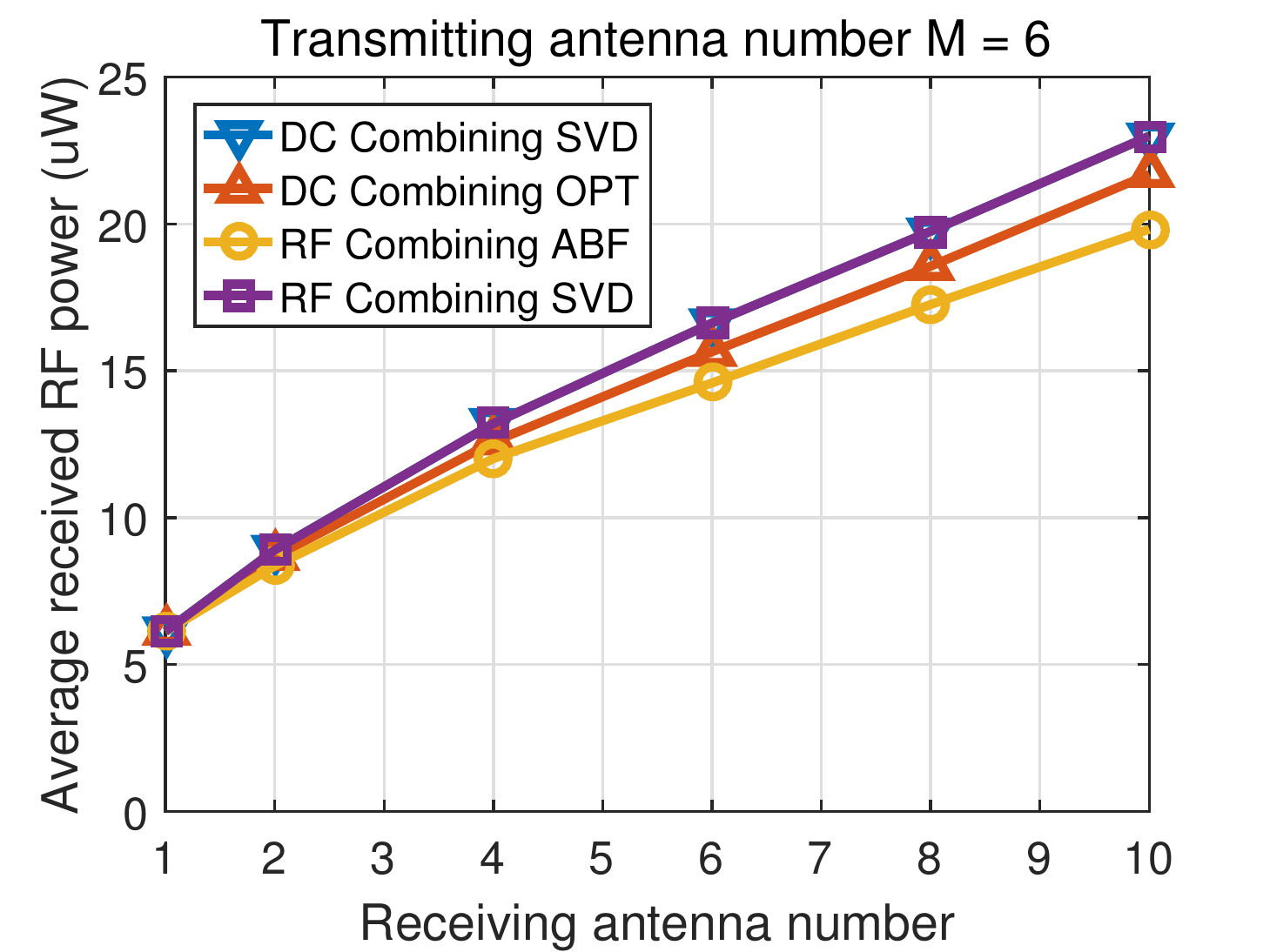}
\par\end{centering}
\begin{centering}
\includegraphics[scale=0.31]{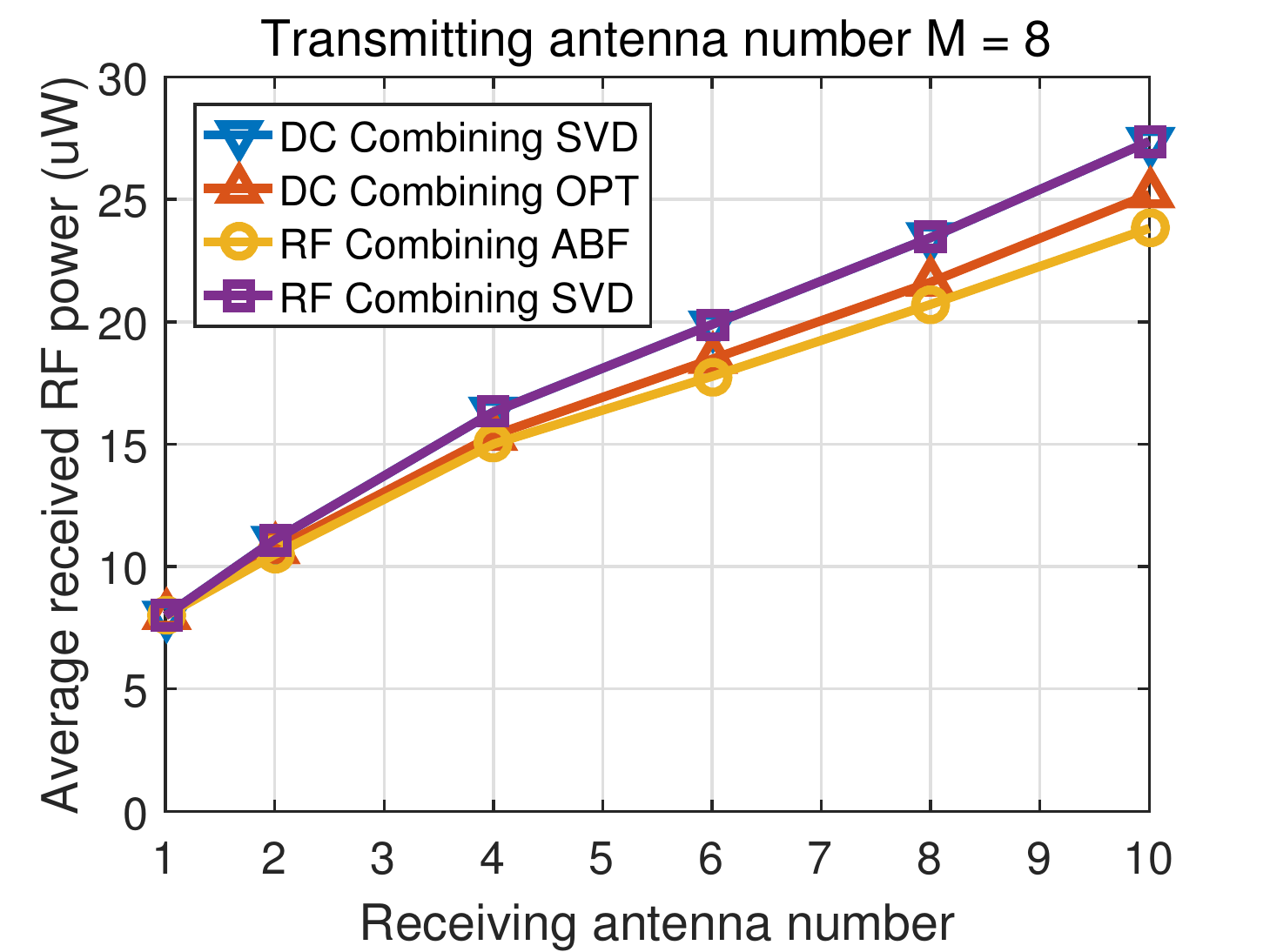}\includegraphics[scale=0.31]{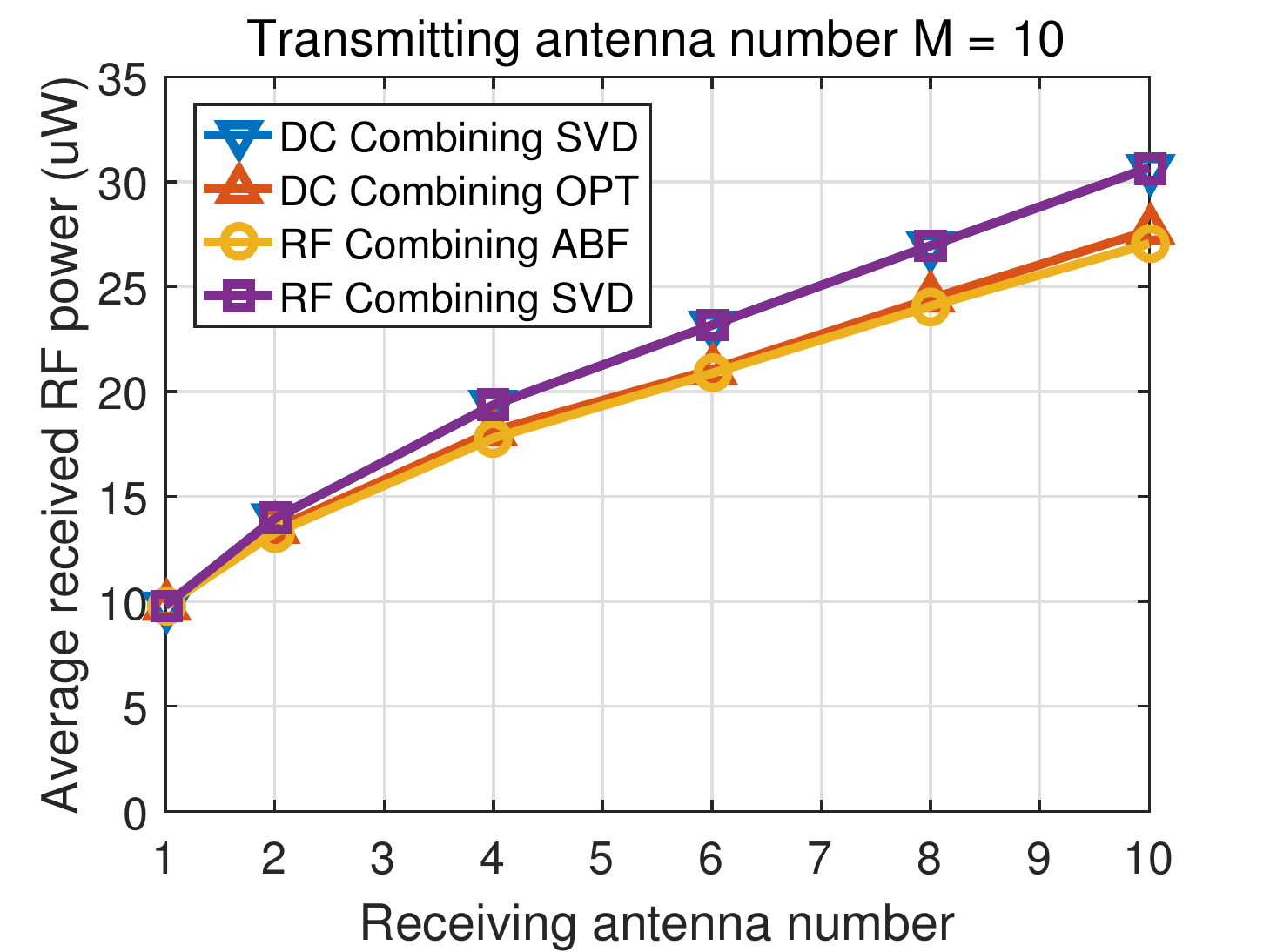}
\par\end{centering}
\caption{\label{fig:Average-received-RF}Average received RF power versus the
number of receive antennas for different numbers of transmit antennas.}
\end{figure}

To get insights into how the rectenna nonlinearity influences the
output DC power, Fig. \ref{fig:Average-received-RF} displays the
received RF power averaged over many channels versus the number of
receive antennas for different numbers of transmit antennas. We make
the following observations. \textsl{First}, the beamforming based
on SVD in DC combining has the same received RF power as the beamforming
based on SVD in RF combining, which is obvious since they all use
the SVD of channel matrix $\mathbf{H}$. \textsl{Second}, for RF combining,
the analog receive beamforming has less received RF power than the
general receive beamforming based on SVD which is due to the constraints
of analog receive beamforming \eqref{eq:Phaseshift-constraint-1},
\eqref{eq:Phaseshift-constraint-2}. \textsl{Third}, for DC combining,
the OPT beamforming has less received RF power than the beamforming
based on SVD.

Based on the above observations from Fig. \ref{fig:Average-output-DC-Rectenna Model}
and Fig. \ref{fig:Average-received-RF}, we can find that achieving
the maximum received RF power does not mean achieving the maximum
output DC power due to the rectenna nonlinearity. Therefore, the rectenna
nonlinearity should be leveraged in WPT to increase the output DC
power. This behavior has been extensively emphasized in \cite{2016_TSP_WPT_Bruno_Waveform},
\cite{2018_MM_WPT_Bruno_1GWPT}, but finds further consequences in
MIMO WPT. Namely, it is concluded from the simulations that two approaches
can be used to leverage the rectenna nonlinearity: 1) optimizing beamforming
by considering the rectenna nonlinearity in DC combining, and 2) using
RF combining which can leverage the rectenna nonlinearity more efficiently.

We also evaluate the performance of Algorithms 1 and 2, which both
use the technique SDR. For Algorithm 1, we find that $\mathbf{W}_{T}^{\star}$
is rank-1 for all tested channel realizations with different $M$
and $Q$. Hence, Algorithm 1 finds a stationary point of the problem
\eqref{eq:OP1-MaxPoutStP-1}-\eqref{eq:OP1-MaxPoutStP-2} so its performance
is guaranteed for all tested channel realizations. For Algorithm 2,
we find that $\mathbf{W}_{R}^{\star}$ is rank-1 for most of the tested
channel realizations with different $M$ and $Q$, but not rank-1
for some channel realizations. So, for each channels realization,
we compute a quantity $R_{1}=\lambda_{\mathrm{max}}\left(\mathbf{W}_{R}^{\star}\right)/\sum_{q=1}^{Q}\lambda_{q}\left(\mathbf{W}_{R}^{\star}\right)$
where $\lambda_{\mathrm{max}}\left(\mathbf{W}_{R}^{\star}\right)$
is the maximum eigenvalue of $\mathbf{W}_{R}^{\star}$ and $\sum_{q=1}^{Q}\lambda_{q}\left(\mathbf{W}_{R}^{\star}\right)$
is the sum of all the eigenvalues of $\mathbf{W}_{R}^{\star}$. For
$\mathbf{W}_{R}^{\star}\succeq0$, $R_{1}=1$ means $\mathbf{W}_{R}^{\star}$
is rank-1, and heuristically we have that $\mathbf{W}_{R}^{\star}$
is close to rank-1 when $R_{1}$ is close to 1. The average $R_{1}$
over channel realizations, denoted as $\bar{R}_{1}$, are summarized
in Table \ref{tab:Evaluations-of-Algorithm}. We find that $\bar{R}_{1}\geq0.94$
for different $M$ and $Q$, showing that $\mathbf{W}_{R}^{\star}$
is highly likely to be rank-1. Besides, we also compute a quantity
$R_{2}=\mathrm{Tr}\left(\mathbf{H}\mathbf{H}^{H}\mathbf{w}_{R}^{\star}\mathbf{w}_{R}^{\star H}\right)/\mathrm{Tr}\left(\mathbf{H}\mathbf{H}^{H}\mathbf{W}_{R}^{\star}\right)$
which shows how tight the SDR is in the problem \eqref{eq:OP10-MaxTrHHWR-SDR-1}-\eqref{eq:OP10-MaxTrHHWR-SDR-4}.
The average $R_{2}$ over channel realizations, denoted as $\bar{R}_{2}$,
are also summarized in Table \ref{tab:Evaluations-of-Algorithm}.
We can find that $\bar{R}_{2}\geq0.997$ for different $M$ and $Q$,
showing that the SDR is very tight. So, Algorithm 2 finds nearly the
global optimal solution of the problem \eqref{eq:OP10-MaxTrHHWR-SDR-1}-\eqref{eq:OP10-MaxTrHHWR-SDR-4}
and its performance is guaranteed for the tested channel realizations.

\begin{table*}[t]
\caption{\label{tab:Evaluations-of-Algorithm}Evaluations of Algorithm 2.}

\centering{}%
\begin{tabular}{|c|c|c|c|c|c|c|c|c|c|c|c|c|}
\hline 
 & \multicolumn{2}{c|}{{\scriptsize{}$Q=1$}} & \multicolumn{2}{c|}{{\scriptsize{}$Q=2$}} & \multicolumn{2}{c|}{{\scriptsize{}$Q=4$}} & \multicolumn{2}{c|}{{\scriptsize{}$Q=6$}} & \multicolumn{2}{c|}{{\scriptsize{}$Q=8$}} & \multicolumn{2}{c|}{{\scriptsize{}$Q=10$}}\tabularnewline
\hline 
\hline 
 & {\scriptsize{}$\bar{R}_{1}$} & {\scriptsize{}$\bar{R}_{2}$} & {\scriptsize{}$\bar{R}_{1}$} & {\scriptsize{}$\bar{R}_{2}$} & {\scriptsize{}$\bar{R}_{1}$} & {\scriptsize{}$\bar{R}_{2}$} & {\scriptsize{}$\bar{R}_{1}$} & {\scriptsize{}$\bar{R}_{2}$} & {\scriptsize{}$\bar{R}_{1}$} & {\scriptsize{}$\bar{R}_{2}$} & {\scriptsize{}$\bar{R}_{1}$} & {\scriptsize{}$\bar{R}_{2}$}\tabularnewline
\hline 
{\scriptsize{}$M=1$} & {\scriptsize{}1.0000} & {\scriptsize{}1.0000} & {\scriptsize{}0.9970} & {\scriptsize{}0.9999} & {\scriptsize{}0.9991} & {\scriptsize{}0.9999} & {\scriptsize{}0.9988} & {\scriptsize{}0.9999} & {\scriptsize{}0.9998} & {\scriptsize{}0.9999} & {\scriptsize{}0.9993} & {\scriptsize{}0.9999}\tabularnewline
\hline 
{\scriptsize{}$M=2$} & {\scriptsize{}1.0000} & {\scriptsize{}1.0000} & {\scriptsize{}0.9983} & {\scriptsize{}0.9999} & {\scriptsize{}0.9976} & {\scriptsize{}0.9999} & {\scriptsize{}0.9946} & {\scriptsize{}0.9999} & {\scriptsize{}0.9847} & {\scriptsize{}0.9997} & {\scriptsize{}0.9716} & {\scriptsize{}0.9989}\tabularnewline
\hline 
{\scriptsize{}$M=4$} & {\scriptsize{}1.0000} & {\scriptsize{}1.0000} & {\scriptsize{}0.9989} & {\scriptsize{}0.9999} & {\scriptsize{}0.9975} & {\scriptsize{}0.9999} & {\scriptsize{}0.9910} & {\scriptsize{}0.9999} & {\scriptsize{}0.9818} & {\scriptsize{}0.9996} & {\scriptsize{}0.9681} & {\scriptsize{}0.9991}\tabularnewline
\hline 
{\scriptsize{}$M=6$} & {\scriptsize{}1.0000} & {\scriptsize{}1.0000} & {\scriptsize{}0.9993} & {\scriptsize{}0.9999} & {\scriptsize{}0.9969} & {\scriptsize{}0.9999} & {\scriptsize{}0.9858} & {\scriptsize{}0.9999} & {\scriptsize{}0.9781} & {\scriptsize{}0.9993} & {\scriptsize{}0.9470} & {\scriptsize{}0.9978}\tabularnewline
\hline 
{\scriptsize{}$M=8$} & {\scriptsize{}1.0000} & {\scriptsize{}1.0000} & {\scriptsize{}0.9995} & {\scriptsize{}0.9999} & {\scriptsize{}0.9974} & {\scriptsize{}0.9999} & {\scriptsize{}0.9861} & {\scriptsize{}0.9999} & {\scriptsize{}0.9717} & {\scriptsize{}0.9993} & {\scriptsize{}0.9524} & {\scriptsize{}0.9982}\tabularnewline
\hline 
{\scriptsize{}$M=10$} & {\scriptsize{}1.0000} & {\scriptsize{}1.0000} & {\scriptsize{}0.9996} & {\scriptsize{}0.9999} & {\scriptsize{}0.9972} & {\scriptsize{}0.9999} & {\scriptsize{}0.9815} & {\scriptsize{}0.9996} & {\scriptsize{}0.9664} & {\scriptsize{}0.9990} & {\scriptsize{}0.9482} & {\scriptsize{}0.9983}\tabularnewline
\hline 
\end{tabular}
\end{table*}

\subsection{Accurate and Realistic Performance Evaluations}

\begin{figure}[t]
\begin{centering}
\includegraphics[scale=0.42]{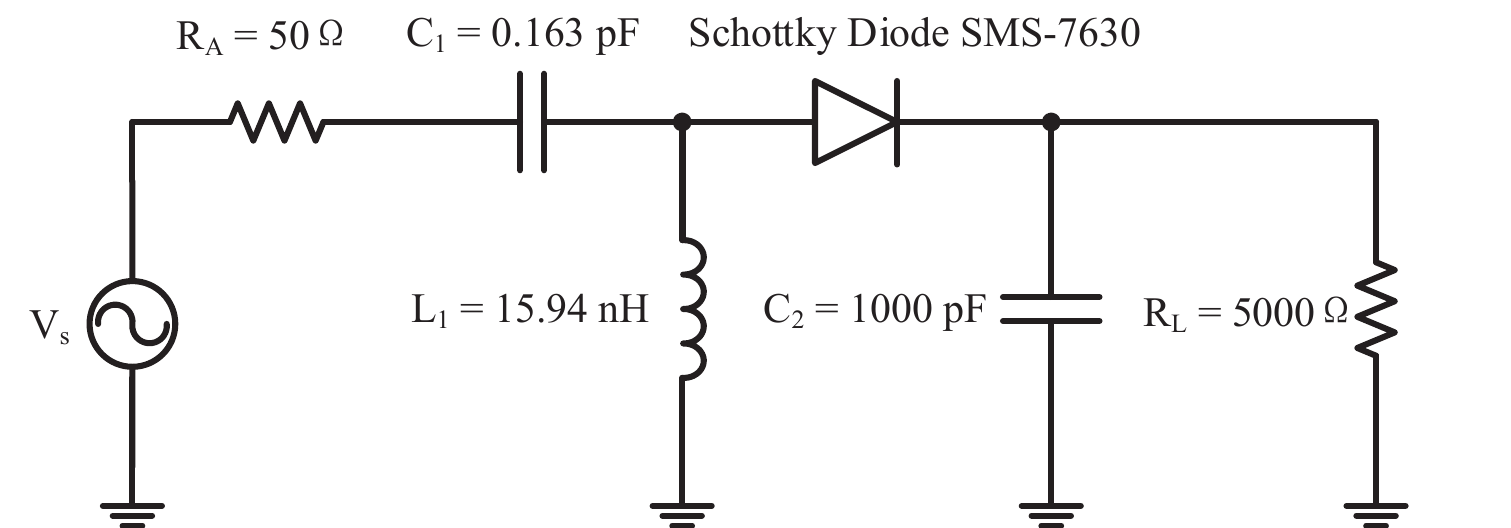}
\par\end{centering}
\caption{\label{fig:Rectenna-ADS}Rectenna with a single diode and L-matching
network used for circuit evaluation in ADS.}
\end{figure}

The second type of evaluations is based on an accurate modeling of
the rectenna in ADS in order to validate the beamforming optimization
with the two combinings and the rectenna nonlinearity model. To that
end, in DC combining the RF signal received by each receive antenna
is used as input to a realistic rectenna as shown in Fig. \ref{fig:Rectenna-ADS}.
Hence $Q$ rectifiers as shown in Fig. \ref{fig:Rectenna-ADS} are
used. While in RF combining, only one rectifier as shown in Fig. \ref{fig:Rectenna-ADS}
is used. The rectenna circuit contains a voltage source, an antenna
impedance, an L-matching network, a Schottky diode SMS-7630, a capacitor
as low-pass filter, and a load resistor. The L-matching network is
used to guarantee a good matching between the rectifier and the antenna
and to minimize the impedance mismatch due to variations in input
RF power level. By using the Harmonic Balance solver with the SPICE
model of SMS-7630 in ADS, the values of the capacitor $C_{1}$ and
the inductor $L_{1}$ in the matching network are optimized to achieve
a good impedance matching.

\begin{figure}[t]
\begin{centering}
\includegraphics[scale=0.31]{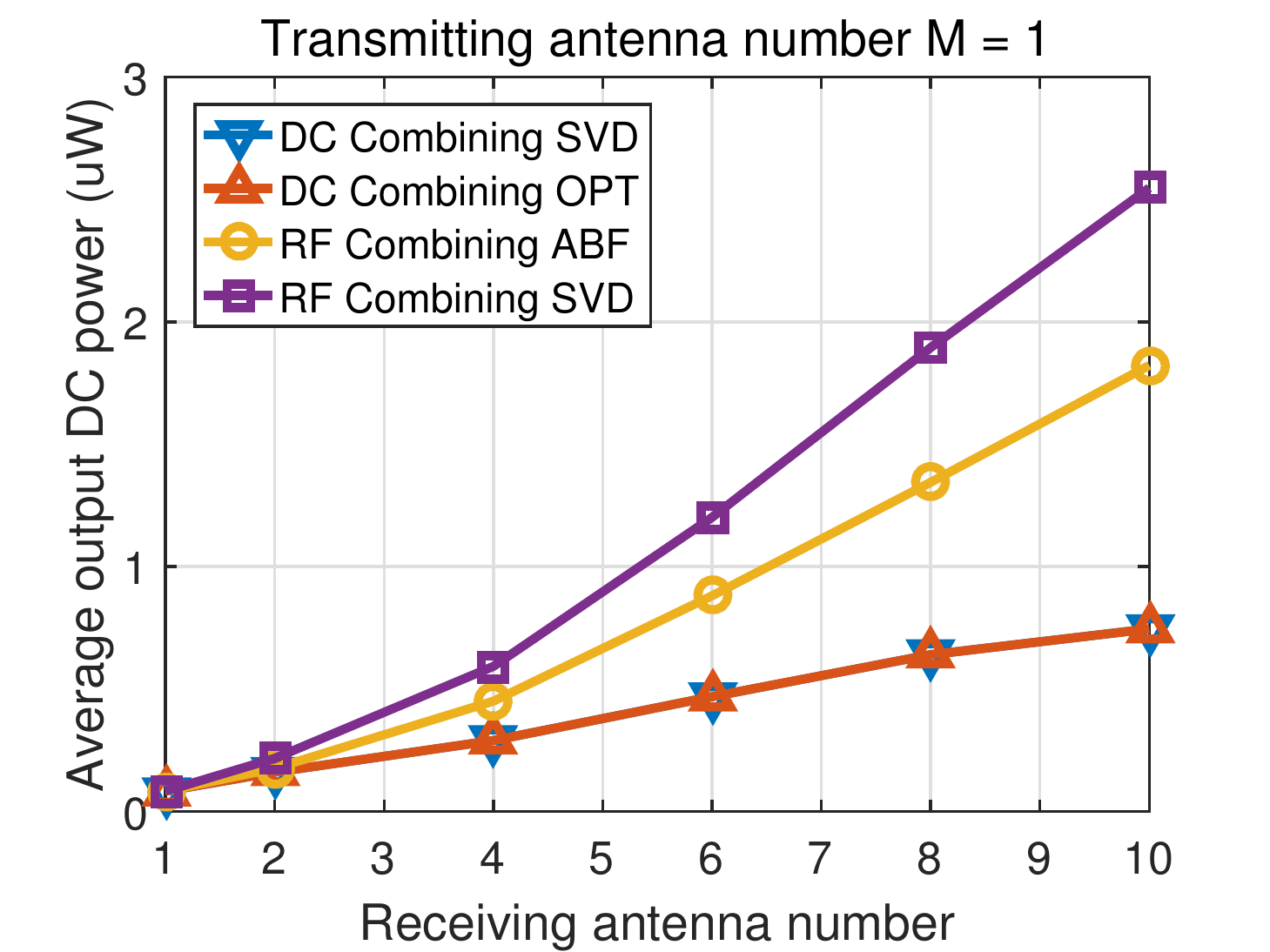}\includegraphics[scale=0.31]{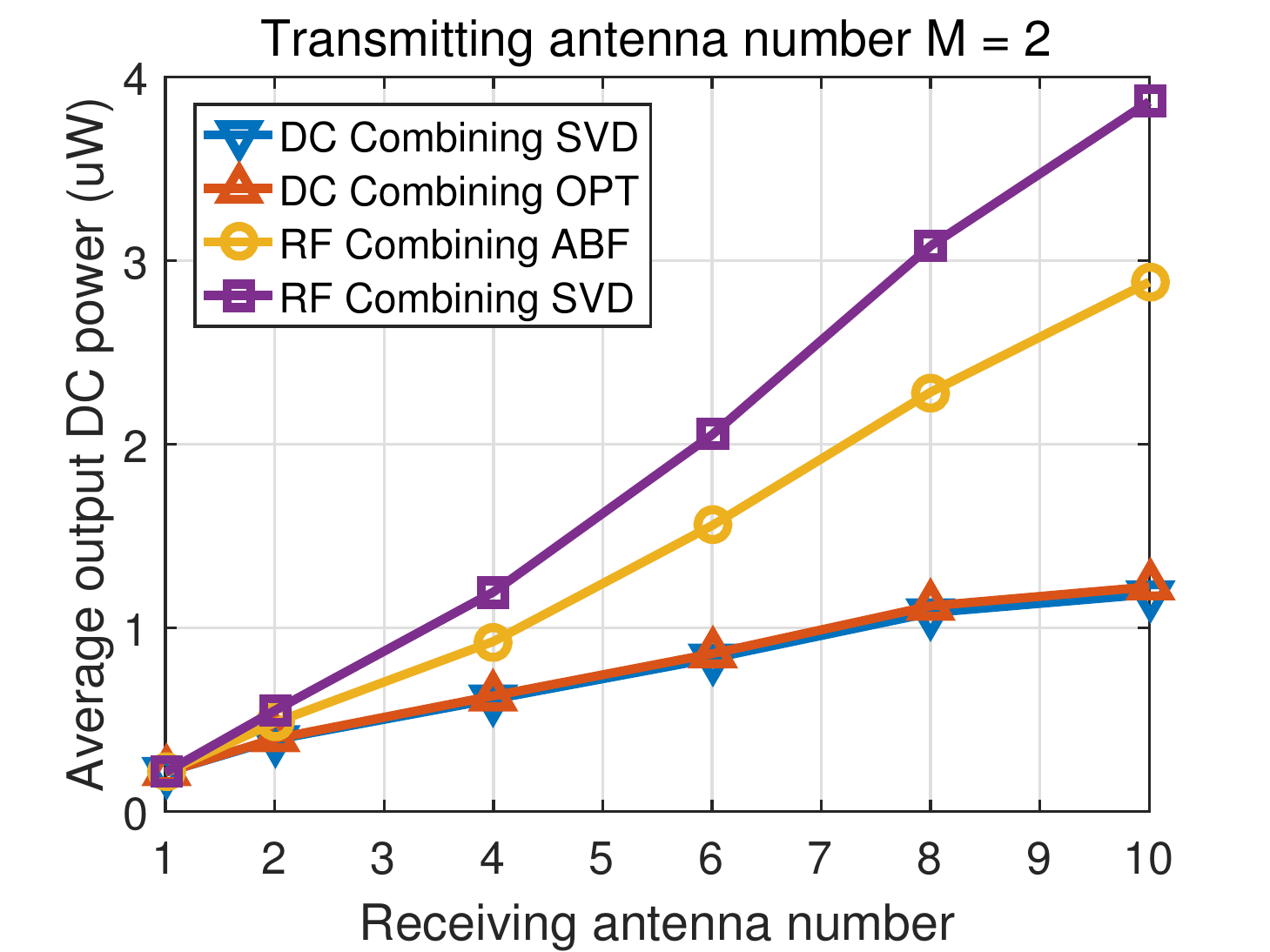}
\par\end{centering}
\begin{centering}
\includegraphics[scale=0.31]{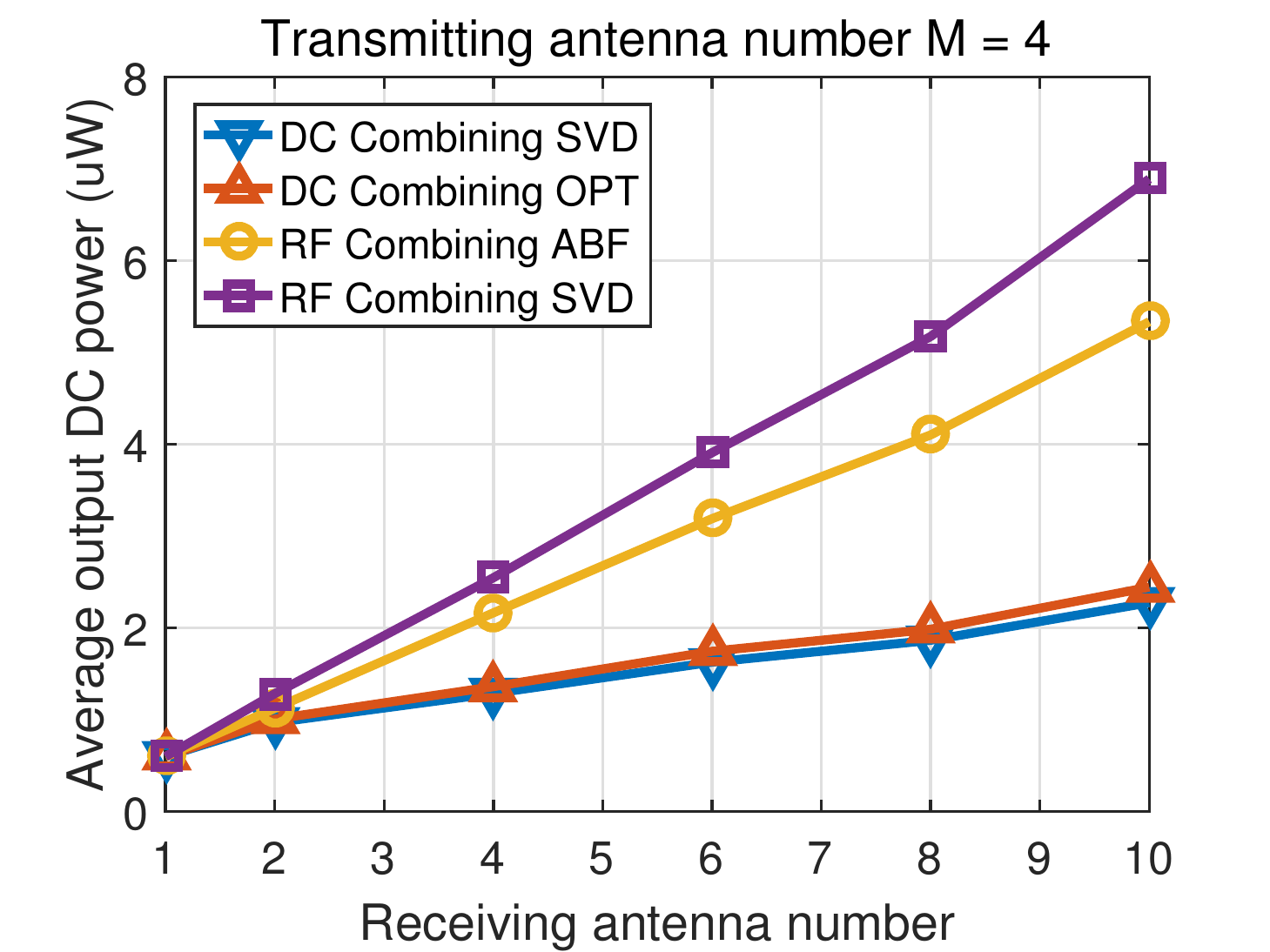}\includegraphics[scale=0.31]{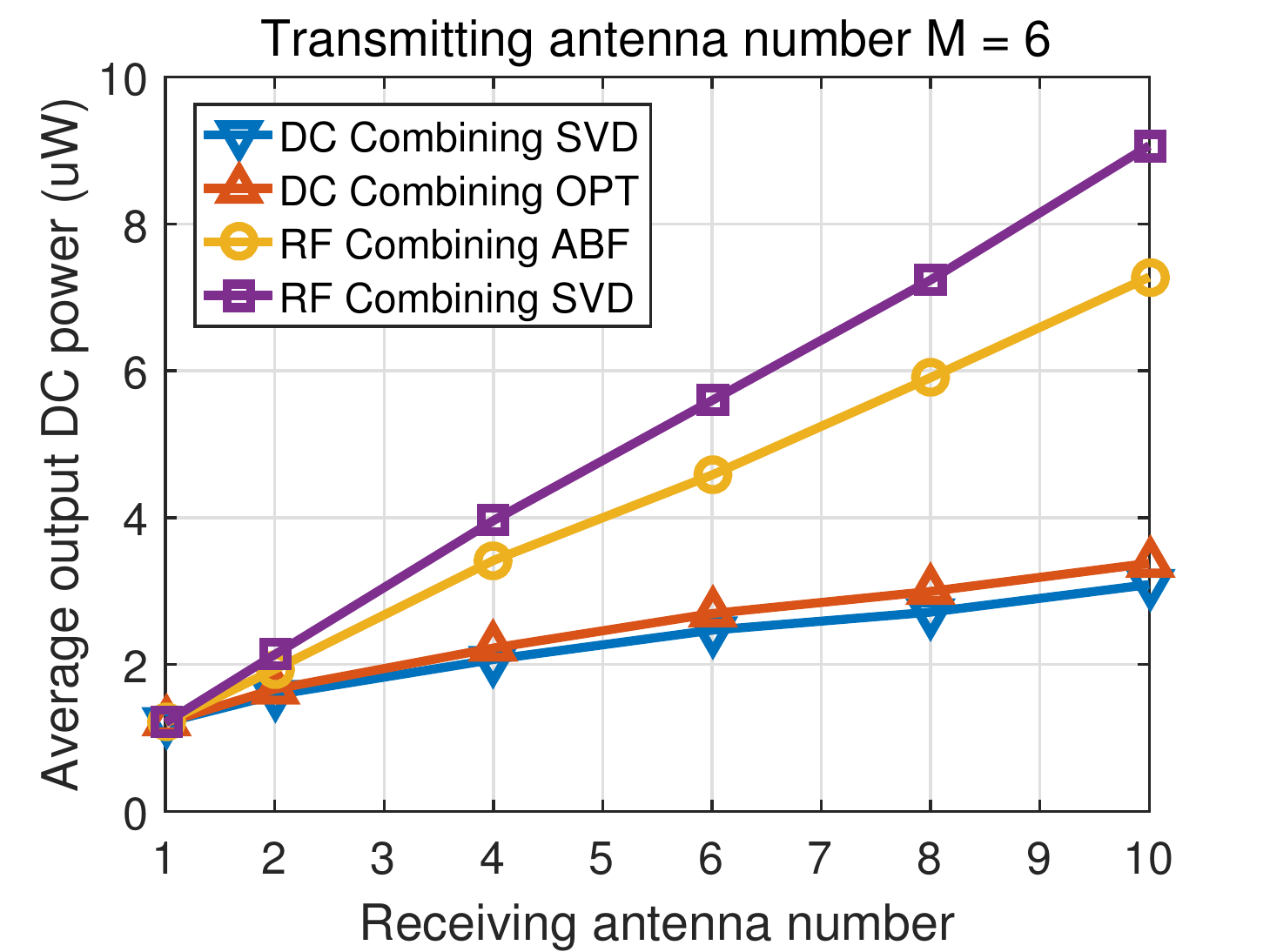}
\par\end{centering}
\begin{centering}
\includegraphics[scale=0.31]{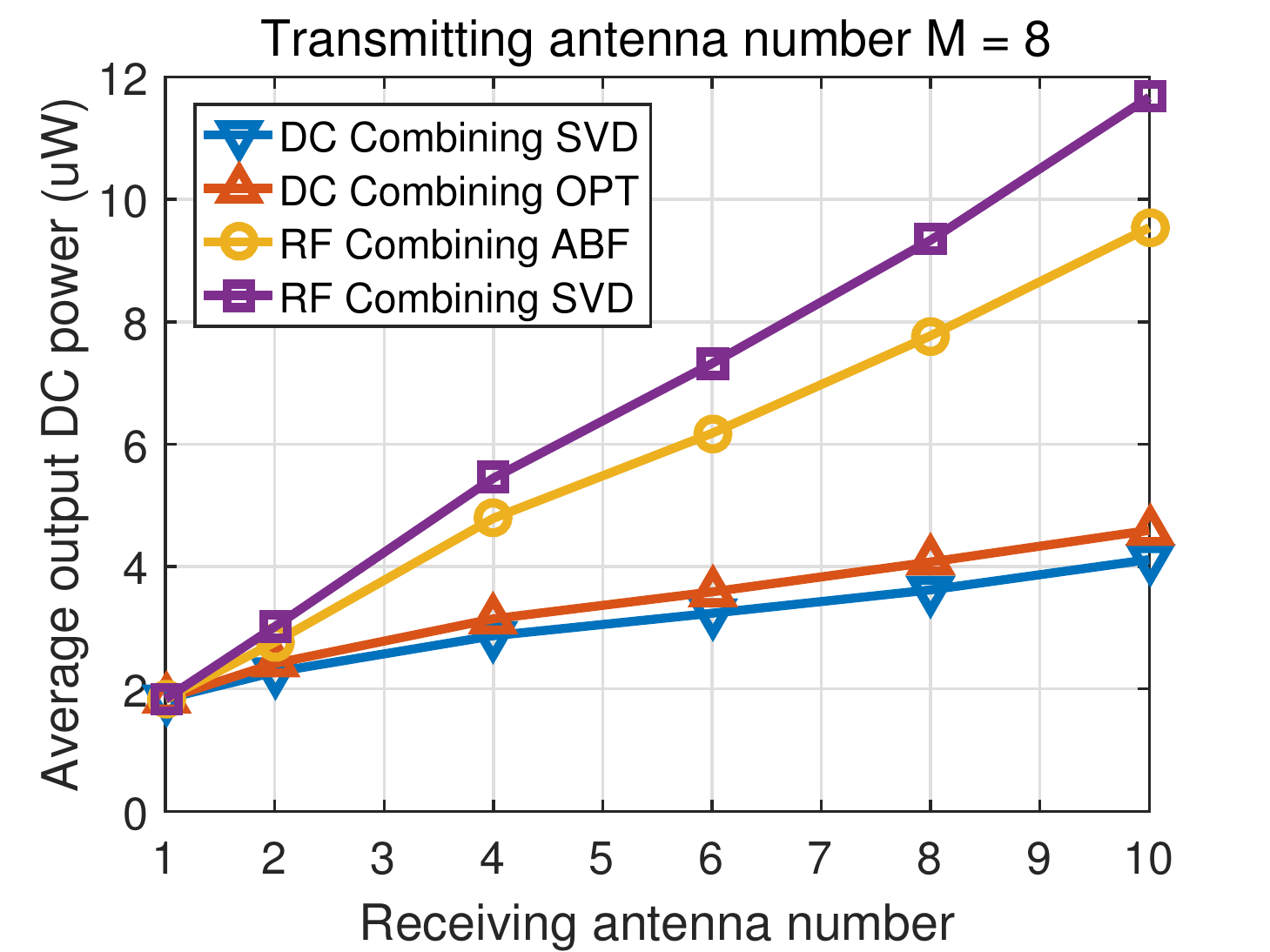}\includegraphics[scale=0.31]{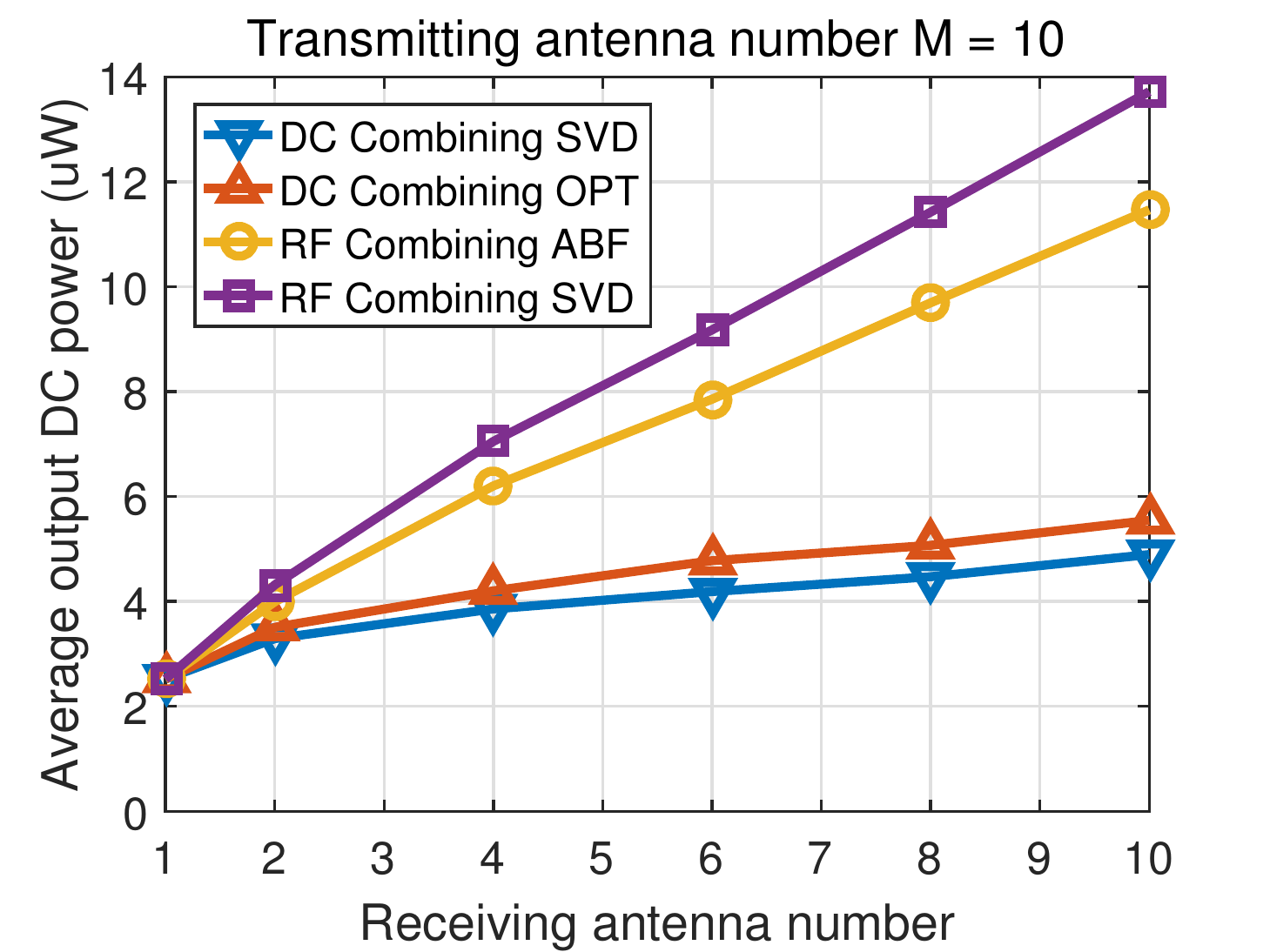}
\par\end{centering}
\caption{\label{fig:Average-output-DC-SPICE}Average output DC power versus
the number of receive antennas for different numbers of transmit antennas
based on the accurate and realistic circuit simulation.}
\end{figure}

We now evaluate the performance of the MIMO WPT system with DC and
RF combinings using the accurate modeling of rectenna in ADS. Again,
we consider two DC combinings (based on SVD and OPT) and two RF combinings
(based on ABF and SVD). Fig. \ref{fig:Average-output-DC-SPICE} displays
the average (over many channel realizations) output DC power solved
by ADS versus the number of receive antennas for different numbers
of transmit antennas. Similar to Fig. \ref{fig:Average-output-DC-Rectenna Model},
we can make the following observations. \textit{First}, the output
DC power increases with the number of transmit/receive antennas in
both DC and RF combinings. \textit{Second}, for DC combining, the
OPT beamforming achieves higher output DC power than the beamforming
based on SVD. \textit{Third}, for RF combining, the general receive
beamforming based on SVD outperforms the analog receive beamforming.
\textit{Fourth}, RF combinings lead to higher average output DC power
than DC combining, especially when the number of receive antennas
goes large. The relative gain of RF combining versus DC combining
can exceed 100\% when the receive antenna number is larger than 6,
and it can reach 240\% (achieved when there is a single transmit antenna
and 10 receive antennas). The explanations for these observations
can be found in the above evaluations. Therefore, the similar observations
in Fig. \ref{fig:Average-output-DC-Rectenna Model} and Fig. \ref{fig:Average-output-DC-SPICE}
confirm the usefulness of the rectenna nonlinearity model and demonstrate
the necessity of leveraging the rectenna nonlinearity in WPT to boost
the output DC power.

\section{Discussions}

In addition to evaluating the performance of DC combining versus RF
combining in MIMO WPT system, it is also worth discussing the impact
of DC combining versus RF combining on the WPT system design.

\subsection{Channel Estimation}

DC combining only needs CSIT for transmit beamforming optimization.
However, RF combining not only needs CSIT but also need CSIR for joint
transmit and receive beamforming optimization. Therefore, channel
estimation is required at the receiver for RF combining, which increases
the complexity of WPT system design. Alternatively, some forms of
communication between the transmitter and the receiver is needed to
tell the receiver about the combining weights to use.

\subsection{Number of Rectifiers}

In DC combining, each receive antenna is connected to a rectifier
so that the number of rectifiers increases with the number of receive
antennas. However, in RF combining, the RF signals from all receive
antennas are firstly RF combined so that only a single rectifier is
needed to rectify the combined RF signal.

\subsection{Combining Network}

DC combining does not need a RF combining circuit. Instead, it needs
a good DC combining circuit such as MIMO switching DC-DC converter
\cite{KKKKI}-\nocite{MIMO_Convertor_ISSCC}\cite{MIMO_Convertor_JSSC}.
Because the RF power input into each rectifier is different in the
MIMO WPT system, the optimal DC bias voltage for maximizing the RF-to-DC
efficiency is different for each rectifier. The MIMO switching DC-DC
converter provides the optimal DC bias voltage for each rectifier
so that the output DC power of all rectifiers can be efficiently combined.
However, using the MIMO switching DC-DC converter increases the complexity
of WPT system design. On the other hand, RF combining only needs a
simple single-input and single-output (SISO) switching DC-DC converter,
but RF combining needs a RF combining circuit to combine all the received
RF signal. The proposed analog receive beamforming consists of multiple
phase shifters and a RF power combiner, and the number of phase shifters
increases with the number of receive antennas.

A comprehensive comparison of DC combining and RF combining is shown
in Table. \ref{tab:Comprehensive-Comparsion-of}. We can conclude
that RF combining achieves high output DC power at the cost of CSIT
and CSIR and RF combining circuit while DC combining is limited by
the low output DC power, multiple rectifiers, and MIMO DC-DC converter
but it only needs CSIT and it does not need a RF combining circuit.

\section{Conclusion and Future Works}

In this paper, we propose a MIMO WPT architecture where the transmitter
and the receiver are equipped with multiple antennas, so as to enhance
the output DC power level. Two combining strategies at the receiver
are introduced, namely DC and RF combining, and the corresponding
beamformers are optimized considering the rectenna nonlinearity.

For DC combining, assuming perfect CSIT and using the rectenna nonlinearity
model, the beamforming for multiple antennas at the transmitter adaptive
to CSI is optimized to maximize the output DC power by solving a non-convex
posynomial maximization problem with SDR. It is also numerically shown
that the proposed algorithm can find a stationary point of the problem
for the tested channel realizations.

\begin{table}[t]
\caption{\label{tab:Comprehensive-Comparsion-of}Comprehensive Comparison of
DC Combining and RF Combining}

\centering{}%
\begin{tabular}{|c|c|c|}
\hline 
 & {\scriptsize{}DC combining} & {\scriptsize{}RF combining}\tabularnewline
\hline 
\hline 
{\scriptsize{}Output DC power} & {\scriptsize{}low} & {\scriptsize{}high}\tabularnewline
\hline 
{\scriptsize{}Channel estimation} & {\scriptsize{}CSIT} & {\scriptsize{}CSIT and CSIR}\tabularnewline
\hline 
{\scriptsize{}Number of rectifiers} & {\scriptsize{}Multiple} & {\scriptsize{}Single}\tabularnewline
\hline 
{\scriptsize{}DC-DC converter} & {\scriptsize{}MIMO} & {\scriptsize{}SISO}\tabularnewline
\hline 
{\scriptsize{}RF Combining Circuit} & {\scriptsize{}No need} & {\scriptsize{}Need}\tabularnewline
\hline 
\end{tabular}
\end{table}

For RF combining, assuming perfect CSIT and CSIR and using the rectenna
nonlinearity model, the beamforming for multiple antennas at both
the transmitter and receiver adaptive to CSI is optimized for output
DC power maximization. The global optimal transmit and receive beamforming
are obtained in closed form. In addition, we propose a practical RF
combining circuit using RF phase shifter and RF power combiner and
also optimize the analog receive beamforming adaptive to CSI by solving
a non-convex optimization problem with SDR. It is also numerically
shown that the proposed algorithm can find nearly the global optimal
solution of the problem for the tested channel realizations.

We also analytically derive the scaling laws of the output DC power
for MISO and SIMO systems with DC and RF combinings as a function
of the number of transmit and receive antennas. Those scaling laws
highlight the benefits of using multiple antennas at the transmitter
and the receiver. They also highlight the benefits of using RF combining
over DC combining since the receive beamforming in RF combining leverages
the rectenna nonlinearity more efficiently.

We also provide two types of performance evaluations for DC and RF
combining in MIMO WPT system. The first is based on the nonlinear
rectenna model while the second is based on realistic and accurate
rectenna simulations in ADS. The two evaluations agree well with each
other, demonstrating the usefulness of the rectenna nonlinearity model.
They also show that the output DC power can be linearly increased
by using multiple rectennas at the receiver and that RF combining
outperforms DC combining in the output DC power level since the receive
beamforming in RF combining leverages the rectenna nonlinearity more
efficiently. The relative gain of RF combining versus DC combining
can exceed 100\% when the receive antenna number is larger than 6,
and it can reach 240\% (achieved when there is a single transmit antenna
and 10 receive antennas). The impact of DC combining versus RF combining
on the WPT system design is also discussed and a comprehensive comparison
of DC and RF combinings is provided.

Future research avenues include considering joint beamforming and
waveform design for MIMO WPT system to further leverage the rectenna
nonlinearity for output DC power enhancement and apply the MIMO WPT
system in simultaneous wireless information and power transfer \cite{2013_TWC_SWIPT_RZhang}
and wireless powered communication \cite{2015_ComMag_WirelessPowerComm}.


\begin{thebibliography}{10}

\bibitem{zorzi2010today}
M.~Zorzi, A.~Gluhak, S.~Lange, and A.~Bassi, ``From today's intranet of things
to a future internet of things: a wireless-and mobility-related view,''
\emph{Wireless Communications, IEEE}, vol.~17, no.~6, pp. 44--51, Jun. 2010.

\bibitem{2004TMTT_EH_Popovic}
J.~A. Hagerty \emph{et~al.}, ``Recycling ambient microwave energy with
broad-{b}and rectenna arrays,'' \emph{IEEE Trans. Microw. Theory Techn.},
vol.~52, no.~3, pp. 1014--1024, Mar. 2004.

\bibitem{2013TMTT_EH_AmbientEH}
M.~Pi{\~n}uela, P.~D. Mitcheson, and S.~Lucyszyn, ``Ambient {RF} energy
harvesting in urban and semi-urban environments,'' \emph{IEEE Trans. Microw.
	Theory Techn.}, vol.~61, no.~7, pp. 2715--2726, Jul. 2013.

\bibitem{2013_ProcIEEE_RF_WSN}
H.~J. {Visser} and R.~J.~M. {Vullers}, ``{RF} energy harvesting and transport
for wireless sensor network applications: Principles and requirements,''
\emph{Proc. IEEE}, vol. 101, no.~6, pp. 1410--1423, June 2013.

\bibitem{2017_TOC_WPT_YZeng_Bruno_RZhang}
Y.~{Zeng}, B.~{Clerckx}, and R.~{Zhang}, ``Communications and signals design
for wireless power transmission,'' \emph{IEEE Trans. Commun.}, vol.~65,
no.~5, pp. 2264--2290, May 2017.

\bibitem{2009_IntConfRFID}
M.~S. {Trotter}, J.~D. {Griffin}, and G.~D. {Durgin}, ``Power-optimized
waveforms for improving the range and reliability of {RFID} systems,'' in
\emph{IEEE Int. Conf. RFID}, April 2009, pp. 80--87.

\bibitem{2011_IMS_Multisine}
A.~S. {Boaventura} and N.~B. {Carvalho}, ``Maximizing {DC} power in energy
harvesting circuits using multisine excitation,'' in \emph{IEEE MTT-S Int.
	Microw. Symp.}, June 2011, pp. 1--4.

\bibitem{2014_MWCL_WPT_Optimal_Waveform}
A.~{Collado} and A.~{Georgiadis}, ``Optimal waveforms for efficient wireless
power transmission,'' \emph{IEEE Microw. Wireless Compon. Lett.}, vol.~24,
no.~5, pp. 354--356, May 2014.

\bibitem{2016_TSP_WPT_Bruno_Waveform}
B.~{Clerckx} and E.~{Bayguzina}, ``Waveform design for wireless power
transfer,'' \emph{IEEE Trans. Signal Process.}, vol.~64, no.~23, pp.
6313--6328, Dec 2016.

\bibitem{2017_TSP_WPT_Bruno_Yang_Large}
Y.~{Huang} and B.~{Clerckx}, ``Large-scale multiantenna multisine wireless
power transfer,'' \emph{IEEE Trans. Signal Process.}, vol.~65, no.~21, pp.
5812--5827, Nov 2017.

\bibitem{2017_AWPL_WPT_Bruno_LowComplexity}
B.~{Clerckx} and E.~{Bayguzina}, ``Low-complexity adaptive multisine waveform
design for wireless power transfer,'' \emph{IEEE Antennas Wireless Propag.
	Lett.}, vol.~16, pp. 2207--2210, 2017.

\bibitem{2017_IEEE_SPAWC_Waveform}
M.~R.~V. {Moghadam}, Y.~{Zeng}, and R.~{Zhang}, ``Waveform optimization for
radio-frequency wireless power transfer,'' in \emph{IEEE Int. Workshop Signal
	Process. Adv. Wireless Commun.}, July 2017, pp. 1--6.

\bibitem{2018_TWC_WPT_Bruno_HYang_LimitedFeedback}
Y.~{Huang} and B.~{Clerckx}, ``Waveform design for wireless power transfer with
limited feedback,'' \emph{IEEE Trans. Wireless Commun.}, vol.~17, no.~1, pp.
415--429, Jan 2018.

\bibitem{2019_JSAC_Waveform_MultipleDevice}
K.~{Kim}, H.~{Lee}, and J.~{Lee}, ``Waveform design for fair wireless power
transfer with multiple energy harvesting devices,'' \emph{IEEE J. Sel. Areas
	Commun.}, vol.~37, no.~1, pp. 34--47, Jan 2019.

\bibitem{2018_TSP_WIPT_Bruno_WIPT}
B.~{Clerckx}, ``Wireless information and power transfer: Nonlinearity, waveform
design, and rate-energy tradeoff,'' \emph{IEEE Trans. Signal Process.},
vol.~66, no.~4, pp. 847--862, Feb 2018.

\bibitem{2019_TOC_SWIPT_nonZeroInput}
M.~{Varasteh}, B.~{Rassouli}, and B.~{Clerckx}, ``{SWIPT} signalling over
frequency-selective channels with a nonlinear energy harvester: Non-zero mean
and asymmetric inputs,'' \emph{IEEE Trans. Commun.}, pp. 1--1, 2019.

\bibitem{2019_TWC_WIPT_AsymmetricModulation}
E.~{Bayguzina} and B.~{Clerckx}, ``Asymmetric modulation design for wireless
information and power transfer with nonlinear energy harvesting,'' \emph{IEEE
	Trans. Wireless Commun.}, pp. 1--1, 2019.

\bibitem{2017_CL_WPBackscatterComm}
B.~{Clerckx}, Z.~{Bayani Zawawi}, and K.~{Huang}, ``Wirelessly powered
backscatter communications: Waveform design and {SNR}-energy tradeoff,''
\emph{IEEE Communications Letters}, vol.~21, no.~10, pp. 2234--2237, Oct
2017.

\bibitem{2019_TWC_MuWPScatteredComm}
Z.~B. {Zawawi}, Y.~{Huang}, and B.~{Clerckx}, ``Multiuser wirelessly powered
backscatter communications: Nonlinearity, waveform design, and {SINR}-energy
tradeoff,'' \emph{IEEE Trans. Wireless Commun.}, vol.~18, no.~1, pp.
241--253, Jan 2019.

\bibitem{ShanpuShen2016_TAP_Impedancematching}
S.~Shen and R.~D. Murch, ``Impedance matching for compact multiple antenna
systems in random {RF} fields,'' \emph{IEEE Trans. Antennas Propag.},
vol.~64, no.~2, pp. 820--825, Feb. 2016.

\bibitem{ShanpuShen2017_AWPL_DPTB}
S.~Shen, C.~Y. Chiu, and R.~D. Murch, ``A dual-port triple-band {L}-probe
microstrip patch rectenna for ambient {RF} energy harvesting,'' \emph{IEEE
	Antennas Wireless Propag. Lett.}, vol.~16, pp. 3071--3074, 2017.

\bibitem{ShanpuShen2018_EuCap_QPDP}
S.~{Shen}, Y.~{Zhang}, C.~{Chiu}, and R.~D. {Murch}, ``A compact quad-port
dual-polarized dipole rectenna for ambient {RF} energy harvesting,'' in
\emph{2018 12th European Conference on Antennas and Propagation}, London,
United Kingdom, Apr. 2018.

\bibitem{ShanpuShen2017_TAP_EHPIXEL}
S.~Shen, C.~Y. Chiu, and R.~D. Murch, ``Multiport pixel rectenna for ambient
{RF} energy harvesting,'' \emph{IEEE Trans. Antennas Propag.}, vol.~66,
no.~2, pp. 644--656, Feb. 2018.

\bibitem{ShanpuShen2019_TMTT_Freqdepend}
S.~{Shen}, Y.~{Zhang}, C.~{Chiu}, and R.~{Murch}, ``An ambient {RF} energy
harvesting system where the number of antenna ports is dependent on
frequency,'' \emph{IEEE Trans. Microw. Theory Tech.}, vol.~67, no.~9, pp.
3821--3832, Sep. 2019.

\bibitem{2011AWPL_EH_InvestRectArray}
U.~Olgun, C.-C. Chen, and J.~L. Volakis, ``Investigation of rectenna array
configurations for enhanced {RF} power harvesting,'' \emph{IEEE Antennas
	Wireless Propag. Lett.}, vol.~10, pp. 262--265, 2011.

\bibitem{2016_TWC_RZhang_MIMOWPT_Limitedfeed}
J.~{Xu} and R.~{Zhang}, ``A general design framework for {MIMO} wireless energy
transfer with limited feedback,'' \emph{IEEE Trans. Signal Process.},
vol.~64, no.~10, pp. 2475--2488, May 2016.

\bibitem{2019_SPL_MIMO_WPT}
G.~{Ma}, J.~{Xu}, Y.~{Zeng}, and M.~R.~V. {Moghadam}, ``A generic receiver
architecture for {MIMO} wireless power transfer with nonlinear energy
harvesting,'' \emph{IEEE Signal Processing Letters}, vol.~26, no.~2, pp.
312--316, Feb 2019.

\bibitem{2019_JSAC_WIPT_Bruno_RZhang_RSchober_DIKim_HVPoor}
B.~{Clerckx}, R.~{Zhang}, R.~{Schober}, D.~W.~K. {Ng}, D.~I. {Kim}, and H.~V.
{Poor}, ``Fundamentals of wireless information and power transfer: From {RF}
energy harvester models to signal and system designs,'' \emph{IEEE J. Sel.
	Areas Commun.}, vol.~37, no.~1, pp. 4--33, Jan 2019.

\bibitem{2013_TWC_SWIPT_RZhang}
R.~{Zhang} and C.~K. {Ho}, ``{MIMO} broadcasting for simultaneous wireless
information and power transfer,'' \emph{IEEE Trans. Wireless Commun.},
vol.~12, no.~5, pp. 1989--2001, May 2013.

\bibitem{2018_TWC_WPT_Bruno_Transmit_Diversity}
B.~{Clerckx} and J.~{Kim}, ``On the beneficial roles of fading and transmit
diversity in wireless power transfer with nonlinear energy harvesting,''
\emph{IEEE Trans. Wireless Commun.}, vol.~17, no.~11, pp. 7731--7743, Nov
2018.

\bibitem{2019_Junghoon_Prototyping}
J.~{Kim}, B.~{Clerckx}, and P.~D. {Mitcheson}, ``Signal and system design for
wireless power transfer : Prototype, experiment and validation,'' \emph{IEEE
	Trans. Wireless Commun.}, pp. 1--1, 2020.

\bibitem{KKKKI}
Y.~H. Lam, W.~H. Ki, and C.~Y. Tsui, ``Single inductor multiple-input
multiple-output switching converter and method of use,'' Aug.~14 2007, {US}
Patent 7, 256, 568.

\bibitem{MIMO_Convertor_ISSCC}
S.~S. Amin and P.~P. Mercier, ``{MISIMO}: A multi-input single-inductor
multi-output energy harvester employing event-driven {MPPT} control to
achieve 89\% peak efficiency and a 60000x dynamic range in 28nm {FDSOI},'' in
\emph{2018 IEEE International Solid State Circuits Conference (ISSCC)}, Feb
2018, pp. 144--146.

\bibitem{MIMO_Convertor_JSSC}
C.~Liu, H.~Lee, P.~Liao, Y.~Chen, M.~Chung, and P.~Chen, ``Dual-source
energy-harvesting interface with cycle-by-cycle source tracking and adaptive
peak-inductor-current control,'' \emph{IEEE Journal of Solid-State Circuits},
pp. 1--10, Oct. 2018.

\bibitem{2010_SPMag_SDR}
Z.~{Luo}, W.~{Ma}, A.~M. {So}, Y.~{Ye}, and S.~{Zhang}, ``Semidefinite
relaxation of quadratic optimization problems,'' \emph{IEEE Signal Process.
	Mag.}, vol.~27, no.~3, pp. 20--34, May 2010.

\bibitem{grant2008cvx}
M.~Grant, S.~Boyd, and Y.~Ye, ``{CVX}: {MATLAB} software for disciplined convex
programming,'' 2008.

\bibitem{chiang2005geometric_programming}
M.~Chiang \emph{et~al.}, ``Geometric programming for communication systems,''
\emph{Foundations and Trends{\textregistered} in Communications and
	Information Theory}, vol.~2, no. 1--2, pp. 1--154, 2005.

\bibitem{Applied_Geometric_Programming}
C.~S. {Beightler} and D.~T. {Philips}, \emph{Applied Geometric
	Programming}.\hskip 1em plus 0.5em minus 0.4em\relax Wiley, 1976.

\bibitem{OperRes_1978_InnerApprox}
B.~R. Marks and G.~P. Wright, ``A general inner approximation algorithm for
nonconvex mathematical programs,'' \emph{Operations research}, vol.~26,
no.~4, pp. 681--683, 1978.

\bibitem{2007_TWC_GP_Daniel}
M.~{Chiang}, C.~W. {Tan}, D.~P. {Palomar}, D.~{O'neill}, and D.~{Julian},
``Power control by geometric programming,'' \emph{IEEE Trans. Wireless
	Commun.}, vol.~6, no.~7, pp. 2640--2651, July 2007.

\bibitem{ReconfiguarablePowerCombiner}
A.~Ocera, R.~V. Gatti, P.~Mezzanotte, P.~Farinelli, and R.~Sorrentino, ``A mems
programmable power divider/combiner for reconfigurable antenna systems,'' in
\emph{2005 European Microwave Conference}, vol.~1.\hskip 1em plus 0.5em minus
0.4em\relax IEEE, 2005, pp. 4--pp.

\bibitem{2012_TSP_Nonsmooth}
A.~H. {Phan}, H.~D. {Tuan}, H.~H. {Kha}, and D.~T. {Ngo}, ``Nonsmooth
optimization for efficient beamforming in cognitive radio multicast
transmission,'' \emph{IEEE Trans. Signal Process.}, vol.~60, no.~6, pp.
2941--2951, 2012.

\bibitem{2018_MM_WPT_Bruno_1GWPT}
B.~{Clerckx}, A.~{Costanzo}, A.~{Georgiadis}, and N.~{Borges Carvalho},
``Toward {1G} mobile power networks: {RF}, signal, and system designs to make
smart objects autonomous,'' \emph{IEEE Microw. Mag.}, vol.~19, no.~6, pp.
69--82, Sep. 2018.

\bibitem{2015_ComMag_WirelessPowerComm}
S.~{Bi}, C.~K. {Ho}, and R.~{Zhang}, ``Wireless powered communication:
opportunities and challenges,'' \emph{IEEE Communications Magazine}, vol.~53,
no.~4, pp. 117--125, April 2015.

\end{thebibliography}
\end{document}